\documentclass[review]{elsarticle}
\pdfoutput=1
\usepackage{lineno,hyperref}
\usepackage{graphicx}
\usepackage[linesnumbered,ruled,lined]{algorithm2e}
\usepackage{geometry}
\usepackage{float}
\usepackage{tikz}
\usepackage{epstopdf}
\usepackage{amssymb}
\usepackage{amsmath}
\usepackage{subcaption}
\usepackage{footnote}
\usepackage{array}
\usepackage{cases}
\usepackage{multirow}
\usepackage{color}
\usepackage{hyperref}
\usepackage{bm}
\graphicspath{ {./Figure/} }
{\color{red}\journal{Elsevier}}
\makeatletter
\input pdf-trans
\newbox\qbox
\def\usecolor#1{\csname\string\color@#1\endcsname\space}
\newcommand\bordercolor[1]{\colsplit{1}{#1}}
\newcommand\fillcolor[1]{\colsplit{0}{#1}}
\newcommand\outline[1]{\leavevmode%
  \def\maltext{#1}%
  \setbox\qbox=\hbox{\maltext}%
  \boxgs{Q q 2 Tr \thickness\space w \fillcol\space \bordercol\space}{}%
  \copy\qbox%
}
\makeatother
\newcommand\colsplit[2]{\colorlet{tmpcolor}{#2}\edef\tmp{\usecolor{tmpcolor}}%
  \def\tmpB{}\expandafter\colsplithelp\tmp\relax%
  \ifnum0=#1\relax\edef\fillcol{\tmpB}\else\edef\bordercol{\tmpC}\fi}
\def\colsplithelp#1#2 #3\relax{%
  \edef\tmpB{\tmpB#1#2 }%
  \ifnum `#1>`9\relax\def\tmpC{#3}\else\colsplithelp#3\relax\fi
}
\bordercolor{black}
\fillcolor{white}
\def\thickness{.3}

\begin{document}
\begin{frontmatter}
		\title{Toward Efficient FSI Modeling in Patient-Specific Arteries: SPH Simulation 
        of Blood Flow in Thin Deformable Vessels}
		\author[myfirstaddress]{Chenxi Zhao}
		\ead{chenxi.zhao@tum.de}
            \author[myfirstaddress]{Dong Wu}
		\ead{dong.wu@tum.de}
            \author[mysecondaddress]{Weiyi Kong}
		\ead{kong@virtonomy.io}
		\author[myfirstaddress]{Oskar J. Haidn}
		\ead{oskar.haidn@tum.de}
		\author[myfirstaddress]{Xiangyu Hu\corref{mycorrespondingauthor}}
		\ead{xiangyu.hu@tum.de}
		\address[myfirstaddress]{School of Engineering and Design, 
			Technical University of Munich, 85748 Garching, Germany}
            \address[mysecondaddress]{Virtonomy GmbH, 80336 Munich, Germany}
		\cortext[mycorrespondingauthor]{Corresponding author. }

\begin{abstract}

Accurate simulation of blood flow in deformable vessels is critical in cardiovascular research for 
understanding disease progression and informing clinical decision-making. However, due to the 
thin-walled nature of arteries, traditional smoothed particle hydrodynamics (SPH) approaches based 
on full-dimensional volume modeling often require extremely fine particle spacing to ensure numerical 
convergence for the solid mechanics. This, in turn, leads to redundant resolution in the fluid domain 
to maintain sufficient kernel support near the fluid-solid interface in fluid-structure interaction 
(FSI) simulations.

To address this limitation, we propose an efficient reduced-dimensional shell-based SPH method for 
modeling thin-walled deformable arteries, and conduct FSI for capturing hemodynamics and arterial 
wall mechanics. Through a series of validation cases, the proposed shell model demonstrates comparable 
accuracy in fluid dynamics to the volume model, while achieving faster convergence in solid mechanics 
and reduced computational cost. We further investigate the influence of wall compliance on flow 
transitions and key hemodynamic indices, highlighting the necessity of FSI modeling over rigid-wall 
assumptions. Finally, the method is applied to two patient-specific vascular geometries, i.e. the 
carotid artery and the aorta, which demonstrates its robustness, efficiency and physiological 
relevance in realistic cardiovascular simulations.
\end{abstract}

\begin{keyword}
smoothed particle hydrodynamics (SPH), thin-walled vessels, shell modeling, hemodynamics, 
fluid-structure interaction (FSI)
\end{keyword}

\end{frontmatter}

\section{Introduction} \label{section: introduction}
Cardiovascular diseases remain the leading cause of mortality worldwide as highlighted by the World 
Health Organization. In recent years, numerical simulations have emerged as powerful tools for 
analyzing hemodynamics and vessel deformations. Compared to experimental approaches, numerical methods 
offer faster predictions, non-invasive evaluation capabilities, and the flexibility to explore a wide 
range of physiological and pathological scenarios. These advantages make computational modeling 
particularly valuable for clinical risk assessment and surgical planning \cite{schwarz2023beyond}. 

A significant amount of research has focused on simulating blood flow within vessels with rigid walls, 
showcasing the applicability of modern computational fluid dynamics (CFD) techniques in 
patient-specific hemodynamic studies. For example, Kaid et al. \cite{kaid2024unveiling} employed 
COMSOL Multiphysics with the finite element method (FEM) to investigate wall shear stress (WSS) 
distributions and other hemodynamic factors in the carotid artery under normal and stenotic conditions. 
They also analyzed the influence of Reynolds number, Womersley number, and arterial geometry on 
flow disruption and stagnation points. Additionally, Laha et al. \cite{laha2024smoothed} 
demonstrated the potential of the smoothed particle hydrodynamics (SPH) method in predicting 
hazards associated with mechanical heart valves within rigid vessels. 
Deyranlou et al. \cite{deyranlou2020numerical} conducted a parametric study using ANSYS CFX with the 
finite volume method (FVM) to evaluate the impact of atrial fibrillation traits on aortic flow. 
Similarly, Singhal et al. \cite{singhal2024hemodynamics} employed ANSYS Fluent with FVM to study the 
left coronary artery, demonstrating that the presence of the ramus intermedius may contribute to 
plaque development in the furcation region and proximal parts of the left anterior descending artery. 
In addition, Djukic et al. \cite{djukic2023validation} compared the Lattice Boltzmann method (LBM) 
with FEM and SPH, revealing the ability of LBM to deliver fast and accurate results for 
patient-specific coronary artery simulations. 

In addition to rigid-wall assumptions, several studies have also investigated the effects of vessel 
wall properties on blood flow parameters \cite{figueroa2006coupled, long2012fluid, reymond2013physiological, roy2024does}.
For instance, Figueroa et al. \cite{figueroa2006coupled} demonstrated significant differences in 
pressure and flow waveforms between rigid and deformable vessel wall solutions, noting a phase lag 
between inlet and outlet flow in vessels with deformable walls. Roy et al. \cite{roy2024does} 
reported that arterial wall and plaque mechanics substantially influence hemodynamic indices such 
as time-averaged wall shear stress (TAWSS), oscillatory shear index (OSI), and fractional flow 
reserve (FFR). Similarly, Brown et al. \cite{brown2012accuracy} reported that the rigid wall approximation 
over-predicts WSS compared to fluid-structure interaction (FSI) models. 
Accounting for wall deformability is crucial for understanding disease progression, such as 
atherosclerosis and aneurysm formation, and predicting the outcomes of medical interventions 
like stenting or bypass surgery. Current mesh-based methods for simulating blood flow in deformable 
vessels can be generally categorized into two main approaches: (1) frequent updates to the fluid and 
structural mesh geometry using formulations such as the Arbitrary Lagrangian-Eulerian (ALE) method, 
which is commonly adopted in the open-source and commercial cardiovascular software 
(lifex-cfd \cite{africa2024lifex}, SimVascular \cite{updegrove2017simvascular} and 
Crimson \cite{arthurs2021crimson}); and (2) direct incorporation of vessel wall boundary effects 
into fluid equations, such as in the coupled momentum method 
(CMM) \cite{figueroa2006coupled, kung2011vitro}. Although the ALE method yields accurate results, 
frequent mesh updates increase computational costs. Methods like CMM struggle to the precision of 
large deformable geometries, limiting its applicability \cite{figueroa2006coupled}. 
On the other hand, mesh-free methods, such as the SPH approach, have gained attention in cardiovascular 
problems in recent years, primarily due to their ability to handle fluid-structure interfaces without 
the need for explicit interface-tracking techniques. For example, Lu et al. \cite{lu2024gpu} developed 
a GPU-accelerated FSI framework that combines incompressible SPH (ISPH) for fluid dynamics with total 
Lagrangian SPH (TLSPH) for solid mechanics. Their method successfully captured blood flow in vessels 
and demonstrated good agreement with ALE-based FSI results from SimVascular. Also, despite the FEM has 
been widely validated for stress and strain analysis in structural mechanics, a pure SPH-based FSI 
framework offers the advantage of strong coupling, thereby eliminating potential data transfer errors 
between separate fluid and solid solvers, which is an issue commonly arising in SPH-FEM 
hybrid approaches.

The SPH method has demonstrated notable success in FSI applications across various fields, including 
ocean engineering \cite{sun2021accurate}, aerospace \cite{oger2020simulations}, and others. 
Specifically, the volume model, as a fully dimensional representation in traditional SPH for solid domain, 
has been extensively adopted and validated. However, achieving numerical convergence with this model 
necessitates multiple layers of particles through the thickness direction. This requirement leads to 
very fine particle spacing in the thin structures like blood vessel walls, resulting in substantially 
increased memory consumption and computational cost. This issue becomes even more pronounced in FSI 
simulations. In addition to the structural domain requiring high spatial resolution, the adjacent fluid 
domain must also be finely discretized to ensure sufficient kernel support for fluid particles near the 
fluid-solid interface, even if such high resolution is not essential for capturing the fluid dynamics 
itself. This redundant resolution introduces computational inefficiencies and renders the entire 
simulation more expensive. To address this limitation, thin-walled structures can be modeled more 
efficiently using a reduced-dimensional SPH shell model, which represents the wall with a single layer 
of particles with the physical wall thickness explicitly assigned in the formulation. This approach 
maintains physical fidelity while significantly reducing the total particle count and computational 
load, as demonstrated in recent studies \cite{wu2024sph, tang2024simulating}. In the context of FSI, 
the shell model decouples fluid resolution from wall thickness, thus enhancing computational efficiency 
without sacrificing accuracy, which will be illustrated in the following case study. 
Moreover, Ref.\cite{djukic2023validation} emphasizes that SPH often involves complex model generation 
processes, including the creation of template particles and the implementation of activation and 
deactivation planes. In this paper, we address these challenges by introducing an easy particle 
generation approach for fluid and solid domains, and the injection and deletion methods of particles 
will also be presented. 

In this study, we carried out a comprehensive process for SPH-based simulations of blood flow in 
vessels using SPHinXsys 
(an open-source library, \href{https://github.com/Xiangyu-Hu/SPHinXsys}{https://github.com/Xiangyu-Hu/SPHinXsys}). 
The remainder of this paper is organized as follows: Section \ref{section: Methodology} outlines the 
numerical methodology adopted in this work. In particular, 
Section \ref{subsection: Particle generation process} introduces a generalized particle generation 
approach for both fluid and solid domains, directly constructed from available standard triangle 
language (STL) or visualization toolkit (VTP) files. 
Section \ref{subsection: Governing equations and SPH discretization} presents the governing equations 
for FSI and their corresponding SPH discretizations. 
Section \ref{subsection: In-/outlet boundary condition implementation} introduces the implementation 
of in-/outlet boundary conditions. Further, the calculated results are shown and analyzed in 
Section \ref{section: Results and discussion}. Detailedly, the accuracy of the imposed boundary 
condition and the shell model are first validated. Then, the influence of wall deformability on 
hemodynamic behavior is investigated through comparisons between rigid and deformable shell models. 
Finally, two patient-specific vascular cases, i.e. the carotid artery and the aorta, are simulated 
under physiologically relevant conditions. These cases demonstrate the versatility, accuracy, and 
computational efficiency of the proposed shell-based SPH framework for modeling complex FSI phenomena 
in thin-walled, deformable vessels. Additionally, the paper concludes with a summary of our findings 
in the last section. 

\section{Methodology} \label{section: Methodology}
\subsection{Particle generation process} \label{subsection: Particle generation process}
\subsubsection{Particle generation for fluid body}
The geometry of cardiovascular structures is commonly provided in STL format, which is widely 
accessible online. Additionally, the Vascular Model Repository 
\cite{wilson2013vascular} (\href{https://www.vascularmodel.com}{https://www.vascularmodel.com}) 
offers approximately 300 computational models of normal and diseased cardiovascular geometries in VTP 
format, which are compatible with SimVascular software. The particle generation method for fluid domain 
defined by closed triangle mesh basically follows the principle of CAD-compatible body-fitted particle 
generator for arbitrarily complex geometry, as described in Ref.\cite{zhu2021cad}. 

The process begins with the construction of a initial lattice particle distribution. This is 
followed by a physics-driven relaxation procedure governed by the transport velocity equation:
\begin{equation}\label{eq: transport-velocity}
    \frac{\mathrm{d} \mathbf{v}}{\mathrm{d}t} = \mathbf{F}_p,
\end{equation}
where the $\mathbf{v}$ represents the advection velocity, and the $\mathbf{F}_p$ denotes the acceleration 
induced by the repulsive pressure force. This force is achieved by applying a constant background pressure 
to ensure an isotropic particle distribution:
\begin{equation}\label{eq: momentum-sph}
    \mathbf{F}_{p,i} = -\frac{2 p^{0} V_i}{m_i}\sum_j \nabla_i W_{ij} V_j,
\end{equation}
Here, $m$ is the particle mass, $V$ is the particle volume, $p^0 = 1$ is the constant background pressure, 
and $\nabla_i W_{ij}$ denotes the gradient of the kernel function 
$W(|\mathbf{r}_{ij}|, h)$ with respect to particle $i$. 
The terms $\mathbf{r}_{ij} = \mathbf{r}_{i} - \mathbf{r}_{j}$ and $h$ refer to the relative position vector 
and the smoothing length, respectively.

To achieve a body-fitted particle distribution, a surface bounding method is applied. This ensures that 
surface particles are positioned such that their centers lie 0.5 times the particle spacing inside the 
geometric boundary surface, thereby ensuring geometric conformity and boundary accuracy.

\subsubsection{Particle generation for solid wall by volume and shell models}
To construct vessel wall geometries from these existing STL/VTP blood flow files, a typical approach 
involves suturing the triangular surfaces and extending the integral surface with a specified thickness in 
the 3D design software. However, this process is challenging and may result in suboptimal wall geometry 
quality. To address this challenge, we generate solid wall particles with the thickness property directly 
within the SPH framework using the input blood flow geometry files.

\begin{figure}[htbp]
    \centering
    \includegraphics[width=15cm]{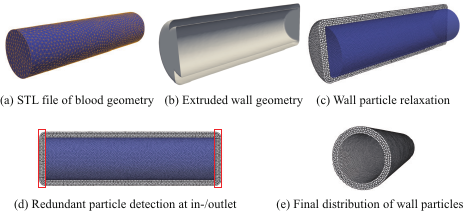}
    \caption{Illustration of wall particle generation by volume model.}
    \label{fig: volume-particle-distribution}
\end{figure}

\begin{figure}[htbp]
    \centering
    \includegraphics[width=15cm]{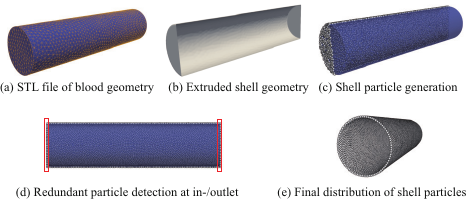}
    \caption{Illustration of wall particle generation by shell model.}
    \label{fig: shell-particle-distribution}
\end{figure}

The vessel wall geometry is constructed using an extrusion technique based on the STL triangle mesh of the 
blood flow geometry, as shown in Fig.\ref{fig: volume-particle-distribution} (a-b) and 
Fig.\ref{fig: shell-particle-distribution} (a-b) for volume and shell models, respectively. In the volume 
model, the extrusion value equals the physical wall thickness, whereas in the shell model, the extrusion 
corresponds to half the shell particle spacing. This results in a fully enclosed wall structure, 
including sealed inlet and outlet surfaces that require post-processing.

For the volume-based model, a lattice distribution of particles is initially generated within the extruded 
wall volume. Particle positions are then relaxed using a physics-driven relaxation procedure combined with 
surface bounding \cite{zhu2021cad} in Fig.\ref{fig: volume-particle-distribution} (c). Subsequently, 
redundant particles located at the in-/outlet are detected and removed, as illustrated in 
Fig.\ref{fig: volume-particle-distribution} (d). Fig.\ref{fig: volume-particle-distribution} (e) is then 
employed in the simulation with volume model as the wall representation.

For the shell-based model, physical wall thickness is assigned directly in the formulation, and the volume 
of each shell particle is defined as the product of the square of the particle spacing and the wall 
thickness. The total number of particles to be distributed over the surface is estimated as:
\begin{equation}\label{eq: shell-particle-num}
    N = \lceil \frac{A_{\text{TS}}}{(dp^0)^2} \rceil,
\end{equation}
where $A_{\text{TS}}$ represents the total surface area of the input STL or VTP geometry, and $dp^0$ is 
the initial particle spacing. The number of particles allocated to each triangular face is proportional to 
its area. If the total number of mesh faces exceeds the intended particle count, a random sampling policy 
is applied to select a subset of faces for particle generation. Within each selected face, the particle 
positions are evenly distributed relative to the triangle vertices, as depicted in 
Fig.\ref{fig: shell-particle-distribution} (c). These initial positions are then relaxed through a 
physics-based relaxation process, with a surface-specific bounding strategy. Unlike relaxation for 
volume model, where particles are placed at a fixed $0.5dp^0$ offset from the bounds, surface relaxation 
constrains shell particles to lie directly on the mesh surface, using nearest-point projection. 
Additionally, particle normals are smoothed within their support domain using the weighted averaging 
technique \cite{wulevel}:
\begin{equation}\label{eq: smoothed-normal-direction}
    \hat{\mathbf{n}_i} = \frac{W^0 \mathbf{n}_i + \sum_j W_{ij}\mathbf{n}_j}{W^0 + \sum_j W_{ij}}, 
\end{equation}
where $W^0 = W(\mathbf{0}, h)$ denotes the kernel weight at zero distance.
After relaxation, inlet and outlet particles are removed to open the shell boundaries, as shown in 
Fig.\ref{fig: shell-particle-distribution} (d). The final shell particle configuration is presented in 
Fig.\ref{fig: shell-particle-distribution} (e).

\subsection{Governing equations and SPH discretizations} \label{subsection: Governing equations and SPH discretization}
\subsubsection{Fluid dynamics method based on Riemann solver}  \label{subsubsection: Fluid dynamics method based on Riemann solver}
In this study, blood is modeled as a weakly compressible Newtonian viscous fluid. The governing equations 
for mass and momentum conservation are expressed as
\begin{equation}\label{eq: mass}
    \frac{\mathrm{d}\rho}{\mathrm{d}t} = -\rho \nabla \cdot \mathbf{v},
\end{equation}
\begin{equation}\label{eq: momentum}
    \frac{\mathrm{d}\mathbf{v}}{\mathrm{d}t} = \frac{1}{\rho} (-\nabla p + \eta \nabla^2 \mathbf{v}) + \mathbf{f},
\end{equation}
where $\rho$, $\mathbf{v}$, $p$, and $\eta$ are the fluid density, velocity, pressure, and dynamic 
viscosity, respectively. $\mathbf{f}$ represents the body force term. In the weakly compressible SPH 
(WCSPH) scheme, the pressure is computed via an artificial equation of state (EoS):
\begin{equation}\label{eq: eos}
    p = c_f^2 (\rho - \rho^0),
\end{equation}
where $c_f = 10|\mathbf{v}|_{\rm max}$ is the numerical sound speed to satisfy the weakly compressible assumption where 
the density variation remains around 1\%, and the superscript $(\bullet)^0$ is the reference value in the 
initial configuration. 

The SPH discretization of continuity and momentum equations with a low-dissipation Riemann solver for the 
blood flow can be written as
\begin{equation}\label{eq: continuity-sph}
    \frac{\mathrm{d}\rho_i}{\mathrm{d}t} = 2\rho_i\sum_{j}(\mathbf{v}_i - 
    \mathbf{v}^\ast)\cdot\nabla_iW_{ij}V_j,
\end{equation}
\begin{equation}\label{eq: momentum-sph}
    \frac{\mathrm{d}\mathbf{v}_i}{\mathrm{d}t} = -2\sum_{j} \frac{p^\ast}{\rho_i}\nabla_iW_{ij}V_j + 
    2\sum_{j}\frac{\eta_{ij}}{\rho_i}
    \frac{\mathbf{v}_{ij}}{r_{ij}}\frac{\partial W_{ij}}{\partial r_{ij}}V_j + \mathbf{f}_i.
\end{equation}
Here, $\nabla_iW_{ij} = (\partial W_{ij} / \partial r_{ij}) \mathbf{e}_{ij}$ 
and the direction vector $\mathbf{e}_{ij} = \mathbf{r}_{ij} / r_{ij}$. 
The intermediate velocity $\mathbf{v}^\ast$ and pressure $p^\ast$ are obtained by solving the Riemann 
problem constructed along the interacting line of each pair of particles \cite{zhang2017weakly}, 
with left ($L$) and right ($R$) states:
\begin{equation}\label{eq: wcsph-riemann}
\left\{\begin{aligned}
    & (\rho_L, U_L, p_L) = (\rho_i, \mathbf{v}_i \cdot \mathbf{e}_{ij}, p_i), \\
    & (\rho_R, U_R, p_R) = (\rho_j, \mathbf{v}_j \cdot \mathbf{e}_{ij}, p_j), \\
\end{aligned}\right.
\end{equation}
where $\mathbf{e}_{ij}$ is the unit vector connecting particles $i$ and $j$. The intermediate states, 
under the assumptions $U^\ast = U_L^\ast = U_R^\ast$ and $p^\ast = p_L^\ast = p_R^\ast$, are computed as
\begin{equation}\label{eq: velocity-pressure-riemann}
\left\{\begin{aligned}
    & U^\ast = \overline{U} + \frac{p_L - p_R}{c(\rho_L+\rho_R)}, \\
    & p^\ast = \overline{p} + \frac{\rho_L\rho_R\beta(U_L-U_R)}{\rho_L+\rho_R}, \\
\end{aligned}\right.
\end{equation}
where $\overline{U}$ and $\overline{p}$ are averages of velocity and pressure, 
and $\beta = {\rm min}(3\max(U_L-U_R, 0), c_f)$ is a dissipation limiter proposed in Ref.\cite{zhang2017weakly}. 
Then the intermediate velocity vector $\mathbf{v}^\ast$ in 
Eq.(\ref{eq: continuity-sph}) is reconstructed by 
$\mathbf{v}^\ast = U^\ast\mathbf{e}_{ij} + (\overline{\mathbf{v}}_{ij} - \overline{U}\mathbf{e}_{ij})$, 
and $\overline{\mathbf{v}}_{ij} = (\rho_i\mathbf{v}_i + \rho_j\mathbf{v}_j) / (\rho_i+\rho_j)$.

According to the latest reverse kernel gradient correction (RKGC) 
method \cite{zhang2025towards, zhang2025corrected} adopted in SPHinXsys, which is conservative and ensures 
the zero- and first-order consistencies, the particle-pair average term in the Riemann solution in the 
momentum equation is modified as
\begin{equation}\label{eq: pij-RKGC}
    \overline{p}_{ij} \Rightarrow \overline{p\mathbb{B}}_{ij} = \frac{1}{2}(p_i\mathbb{B}_j+p_j\mathbb{B}_i),
\end{equation}
where $\mathbb{B}_i = 
(-\sum_{j}\mathbf{r}_{ij}\otimes \nabla_iW_{ij}V_j)^{-1}$.

In addition, to mitigate particle clumping and void regions in the SPH method, the transport velocity 
formulation (TVF) \cite{adami2013transport, zhang2017generalized} is applied. The particle positions are 
updated using
\begin{equation}\label{eq: transport-v}
    \frac{\mathrm{d}\mathbf{r}_i}{\mathrm{d}t} = \widetilde{\mathbf{v}}_i,
\end{equation}
where $\widetilde{\mathbf{v}}_i$ is the advection velocity. Recent work by Zhang et al. 
\cite{zhang2025towards} simplifies the displacement correction as
\begin{equation}\label{eq: transport-v}
     \Delta \widetilde{\mathbf{r}}_i = 0.2h^2\nabla_iW_{ij}V_j,
\end{equation}
which is applied in combination with the momentum velocity to correct zero-order integration 
errors by adjusting particle positions.

\subsubsection{Solid dynamics with volume model}
For solid mechanics with full-dimensional volume model, the total Lagrangian formulation is employed. 
The mass and momentum conservation equations are
\begin{equation}\label{eq: solid-density}
    \rho = \rho^0 \frac{1}{\rm det(\mathbb{F})},
\end{equation}
\begin{equation}
     \frac{\mathrm{d}\mathbf{v}}{\mathrm{d}t} = \frac{1}{\rho^0}\nabla^0\cdot\mathbb{P}^\mathrm{T} + 
     \mathbf{f}^{f}.
\end{equation}
Here, the force term $\mathbf{f}^{f} = \mathbf{f}^{f:p} + \mathbf{f}^{f:v}$ includes both pressure and 
viscous contributions from fluid forces. $\mathbb{F}$ is the deformation tensor, and $\mathbb{P} = \mathbb{F}\mathbb{S}$ is the first 
Piola-Kirchhoff stress tensor, with $\mathbb{S}$ being the second Piola-Kirchhoff stress tensor. 
For a linearly elastic and isotropic material, the second Piola-Kirchhoff stress tensor $\mathbb{S}$ is defined as
\begin{equation}\label{eq: 2nd-stress}
\begin{split}
     \mathbb{S} & = 
     K {\rm tr}\left( \mathbb{E} \right) \mathbb{I} + 2G\left(\mathbb{E} - \frac{1}{3} {\rm tr}\left( \mathbb{E} \right) \mathbb{I}\right)\\
     & = \lambda {\rm tr}\left( \mathbb{E} \right) \mathbb{I} + 2 \mu \mathbb{E},
\end{split}
\end{equation}
where $\mathbb{E} = \frac{1}{2} (\mathbb{F}^T \mathbb{F} - \mathbb{I})$ is the Green-Lagrange strain tensor. 
$K = \lambda + \frac{2}{3}\mu$ is the bulk modulus, and
$G = \mu$ is the shear modulus, with
$\lambda$ and $\mu$ representing Lamé parameters, related to the Young’s modulus $E$ and Poisson’s ratio $\nu$ by
\begin{equation}\label{eq: Young-modulus}
     E = 2G(1+2\nu) = 3K(1-2\nu).
\end{equation}

The total Lagrangian formulation is implemented in SPHinXsys using an initial reference configuration. 
This allows neighboring particle relationships to remain fixed throughout the simulation, 
ensuring efficient computation of deformation and stress. The discretized equations are
\begin{equation}\label{eq: solid-density}
    \rho_a = \rho^0_a \frac{1}{\rm det(\mathbb{F})},
\end{equation}
\begin{equation}\label{eq: solid-momentum}
    \frac{\mathrm{d}\mathbf{v}_a}{\mathrm{d}t} = 
    \frac{1}{\rho_a} \sum_{b} (\mathbb{P}_a\mathbb{B}_a^0+\mathbb{P}_b\mathbb{B}_b^0)
    \nabla_a^0W_{ab}V_b + \mathbf{f}_a^{f:p} + \mathbf{f}_a^{f:v}.
\end{equation}
Here, subscript $a$ refers to a solid particle. $\mathbb{B}_a^0$ is the correction matrix for spatial 
homogeneity, defined as
$\mathbb{B}^0_a = 
\big{(}\sum_{b}V_b(\mathbf{r}^0_b-\mathbf{r}^0_a)\otimes \nabla_a^0W_{ab} \big{)^{-1}}$. 
The deformation tensor $\mathbb{F}$ is updated as
\begin{equation}\label{eq: deformation-tensor}
    \mathbb{F}_a = 
    \bigg{(}\sum_{b}(\mathbf{u}_b-\mathbf{u}_a)\otimes \nabla_a^0W_{ab}V_b \bigg{)}\mathbb{B}_a^0 + \mathbb{I}.
\end{equation}

To enhance stability, a Kelvin-Voigt (KV) type damping \cite{zhang2022artificial} is adopted, 
incorporating an artificial damping stress into the Kirchhoff stress:
\begin{equation}\label{eq: kv-damping}
    \mathbb{S_D} = \frac{a\rho_s c_s h_s}{2}(\frac{\mathrm{d}\mathbb{F}}{\mathrm{d}t})^\mathrm{T} \mathbb{F} + \mathbb{F}^\mathrm{T} \frac{\mathrm{d}\mathbb{F}}{\mathrm{d}t},
\end{equation}
where constant parameter $a = 0.5$, $c_s = \sqrt{K / \rho_s}$ and $K$ is bulk modules as shown in Eq.\ref{eq: Young-modulus}.

\subsubsection{Fluid-structure interaction} \label{subsubsection: fluid-strcuture interaction}
The smoothing length for fluid and solid discretization are expressed as $h_f$ and $h_s$, and $h_f \ge h_s$. 
For this study, $h_f = 1.3dp^0$ and $h_s = 1.15dp^0$. The forces exerted by the solid walls on the fluid 
are integrated into the fluid’s momentum equation \cite{zhang2021multi}:
\begin{equation}\label{eq: fsi-pressure-on-fluid}
    \mathbf{f}^{s:p}_i(h_f) = -2\sum_{a}\frac{p^\ast}{\rho_i}\nabla_iW(\mathbf{r}_{ia}, h_f)V_a,
\end{equation}
\begin{equation}\label{eq: fsi-viscous-on-fluid}
    \mathbf{f}^{s:v}_i(h_f) = 
    2\sum_{a}\frac{\eta_{ia}}{\rho_i}\frac{\mathbf{v}_i - \mathbf{v}_a^d}{|\mathbf{r}_{ia}| 
    + 0.01h}\frac{\partial W(\mathbf{r}_{ia}, h_f)}{\partial r_{ia}}V_a,
\end{equation}
where subscript $i$ represents the target fluid particle and $a$ represents its neighboring solid particles, 
$p^\ast = \frac{\rho_i p_a^d + \rho_a^d p_i}{\rho_i+\rho_a^d}$ is the solution to the one-sided Riemann 
problem for fluid-solid interactions. $p_a^d$ and $\mathbf{v}_a^d$ are the imaginary pressure and velocity 
of solid particles calculated by imposing the no-slip boundary condition at the fluid-structure interface:
\begin{equation}\label{eq: imaginary-pressure-velocity}
\left\{\begin{aligned}
    & p_a^d = p_i + \rho_i \max \left( 0, \mathbf{g} - \frac{\mathrm{d}\mathbf{v}_a}{\mathrm{d}t} \right) \cdot \mathbf{r}_{ia}, \\
    & \mathbf{v}_a^d = 2\mathbf{v}_i - \mathbf{v}_a. \\
\end{aligned}\right.
\end{equation}

The forces exerted by the fluid on the solid walls are equal and opposite:
\begin{equation}\label{eq: forces-on-solid}
\left\{\begin{aligned}
    & \mathbf{f}^{f:p} = - \mathbf{f}^{s:p}, \\
    & \mathbf{f}^{f:v} = - \mathbf{f}^{s:v}. \\
\end{aligned}\right.
\end{equation}

Time step sizes are determined by the CFL condition and are tailored separately for the fluid and 
solid phases. 

Specifically, the fluid domain employs a dual-criteria time stepping \cite{zhang2020dual}. The advection criterion 
$\Delta t_{ad}$, which controls the update of the neighbor particle list and the corresponding kernel 
weights and gradients, is defined as
\begin{equation}\label{eq: advection-time}
    \Delta t_{ad} = \mathrm{CFL}_{ad} {\rm min} \left( \frac{h}{|\mathbf{v}|_{\rm max}}, \frac{\rho h^2}{\eta} \right),
\end{equation}
with $\mathrm{CFL}_{ad} = 0.25$. The particle density will be re-initialized \cite{rezavand2022generalized} 
at each advection step with
\begin{equation}\label{eq: fluid-density}
    \rho_i = \rho^0_i \frac{\sum_j W(\mathbf{r}_{ij}, h_f)}{\sum_j W^0(\mathbf{r}_{ij}, h_f)}
\end{equation}
to avoid density/volume error accumulation during long-term simulations. 
The acoustic criterion $\Delta t_{ac}$ determines the time integration of the particle density, 
position and velocity, calculated by
\begin{equation}\label{eq: acoustic-time}
    \Delta t_{ac} = \mathrm{CFL}_{ac} \frac{h}{c_f + |\mathbf{v}|_{\rm max}}.
\end{equation}
Here, $\mathrm{CFL}_{ac} = 0.6|\mathbf{v}|_{\rm max}$ is the acoustic CFL number and $\eta$ means the dynamic viscosity.

In SPHinXsys, the position-based Verlet scheme is employed. Within one advection time step $\Delta t_{ad}$, 
multiple acoustic time steps $\Delta t_{ac}$ are executed for pressure relaxation until $\Delta t_{ad}$ is 
reached. The first half-step velocity in the $n$-th acoustic time step is updated as
\begin{equation}\label{eq: midpoint-fluid-velocity}
    \mathbf{v}^{n+\frac{1}{2}}_i = 
    \mathbf{v}^{n}_i + \frac{\Delta t_{ac}}{2}(\frac{\mathrm{d}\mathbf{v}_i}{\mathrm{d}t})^n.
\end{equation}

Then the updated velocity at the midpoint is applied to obtain the particle position and density in the 
meantime for the next acoustic time step
\begin{equation}\label{eq: update-fluid}
\left\{\begin{aligned}
    & \mathbf{r}^{n+1}_i = \mathbf{r}^{n}_i + \Delta t_{ac}\mathbf{v}_i^{n+\frac{1}{2}}, \\
    & \rho^{n+1}_i = \rho^{n}_i + 
    \frac{\Delta t_{ac}}{2}(\frac{\mathrm{d}\rho_i}{\mathrm{d}t})^{n+\frac{1}{2}}. \\
\end{aligned}\right.
\end{equation}

At last, the velocity of the particle $i$ at the end of this acoustic time step is obtained by
\begin{equation}\label{eq: update-fluid-velocity}
    \mathbf{v}^{n+1}_i = 
    \mathbf{v}^{n}_i + \frac{\Delta t_{ac}}{2}(\frac{\mathrm{d}\mathbf{v}_i}{\mathrm{d}t})^{n+1}.
\end{equation}

Note that the transport velocity formulation for fluid dynamics introduced in 
Section \ref{subsubsection: Fluid dynamics method based on Riemann solver} is implemented once to correct the fluid 
particle positions during each advection time step.

For solid mechanics, the time step size is
\begin{equation}\label{eq: solid-time}
    \Delta t_s = 0.6 {\rm min} \left( \frac{h_s}{c_s+|\mathbf{v}|_{\rm max}}, 
    \sqrt{\frac{h_s}{|\frac{\mathrm{d}\mathbf{v}}{\mathrm{d}t}|_{\rm max}}} \right).
\end{equation}

Further, the structure time stepping is coupled with the dual-criteria time stepping for the FSI problem.
For the time integration of solid equations, generally $\Delta t_s < \Delta t_{ac}$. 
Index $x = 0, 1, ... k-1$ is utilized within one acoustic time step of fluid integration 
with $k = \big[\frac{\Delta t_{ac}}{\Delta t^s} \big ] + 1$. The deformation tensor, density and particle 
position are updated to the midpoint of $x$-th time step as
\begin{equation}\label{eq: midpoint-solid}
\left\{\begin{aligned}
    & \mathbb{F}^{x+\frac{1}{2}}_a = 
    \mathbb{F}^{x}_a + \frac{\Delta t_{s}}{2}\frac{\mathrm{d}\mathbb{F}_{a}}{\mathrm{d}t}, \\
    & \rho_a^{x+\frac{1}{2}} = \rho^0_a \frac{1}{J}, \\
    & \mathbf{r}_a^{x+\frac{1}{2}} = \mathbf{r}_a^x + \frac{\Delta t_{s}}{2} \mathbf{v}_a^x. \\
\end{aligned}\right.
\end{equation}

After that, the velocity of solid particle $a$ is updated to the next time step
\begin{equation}\label{eq: update-solid-velocity}
    \mathbf{v}^{x+1}_a = \mathbf{v}^{x}_a + \Delta t_s\frac{\mathrm{d}\mathbf{v}_{a}}{\mathrm{d}t}.
\end{equation}

Finally, the deformation tensor and position of solid particles are updated to the new time step by
\begin{equation}\label{eq: update-solid}
\left\{\begin{aligned}
    & \mathbb{F}^{x+1}_a = \mathbb{F}^{x+\frac{1}{2}}_a 
    + \frac{\Delta t_{s}}{2}\frac{\mathrm{d}\mathbb{F}_{a}}{\mathrm{d}t}, \\
    & \rho_a^{x+1} = \rho^0_a \frac{1}{J}, \\
    & \mathbf{r}_a^{x+1} = \mathbf{r}_a^{x+\frac{1}{2}} + \frac{\Delta t_{s}}{2} \mathbf{v}_a^{x+1}. \\
\end{aligned}\right.
\end{equation}

\subsubsection{Fluid-shell interaction}
\begin{figure}[htbp]
    \centering
    \includegraphics[width=12cm]{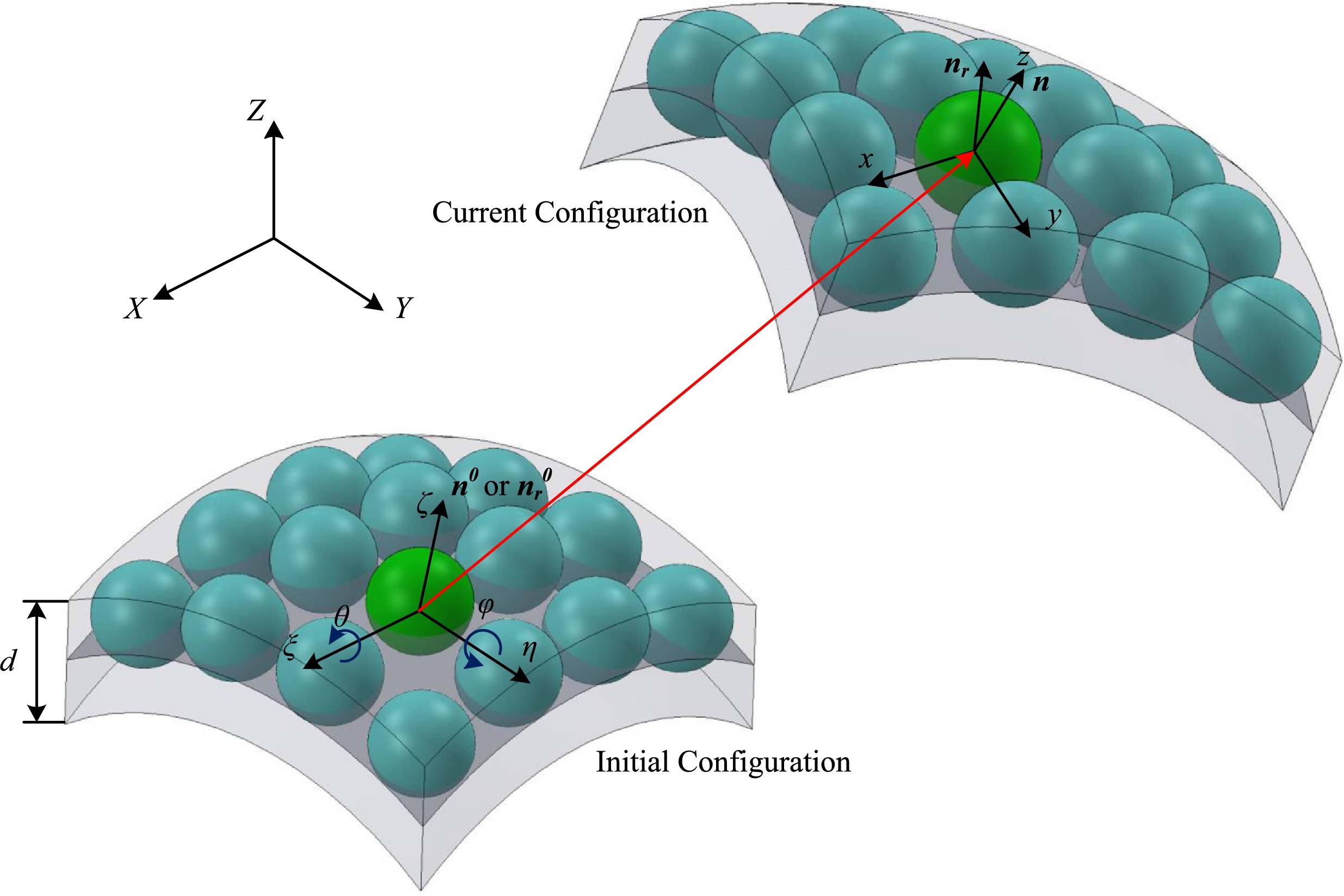}
    \caption{Illustration of 3D shell model \cite{wu2024sph}.}
    \label{fig: shell-scheme}
\end{figure}

The kinematics of the shell are formulated following the approach proposed in Ref.\cite{wu2024sph} based 
on Uflyand-Mindlin plate theory. In the 3D representation, each material point is given by five degrees 
of freedom: three translational components $\mathbf{u}^L = \{ u^L, v^L, w^L \}^ \text{T}$ and two 
rotations $\theta^L = \{\theta^L, \varphi^L \}$. Here, the superscript $(\bullet)^L$ denotes quantities 
expressed in the initial local coordinate system $\bm{\xi} = \{ \xi, \eta, \zeta \}$, as illustrated in 
Fig.\ref{fig: shell-scheme}. The pseudo-normal vector is defined by $\mathbf{n}^L = \{ n_1^L, n_2^L, n_3^L \}^ \text{T}$, 
with its initial configuration given by $\mathbf{n}^{0, L} = \{ 0, 0, 1 \}^ \text{T}$. 
For 2D problems, three degrees of freedom are considered, consisting of two translations $\mathbf{u}^L = \{ u^L, v^L \}^ \text{T}$ 
and one rotation $\theta^L = \{ \varphi^L \}$. In this section, we mainly use 3D formulations to illustrate the shell model. 
Additional details, including the 2D representation, can be found in Ref.\cite{wu2024sph}, 
which is consistent with the SPHinXsys framework.

The local position $\mathbf{r}^L$ of a material point is expressed as
\begin{equation}\label{eq: shell-local-position}
    \mathbf{r}^L ( \xi, \eta, \chi, t) = \mathbf{r}_m^L ( \xi, \eta, t) + \chi \mathbf{n}^L( \xi, \eta, t),
\end{equation}
where the subscript $(\bullet)_m$ refers to the mid-surface, and $\chi \in [-d/2, d/2]$ denotes the thickness coordinate. 
The local displacement $\mathbf{u}^L$ is obtained by
\begin{equation}\label{eq: shell-local-displacement}
    \mathbf{u}^L ( \xi, \eta, \chi, t) = \mathbf{u}_m^L ( \xi, \eta, t) + \chi \Delta \mathbf{n}^L( \xi, \eta, t),
\end{equation}
with $\Delta \mathbf{n}^L = \mathbf{n}^L - \mathbf{n}^{0, L}$. 
The local deformation gradient tensor is defined as
\begin{equation}\label{eq: shell-local-deformation-gradient}
    \mathbb{F}^L = \nabla^{0, L} \mathbf{r}^L + \nabla^{0, L} \mathbf{n}^L - \nabla^{0, L} \mathbf{n}^{0, L},
\end{equation}
where $\nabla^{0, L} \equiv \partial / \partial \bm{\xi}$ is the gradient operator in the initial local configuration.

The Eulerian Almansi strain $\outline{$\mathbb{\epsilon}$}$ is computed from the deformation gradient $\mathbb{F}$ as
\begin{equation}\label{eq: shell-strain}
    \outline{$\mathbb{\epsilon}$}  = \frac{1}{2} (\mathbb{I} - \mathbb{F}^{-\text{T}} \mathbb{F}^{-1}).
\end{equation}
The corresponding Cauchy stress tensor $\outline{$\mathbb{\sigma}$}$ follows the isotropic linear elastic constitutive relation
\begin{equation}\label{eq: shell-constitutive}
    \outline{$\mathbb{\sigma}$} = \lambda {\rm tr}(\outline{$\mathbb{\epsilon}$}) \mathbb{I} + 2 \mu \outline{$\mathbb{\epsilon}$}, 
\end{equation}
which is analogous to the expression of the second Piola-Kirchhoff stress tensor given in Eq.\ref{eq: 2nd-stress}.

By introducing the orthogonal transformation matrix $\mathbb{Q}$ from the global coordinates to the current 
local coordinate system, the local strain tensor $\outline{$\mathbb{\epsilon}$}^l$ is obtained as
\begin{equation}\label{eq: shell-current-strain}
    \outline{$\mathbb{\epsilon}$}^l = \mathbb{Q} ( \mathbb{Q}^0)^{\text{T}} \outline{$\mathbb{\epsilon}$}^L \mathbb{Q}^0 \mathbb{Q}^{\text{T}},
\end{equation}
where $\mathbb{Q}^0$ denotes the transformation from the global coordinates to the initial local system. 
Then to satisfy the plane-stress condition, the strain component in the thickness direction is corrected as
\begin{equation}\label{eq: shell-corrected-strain-3D}
    \overline{\epsilon}^l_{zz} = \frac{-\nu (\epsilon^l_{xx} + \epsilon^l_{yy})}{1 - \nu},
\end{equation}
with $\nu$ denoting the Poisson’s ratio. 

Substituting the corrected strain $\overline{\outline{$\mathbb{\epsilon}$}}^l$ into Eq.\ref{eq: shell-constitutive} 
yields the corrected local Cauchy stress  $\overline{\outline{$\mathbb{\sigma}$}}^l$. To account for transverse 
shear correction, the shear stress components are further modified as
\begin{equation}\label{eq: shell-corrected-stress-3D}
    \overline{\sigma}^l_{xz} = \overline{\sigma}^l_{zx} = \frac{5}{6} \  \sigma^l_{xz}, \ \overline{\sigma}^l_{yz} = \overline{\sigma}^l_{zy} = \frac{5}{6} \ \sigma^l_{yz}.
\end{equation}

Mass conservation follows Eq.\ref{eq: solid-density}, with the Jacobian determinant given by $J = \text{det}(\mathbb{F})$. 
The momentum and angular momentum conservation equations in SPH discretization are expressed as
\begin{equation}\label{eq: shell-momentum}
    d \rho^0_a \Ddot{\mathbf{u}}_{m,a} = \sum_b (J_{m,a} \mathbb{N}_a (\mathbb{F}_{m, a})^{-\text{T}} \tilde{\mathbb{B}}^{0, \mathbf{r}}_a + J_{m,b} \mathbb{N}_b (\mathbb{F}_{m, b})^{-\text{T}} \tilde{\mathbb{B}}^{0, \mathbf{r}}_b)\nabla_a^0W_{ab}V^0_b
\end{equation}
and
\begin{equation}\label{eq: shell-angular-momentum}
    \frac{d^3}{12} \rho^0_a \Ddot{\mathbf{n}}_{a} = \sum_b (J_{m,a} \mathbb{M}_a (\mathbb{F}_{m, a})^{-\text{T}} \tilde{\mathbb{B}}^{0, \mathbf{n}}_a + J_{m,b} \mathbb{M}_b (\mathbb{F}_{m, b})^{-\text{T}} \tilde{\mathbb{B}}^{0, \mathbf{n}}_b)\nabla_a^0W_{ab}V^0_b + J_{m,a} (\mathbb{Q}^0_a)^{\text{T}} \mathbf{q}^l_a,
\end{equation}
where $\mathbb{F}_m = (\mathbb{Q}^0)^{\text{T}} \mathbb{F}^L_m \mathbb{Q}^0$ and $\tilde{\mathbb{B}}^0_a = (\mathbb{Q}^0_a)^{\text{T}} \mathbb{G} \mathbb{B}^{0, L}_a \mathbb{G}^{\text{T}} \mathbb{Q}^0_a$. 
The stress and moment resultants in global coordinates are obtained 
as $\mathbb{N} = \mathbb{Q}^{\text{T}} \mathbb{N}^l \mathbb{Q}$ and $\mathbb{M} = \mathbb{Q}^{\text{T}} \mathbb{M}^l \mathbb{Q}$, 
where the local resultants $\mathbb{N}^l$ and $\mathbb{M}^l$ are computed by integration of the corrected local stress $\overline{\outline{$\mathbb{\sigma}$}}^l$.

Different from the full-dimensional kernel, whose partition-of-unity is enforced with respect to the volume measure, 
the reduced-dimensional kernel for shells in the above momentum equations enforces the unit integral on the reduced 
manifold (obtained by analytically integrating through the thickness). Consequently, the “particle volume” $V$ 
in the discrete summations denotes the measure of the reduced space: it is the length for 2D problems (line manifold) 
and the area for 3D  problems (surface manifold). The reduced kernel thus differs from its full-dimensional 
counterpart only in the normalizing constant, while the polynomial shape remains identical. Specifically, for the 
reduced fifth-order Wendland kernel
\begin{equation}\label{eq: shell-kernel}
    W(q, h) = \alpha 
    \left\{\begin{aligned}
    & (1+2q)(1-q/2)^4   & \text{if } 0 \leq q \leq 2 \\
    & 0   & \text{otherwise}
    \end{aligned}\right.
\end{equation}
the constants are $\alpha = \frac{3}{4h}$ for 2D and $\alpha = \frac{7}{4 \pi h^2}$ for 3D. For comparison, the 
full-dimensional Wendland kernel uses $\frac{7}{4 \pi h^2}$ and $\frac{21}{16 \pi h^3}$ in 2D and 3D problems, respectively.

Time integration for solid mechanics with reduced-dimensional shell model is also performed using the position-based 
Verlet scheme. At the beginning of each time step, besides the deformation tensor and particle position in 
Eq.\ref{eq: midpoint-solid}, the rotation angles and pseudo-normal vector are also updated to the midpoint of $x$-th 
time step as
\begin{equation}\label{eq: midpoint-shell}
\left\{\begin{aligned}
    & \mathbb{F}^{L, x+\frac{1}{2}} = 
    \mathbb{F}^{L, x} + \frac{\Delta t_{s}}{2} \dot{\mathbb{F}}^{L, x}, \\
    & \mathbf{r}_m^{x+\frac{1}{2}} = \mathbf{r}_m^x + \frac{\Delta t_{s}}{2} \dot{\mathbf{u}}_m^x, \\
    & \bm{\theta}^{L, x+\frac{1}{2}} = \bm{\theta}^{L, x} + \frac{\Delta t_{s}}{2} \dot{\bm{\theta}}^{L, x}, \\
    & \mathbf{n}^{L, x+\frac{1}{2}} = \mathbf{n}^{L, x} + \frac{\Delta t_{s}}{2} \dot{\mathbf{n}}^{L, x}. \\
\end{aligned}\right.
\end{equation}

With $\mathbb{F}^{L, x+\frac{1}{2}}$, the corrected Almansi strain $\outline{$\mathbb{\epsilon}$} ^{l, x+\frac{1}{2}}$ 
and corrected Cauchy stress $\outline{$\mathbb{\sigma}$} ^{l, x+\frac{1}{2}}$ are obtained from Eq.\ref{eq: shell-strain} 
to Eq.\ref{eq: shell-corrected-stress-3D}. By integrating the corrected Cauchy stress across the shell thickness, 
the momentum and stress resultants $\mathbb{M}^l$ and $\mathbb{N}^l$, together with transverse shear vector $\mathbf{q}^l$, 
are determined. These quantities are subsequently employed in the conservation equations to solve for the translational 
acceleration $\Ddot{\mathbf{u}}^{x+1}_{m}$ of the mid-surface and the angular acceleration $\Ddot{\mathbf{n}}^{x+1}$ of the 
pseudo-normal vector. After transforming $\Ddot{\mathbf{n}}^{x+1}$ from the global coordinate system into the initial 
local system $\Ddot{\mathbf{n}}^{L, x+1}$, the angular acceleration $\Ddot{\bm{\theta}}^{L, x+1}$ is obtained through 
the kinematic relation between the pseudo-normal vector ${\mathbf{n}}^{L}$ and the rotation angle ${\bm{\theta}}^{L}$. 
The translational and rotational velocities are then updated as
\begin{equation}\label{eq: update-shell-velocity}
\left\{\begin{aligned}
    & \dot{\mathbf{u}}_m^{x+1} = \dot{\mathbf{u}}_m^x + \Delta t_{s} \Ddot{\mathbf{u}}_m^{x+1}, \\
    & \dot{\bm{\theta}}^{L, x+1} = \dot{\bm{\theta}}^{L, x} + \Delta t_{s} \Ddot{\bm{\theta}}^{L, x+1},  \\
\end{aligned}\right.
\end{equation}
while the rate of change of the pseudo-normal vector  $\dot{\mathbf{n}}^{L, x+1}$ is updated consistently 
from $\bm{\theta}^{L, x+1}$ and $\dot{\bm{\theta}}^{L, x+1}$.

Finally, the change rate of the deformation gradient tensor for particle $a$ $\dot{\mathbb{F}}_a^{L, x+1}$ is updated 
according to 
\begin{equation}
    \dot{\mathbb{F}}_a^L = \nabla^{0, L} \dot{\mathbf{u}_a^L} = \nabla^0 \dot{\mathbf{u}_{m, a}^L} + \chi \nabla^0 \dot{\mathbf{n}}_a^L,
\end{equation}
where the gradients of the mid-surface velocity and of the pseudo-normal are given by the corrected SPH formulation as
\begin{equation}
\left\{\begin{aligned}
    & \nabla^0 \dot{\mathbf{u}}_{m, a}^L = \mathbb{Q}^0_a (\sum_b \dot{\mathbf{u}}_{m, ab} \otimes \nabla_a^0W_{ab}V^0_b) \tilde{\mathbb{B}}^{0, \mathbf{r}}_a (\mathbb{Q}^0_a)^{\text{T}}, \\
    & \nabla^0 \dot{\mathbf{n}}_{a}^L = \mathbb{Q}^0_a (\sum_b \dot{\mathbf{n}}_{ab} \otimes \nabla_a^0W_{ab}V^0_b) \tilde{\mathbb{B}}^{0, \mathbf{r}}_a (\mathbb{Q}^0_a)^{\text{T}},  \\
\end{aligned}\right.
\end{equation}
ensuring both consistency and strong-form correction. 
The state variables are then advanced to the new time step as
\begin{equation}\label{eq: update-shell}
\left\{\begin{aligned}
    & \mathbb{F}^{L, x+1} = 
    \mathbb{F}^{L, x+\frac{1}{2}} + \frac{\Delta t_{s}}{2} \dot{\mathbb{F}}^{L, x+1}, \\
    & \rho^{x+1} = (J_m^{x+1})^{-1} \rho^0, \\
    & \mathbf{r}_m^{x+1} = \mathbf{r}_m^{x+\frac{1}{2}} + \frac{\Delta t_{s}}{2} \dot{\mathbf{u}}_m^{x+1}, \\
    & \bm{\theta}^{L, x+1} = \bm{\theta}^{L, x+\frac{1}{2}} + \frac{\Delta t_{s}}{2} \dot{\bm{\theta}}^{L, x+1}, \\
    & \mathbf{n}^{L, x+1} = \mathbf{n}^{L, x+\frac{1}{2}} + \frac{\Delta t_{s}}{2} \dot{\mathbf{n}}^{L, x+1}. \\
\end{aligned}\right.
\end{equation}

For the numerical stability, the time step $\Delta t_s$ for shell model is given by
\begin{equation}\label{eq: shell-time}
    \Delta t_s = 0.6 {\rm min}(\Delta t_{s1}, \Delta t_{s2}, \Delta t_{s3}),
\end{equation}
with
\begin{equation}
\left\{\begin{aligned}
    & \Delta t_{s1} = {\rm min} \left( \frac{h_s}{c_s+|\dot{\mathbf{u}}_m|_{\rm max}}, \sqrt{\frac{h_s}{|\Ddot{\mathbf{u}}_m|_{\rm max}}} \right), \\
    & \Delta t_{s2} = {\rm min} \left( \frac{h_s}{c_s+|\dot{\rm{\theta}}_m|_{\rm max}}, \sqrt{\frac{h_s}{|\Ddot{\rm{\theta}}_m|_{\rm max}}} \right), \\
    & \Delta t_{s3} = h_s \left( \frac{\rho (1-\nu^2) / E}{2 + (\pi^2/12) (1-\nu) [1+1.5(h_s/d)^2]} \right) ^{1/2}. \\
\end{aligned}\right.
\end{equation}

\begin{figure}[htbp]
    \centering
    \includegraphics[width=10cm]{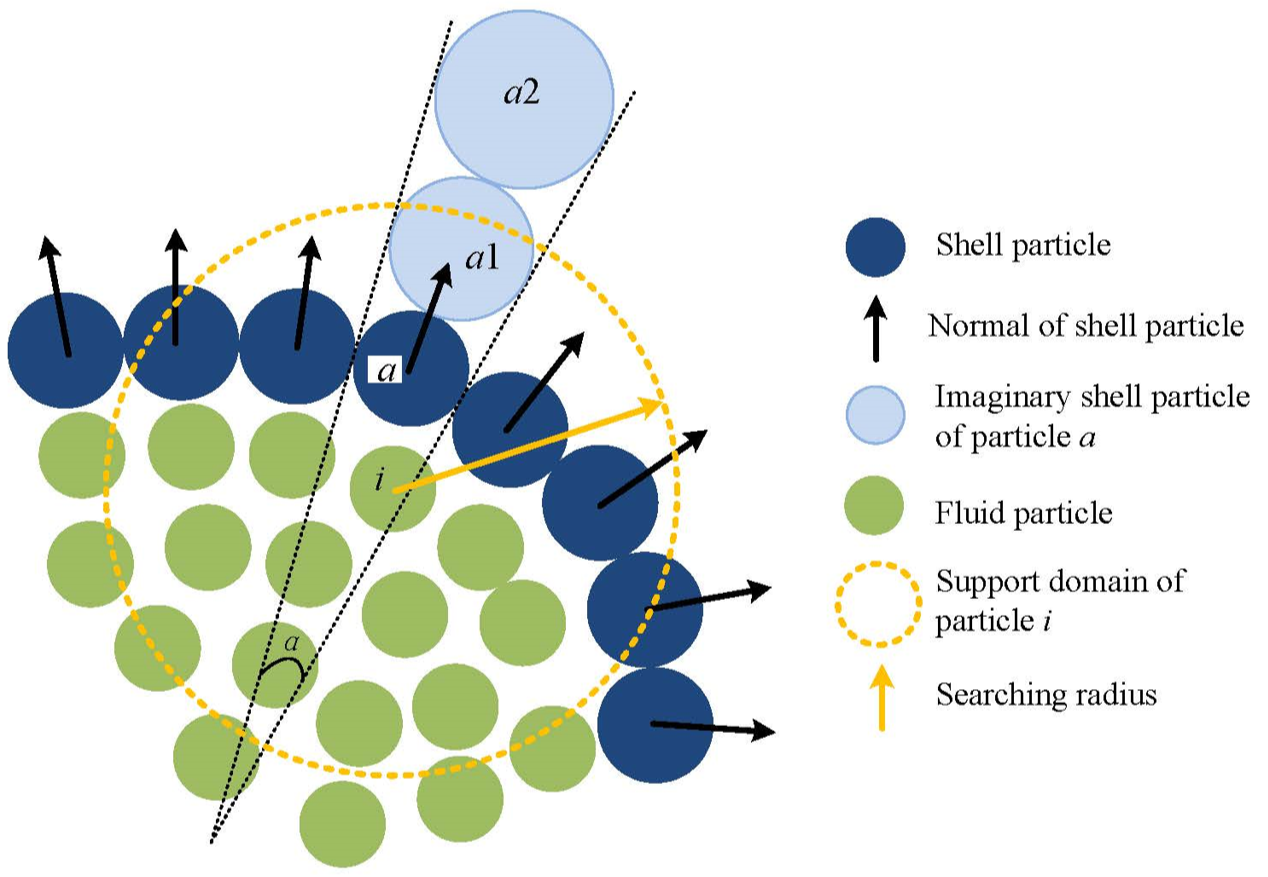}
    \caption{Illustration of the projection method for fluid-shell interaction.}
    \label{fig: fluid-shell-interaction-illustration}
\end{figure}

In fluid-structure interaction involving thin shells, it is essential to properly capture the shell’s thickness effect. 
To this end, the projection method is adopted, as illustrated in Fig.\ref{fig: fluid-shell-interaction-illustration}. 
In this method, a layer of virtual particles is generated along the shell boundary to represent its reduced-dimensional geometry. 
The interaction between fluid particles and these virtual boundary particles is evaluated not through direct volume 
integration but via a projection procedure. Specifically, the overlapping volume between a fluid particle and a virtual 
boundary particle is projected into an equivalent area in three dimensions (or length in two dimensions). This projected 
measure is then employed to correct the kernel function or particle interaction formulation, thereby ensuring that the 
density and momentum equations consistently reflect the reduced dimensionality of the shell. Consequently, the fluid-shell 
coupling is accurately described across the interface, with the corrected kernel providing precise force transfer between 
the fluid and shell domains.

For each shell particle $a$, a set of virtual particles is placed along its local normal direction. 
The equivalent projected area of the $k$-th virtual particle is defined, for the 2D case, as
\begin{equation}\label{eq: 2d-shell-area}
    A^k_a = A_a (1 + k \cdot \chi_a \cdot dp_s),
\end{equation}
where $\chi_a = \nabla \cdot \mathbf{n}_a$ denotes the curvature-related term of the shell mid-surface. Moreover, for 
3D problems, the projected area is computed as
\begin{equation}\label{eq: 3d-shell-area}
    A^k_a = A_a (1 + k \cdot \chi_{a1} \cdot dp_s)(1 + k \cdot \chi_{a2} \cdot dp_s),
\end{equation}
where $\chi_{a1} = M + \sqrt{M^2 - K}$ and $\chi_{a2} = M - \sqrt{M^2 - K}$ are the principal curvatures of the shell, 
with the mean curvature $M = \chi_{a1} + \chi_{a2} = \frac{1}{2}\nabla \cdot \mathbf{n}_a$ and the Gaussian 
curvature $K = \chi_{a1} \cdot \chi_{a2} = \frac{1}{2}[(\nabla \cdot \mathbf{n}_a)^2 - \sum_m \sum_n (\frac{\partial n_n}{\partial X_m} \frac{\partial n_m}{\partial X_n})]$. 
This treatment ensures that the geometric curvature of the shell surface is properly incorporated in the projection 
when interacting with the surrounding fluid particles.

During fluid density re-initialization of Eq.\ref{eq: fluid-density}, the density of a fluid particle $i$ is re-defined as
\begin{equation}\label{eq: fluid-shell-density}
    \rho_i = \rho^0_i \frac{\sum_j W(\mathbf{r}_{ij}, h_f) + \sum_a \overline W(\mathbf{r}_{ia}, h_f) \frac{V^0_a}{V^0_i}}{\sum_j W^0(\mathbf{r}_{ij}, h_f)},
\end{equation}
where $\overline W(\mathbf{r}_{ia}, h_f)$ is the projection-corrected kernel function, computed by summing over 
all imaginary particles:
\begin{equation}\label{eq: fluid-shell-kernel}
    \overline W(\mathbf{r}_{ia}, h_f) = \frac{1}{A^0_ad^0_a} \sum_k W(\mathbf{r}^k_{ia}, h_f) A^k_a dp_s.
\end{equation}
Here, $A^0_a$ and $d^0_a$ denote the reference area and thickness of the shell particle, respectively. This correction 
guarantees kernel consistency when fluid particles are located near the shell surface.

In the momentum conservation equation, the interaction between a fluid particle $i$ and a shell particle $a$ is 
introduced via the projection-corrected kernel as Eq.\ref{eq: fluid-shell-kernel} and its gradient as Eq.\ref{eq: fluid-shell-kernel-gradient}. 
\begin{equation}\label{eq: fluid-shell-kernel-gradient}
    \frac{\partial \overline W(\mathbf{r}_{ia}, h_f)}{\partial r_{ia}} = \frac{1}{A_a} \sum_k \frac{\partial W(\mathbf{r}^k_{ia}, h_f) }{\partial r^k_{ia}} A^k_a.
\end{equation}
Accordingly, the interaction forces acting on the fluid are given by
\begin{equation}\label{eq: fluid-shell-pressure-on-fluid}
    \mathbf{f}^{s:p}_{ia}(h_f) = -2 \frac{p^\ast}{\rho_i} \frac{\partial \overline W(\mathbf{r}_{ia}, h_f)}{\partial r_{ia}} \overline{\mathbf{e}} _{ia} V_a,
\end{equation}
for pressure contribution; and 
\begin{equation}\label{eq: fluid-shell-viscous-on-fluid}
    \mathbf{f}^{s:v}_{ia}(h_f) = 
    2\frac{\eta_{ia}}{\rho_i}\frac{\mathbf{v}_i - \mathbf{v}_a^d}{|\mathbf{r}_{ia}| 
    + 0.01h}\frac{\partial \overline W(\mathbf{r}_{ia}, h_f)}{\partial r_{ia}}V_a,
\end{equation}
for viscous contribution. $\overline{\mathbf{e}}_{ia}$ is the weighted average direction vector of the imaginary particles derived as
\begin{equation}\label{eq: fluid-shell-direction-vector}
    \overline{\mathbf{e}}_{ia} = \frac{ \sum_k \frac{\partial W(\mathbf{r}^k_{ia}, h_f) }{\partial r^k_{ia}} \mathbf{e}^k_{ia} A^k_a}{ \sum_k \frac{\partial W(\mathbf{r}^k_{ia}, h_f) }{\partial r^k_{ia}} A^k_a}.
\end{equation}

For the force acting on the shell, an equivalent kernel is applied with $W_{ai} = -W_{ia}$. The remaining coupling 
procedures in the FSI framework follow the same formulation as the volume-based model discussed in 
Section \ref{subsubsection: fluid-strcuture interaction}.

\subsection{In-/outlet boundary condition implementation} \label{subsection: In-/outlet boundary condition implementation}
To impose velocity and pressure boundary conditions at the inlet and outlet(s), we adopt the four-layer 
bidirectional buffer approach proposed in Ref.\cite{zhang2024generalized}. This method allows for the 
dynamic injection and deletion of fluid particles within a single buffer, which is suitable for the 
potential backflow phenomenon with pressure boundary condition.

At the inlet, a time-dependent velocity profile is imposed on the buffer particles using the coordinate 
transformation method introduced in Ref.\cite{zhang2024generalized}. At the outlet(s), various pressure 
boundary conditions, including constant pressure, resistance model and Windkessel model, are implemented 
following the method described in Ref.\cite{zhang2025dynamical}. The pressure gradient in 
Eq.(\ref{eq: momentum}) at the near-boundary particle $i$ is calculated by
\begin{equation}\label{eq: outlet-pressure-gradient}
    \nabla p_i = 2\sum_{j}p^\ast\nabla_iW_{ij}V_j - 2 p_\mathrm{target} \sum_{j} \nabla_iW_{ij}V_j,
\end{equation}
where $p_\mathrm{target}$ is the prescribed outlet pressure. The Riemann-based discretized momentum 
equation Eq.(\ref{eq: momentum-sph}) at the pressure boundary is then modified by
\begin{equation}\label{eq: momentum-sph-pressure-bc}
    \frac{\mathrm{d}\mathbf{v}_i}{\mathrm{d}t} = -2\sum_{j} \frac{p^\ast}{\rho_i}\nabla_iW_{ij}V_j + 2p_\mathrm{target}\sum_{j}\frac{1}{\rho_i}\nabla_iW_{ij}V_j + 
    2\sum_{j}\frac{\eta_{ij}}{\rho_i}
    \frac{\mathbf{v}_{ij}}{r_{ij}}\frac{\partial W_{ij}}{\partial r_{ij}}V_j + \mathbf{f}_i.
\end{equation}

The resulting velocity of buffer particles is projected onto the normal direction of the pressure boundary 
domain. The density of newly added buffer particles is computed using the equation of state:
\begin{equation}\label{eq: pressure-bc-EoS}
    \rho_i = \rho^0 + \frac{p_\mathrm{target}}{c_f^2}.
\end{equation}

\section{Results and discussion}  \label{section: Results and discussion}
\subsection{Validation for in-/outlet boundary condition implementation}
\begin{figure}[htbp]
    \centering
    \includegraphics[width=7cm]{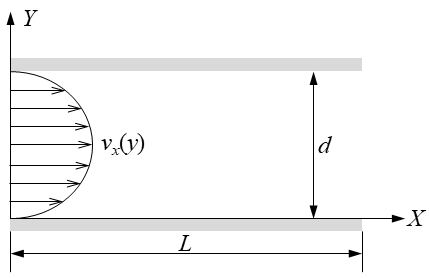}
    \caption{Poiseuille flow in 2D channel: schematic illustration.}
    \label{fig: channel-schematic}
\end{figure}

As a canonical test case, the two-dimensional Poiseuille flow is examined first to ensure the accuracy of 
the boundary conditions. The simulation setup is depicted in Fig.\ref{fig: channel-schematic}, where the 
flow is driven by a constant pressure gradient between two stationary plates. The velocity profile over 
time is analytically derived as follows
\begin{equation}\label{eq: poiseuille-flow-analytical}
  v_x(y,t) = \frac{\Delta P}{2 \eta L}y(y-d) + \sum _{n = 0} ^{\infty} \frac{4 \Delta P d^2}{\eta L \pi^3 (2n+1)^3} \sin{\left(\frac{\pi y}{d} (2n+1)\right)}\exp \left(-\frac{(2n+1)^2 \pi^2 \eta}{\rho d^2} t\right),
\end{equation}
where $y \in (0, d)$ with $d = 0.001 \text{ m}$ representing the gap between the plates, 
$\Delta P = 0.1 \text{ Pa}$ as the pressure difference, and $L = 0.004 \text{ m}$ as the length over which 
the pressure drops. The inlet pressure setting in the Poiseuille flow case is replaced by the steady 
velocity with parabolic distribution corresponding to the steady component of 
Eq.\ref{eq: poiseuille-flow-analytical} here, while the outlet pressure is maintained 
at $P_\text{out} = 0.1 \text{ Pa}$. The dynamic viscosity is calculated from the formula 
$\eta = \sqrt{\rho d^3 \Delta P /(8LRe)}$, where $\rho_f = 1000 \text{ kg}/\text{m}^3$ is the fluid density 
and $Re = 50$ denotes the Reynolds number. The artificial sound speed is set to $c^0_f = 10v_{x}^{\max}$, 
with $v_{x}^{\max} = d^2 \Delta P / (8 \eta L)$. In particular, the wall boundary with 2D volume model and 
shell model are modeled as rigid body for testing fluid dynamics simulation with the implementation of 
in-/outlet boundary conditions.

\begin{figure}[htbp]
    \centering
    \includegraphics[width=9cm]{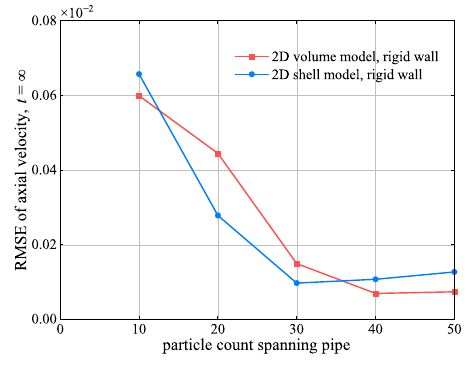}
    \caption{Poiseuille flow in 2D channel (rigid wall): convergence study.}
    \label{fig: 2d-channel-vipo-convergence}
\end{figure}

To ascertain the convergence of the solution, simulations of varying resolutions are performed, 
specifically targeting the velocity profile at the midsection along the streamwise direction. Measurement 
points are aligned radially at this midsection, and the Root Mean Squared Error (RMSE), defined as 
$\text{RMSE} = \sqrt{\sum_{n = 1}^N (v_x(y_n, t) - \hat{v}_x(y_n, t)^2/N}$, is employed to assess 
discrepancies between the SPH simulation results $v_x(y_n, t)$ and the analytical solution 
$\hat{v}_x(y_n, t)$ at $t = \infty$. As illustrated in Fig.~\ref{fig: 2d-channel-vipo-convergence}, 
increasing the particle count across the pipe cross-section can reduce the RMSE in axial velocity of SPH 
results, and both the volume and shell models as wall boundary exhibit comparable convergence behavior when 
only considering the fluid dynamics. Based on these results, a resolution of $dp^0 = d / 30$ is selected 
for subsequent simulations to balance the computational accuracy and efficiency, achieving an RMSE below 
0.02\%.

\begin{figure}[htbp]
    \centering
    \includegraphics[width=9cm]{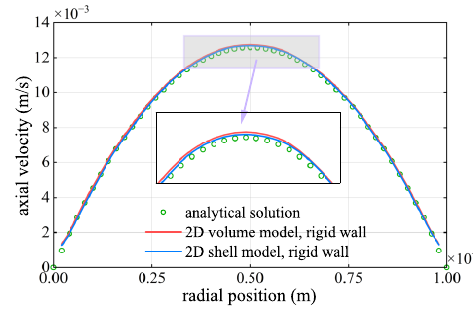}
    \caption{Poiseuille flow in 2D channel (rigid wall): axial velocity distribution along the radial direction at the midsection.}
    \label{fig: 2d-channel-vipo-axial-vel}
\end{figure}

\begin{figure}[htbp]
    \centering
    \includegraphics[width=15cm]{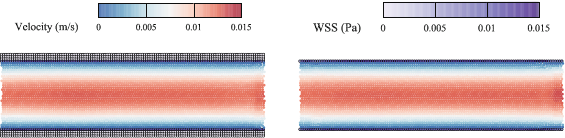}
    \caption{Poiseuille flow in 2D channel (rigid wall): velocity and WSS contours. Left: volume model; 
    right: shell model.}
    \label{fig: 2d-channel-vipo-vel-wss-contour}
\end{figure}

\begin{figure}[htbp]
    \centering
    \includegraphics[width=10cm]{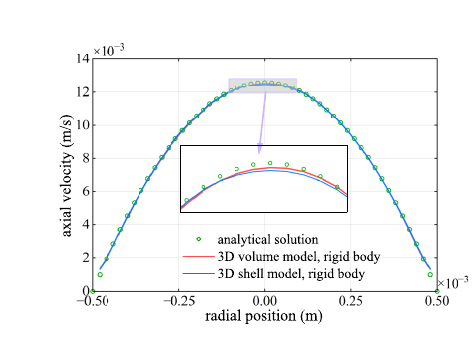}
    \caption{Poiseuille flow in 3D cylinder (rigid wall): axial velocity distribution along the radial direction
    at the midsection.}
    \label{fig: 3d-cylinder-vipo-axial-vel}
\end{figure}

Fig.\ref{fig: 2d-channel-vipo-axial-vel} presents the axial velocity profiles along the radial direction 
at the midsection of the 2D channel. It is evident that the fluid dynamics results obtained using both the 
2D volume model and the shell model as wall boundary closely match the analytical solution. The relative 
errors in the maximum axial velocity are 1.14\% for the volume model and 0.65\% for the shell model. 
Additionally, Fig.\ref{fig: 2d-channel-vipo-vel-wss-contour} shows the steady-state velocity field and WSS 
distributions for both boundary representations. The results reveal a high degree of agreement between the 
volume and shell models, further demonstrating the consistency of the shell-based approach in fluid 
dynamics simulations. Also, to extend the validation to three dimensions, the two parallel plates in the 2D 
case are replaced with a rigid cylindrical pipe, discretized with 30 particles along its diameter. 
Fig.\ref{fig: 3d-cylinder-vipo-axial-vel} compares the axial velocity profiles at the cylinder's midsection 
among the SPH 3D volume model, the shell model and the analytical solutions. The results demonstrate that 
the 3D shell model achieves a level of accuracy of fluid dynamics simulation comparable to that of the 3D 
volume model, with relative errors in the maximum axial velocity of 0.48\% and 0.79\% for the 3D volume and 
shell models, respectively.

Several simplified outlet boundary condition models are commonly employed in hemodynamic simulations, 
including resistance-type, Windkessel, and impedance boundaries. The resistance model establishes a linear 
relationship between the pressure and the flow rate at the outlet, efficiently representing the downstream 
vascular resistance, and is expressed as
\begin{equation}\label{eq: resistance-model}
    p = p^0 + QR,
\end{equation}
where $p^0$ is the base pressure, $Q$ is the volume flow rate at the outlet and $R$ is the resistance parameter.

The flow rate $Q$ is computed as an average value over a predefined time period to ensure simulation 
stability, instead of using the transient flow rate. In the present work, the flow rate $Q$ is determined 
by the cumulative volume of particles added or deleted by the bidirectional buffer at the outlet. 
Specifically, particles deleted by the buffer are recorded as positive contributions, while particles 
generated are recorded as negative contributions. This approach yields a more accurate flow rate over the 
selected period than relying on the average cross-sectional velocity. 

\begin{figure}[htbp]
    \centering
    \begin{subfigure}[b]{0.48\textwidth}
        \includegraphics[width=\textwidth]{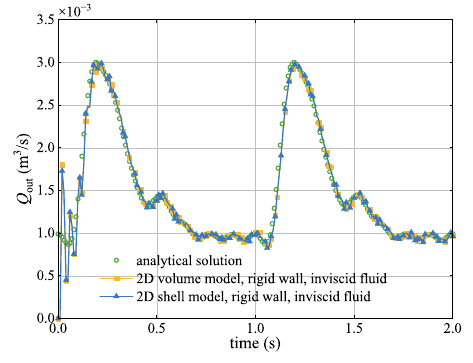}
        \caption{}
        \label{fig: 2d-resistance-inviscid-Q}
    \end{subfigure}
    \begin{subfigure}[b]{0.48\textwidth}
        \includegraphics[width=\textwidth]{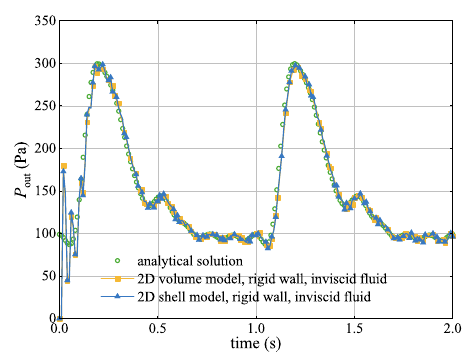}
        \caption{}
        \label{fig: 2d-resistance-inviscid-P}
    \end{subfigure}
    \caption{Verification of resistance boundary implementation in a 2D rigid channel: 
    comparison of (a) outlet volume flow rate and (b) outlet pressure between SPH simulations with 
    volume- and shell-based wall models and the analytical solution.}
    \label{fig: 2d-resistance-inviscid}
\end{figure}

To validate the correct implementation of the resistance pressure boundary in the SPH code, we compared SPH 
results using inviscid fluid with density of $\rho_f = 1000 \text{ kg}/\text{m}^3$ and plug flow inlet 
velocity with analytical solutions. A two-dimensional channel flow is employed, with a domain height of 
$d = 0.00635 \text{ m}$ and length $L = 0.03175 \text{ m}$, discretized using 30 particles across its 
height. A pulsatile inlet velocity is prescribed with a period of $T = 1 \text{s}$, and its time-varying 
profile is expressed as
\begin{equation}\label{eq: resistance-vinlet}
  v_{x, \text{avg}} = 0.2339+\sum_{i=1}^{8}[a_i\cos(\omega it) + b_i \sin(\omega it)],
\end{equation}
where the coefficients are
\begin{align*}\label{eq: resistance-coefficients}
  a &= [-0.0176, \ -0.0657, \ -0.0280, \ 0.0068, \ 0.0075, \ 0.0115, \ 0.0040, \ 0.0035],\\
  b &= [0.1205, \ 0.0171, \ -0.0384, \ -0.0152, \ -0.0122, \ 0.0002, \ 0.0033, \ 0.0060],\\
  \omega &= 2 \pi / T.
\end{align*}
A resistance boundary condition in Eq.\ref{eq: resistance-model} 
($R = 10^5 \text{ kg} \cdot \text{m}^{-4}\text{s}^{-1}$) is imposed at the outlet. As shown in 
Fig.\ref{fig: 2d-resistance-inviscid}, the outlet flow rate and pressure predicted by the SPH simulations 
with both the 2D volume model and rigid shell model as wall boundaries show excellent agreement with the 
analytical solution, confirming the accurate implementation of the resistance boundary condition.

For more complex boundary conditions, the Windkessel model accounts for both the resistive and compliant 
properties of the downstream vascular system. This model effectively buffers the pulsatile nature of 
cardiac output, generating a more continuous and physiologically realistic pressure-flow profile. 
The outlet pressure is governed by the following ordinary differential equation (ODE):
\begin{equation}\label{eq: windkessel-model}
    \frac{\mathrm{d}p}{\mathrm{d}t} + \frac{p}{CR_d} = \frac{R_p+R_d}{CR_d}Q +R_p\frac{\mathrm{d}Q}{\mathrm{d}t},
\end{equation}
where $R_p$ and $R_d$ are the proximal and distal resistances, respectively, and $C$ is the vascular 
compliance. Parameter estimation for the RCR model can follow the principle outlined in 
Ref.\cite{deyranlou2020numerical} if not given.

To solve Eq.(\ref{eq: windkessel-model}) numerically, a modified Euler’s method \cite{lu2024gpu} is 
employed to ensure stability and accuracy. Specifically, the predicted pressure $p^\prime$ at the predictor 
step is computed by the pressure and its change rate at the $n$-th step:
\begin{equation}\label{eq: p-star}
   p^\prime = p^n + \Delta t(\frac{\mathrm{d}p}{\mathrm{d}t})^n,
\end{equation}
where the pressure change rate is defined as
\begin{equation}\label{eq: p-change-rate}
  (\frac{\mathrm{d}p}{\mathrm{d}t})^n = -\frac{p^n}{CR_d} + \frac{R_p+R_d}{CR_d}Q^n + R_p\frac{Q^n-Q^{n-1}}{\Delta t}.
\end{equation}

After that, the pressure in the corrector step is updated by
\begin{equation}\label{eq: p-next}
   p^{n+1} = p^n + \frac{1}{2}\Delta t[(\frac{\mathrm{d}p}{\mathrm{d}t})^n + (\frac{\mathrm{d}p}{\mathrm{d}t})^\prime].
\end{equation}
Here, the predicted pressure change rate $(\frac{\mathrm{d}p}{\mathrm{d}t})^\prime$ is defined as
\begin{equation}\label{eq: p-change-rate-star}
  (\frac{\mathrm{d}p}{\mathrm{d}t})^\prime = -\frac{p^\prime}{CR_d} + \frac{R_p+R_d}{CR_d}Q^n + R_p\frac{Q^n-Q^{n-1}}{\Delta t}.
\end{equation}

Similarly, to verify the Windkessel boundary implementation in a two-dimensional flow, we employed the 
same geometry, fluid properties, and inlet velocity profile as in the resistance model validation. 
The three-element Windkessel parameters for this setup are as proximal resistance of 
$R_p = 1.52 \times 10^6 \text{ kg} \cdot \text{m}^{-4}\text{s}^{-1}$, compliance 
of $C = 1.96 \times 10^{-7} \text{ m}^4\text{s}^2 \cdot \text{kg}^{-1}$ and distal resistance 
of $R_d = 6.85 \times 10^6 \text{ kg} \cdot \text{m}^{-4}\text{s}^{-1}$. The outlet pressure at the 
initial time is prescribed as $80 \text{ mmHg}$, corresponding to the lower limit of normal human blood pressure. 
Since WCSPH relies on pressure gradients for flow driving, the outlet pressure in simulations is offset by 
subtracting $80 \text{ mmHg}$ for numerical stability, while the Windkessel model is solved using absolute 
pressure values. For post-processing of SPH results, the pressure field is accordingly shifted back by adding $80 \text{ mmHg}$.

Unlike the resistance boundary, which relates pressure solely to the instantaneous flow rate, the Windkessel 
model also incorporates the time derivative of flow $\mathrm{d}Q/\mathrm{d}t$. Therefore, to ensure 
accurate initial conditions without integrating flow over time, the initial outlet flow rate is estimated by 
multiplying the instantaneous inlet velocity by the outlet cross-sectional area. Under the assumption of 
weakly compressible fluid, this provides a reasonable approximation, as the inlet and outlet flow rates are 
nearly balanced. For cases involving multiple outlets, the initial outlet flow rates are distributed 
proportionally based on their cross-sectional areas. Although omitting this initialization still leads to 
convergence after several cardiac cycles, proper specification of the initial outlet flow rate can 
significantly reduce the number of cycles required to reach a steady periodic state. Additionally, directly 
prescribing the initial outlet pressure (e.g., $80 \text{ mmHg}$) further accelerates convergence. This approach avoids 
the prolonged transient behavior observed in previous studies such as Ref.\cite{lu2024gpu}, where up to 8 to 
9 cycles were required to achieve stability.

\begin{figure}[htbp]
    \centering
    \begin{subfigure}[b]{0.48\textwidth}
        \includegraphics[width=\textwidth]{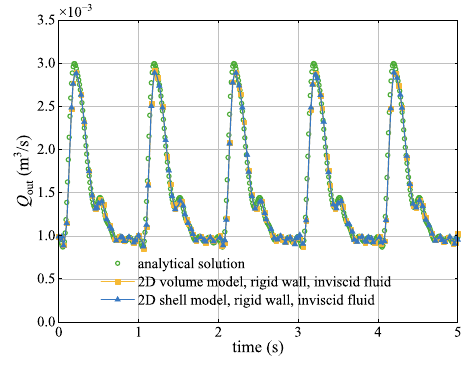}
        \caption{}
        \label{fig: 2d-windkessel-inviscid-Q}
    \end{subfigure}
    \begin{subfigure}[b]{0.48\textwidth}
        \includegraphics[width=\textwidth]{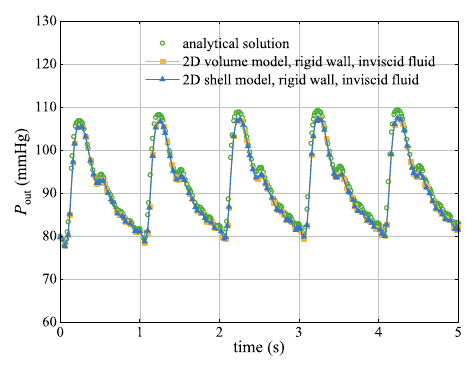}
        \caption{}
        \label{fig: 2d-windkessel-inviscid-P}
    \end{subfigure}
    \caption{Verification of Windkessel boundary implementation in a 2D rigid channel: comparison of (a) 
    outlet volume flow rate and (b) outlet pressure between SPH simulations with volume- and shell-based 
    wall models and the analytical solution.}
    \label{fig: 2d-windkessel-inviscid}
\end{figure}

As shown in Fig.\ref{fig: 2d-windkessel-inviscid}, the outlet flow rate and pressure predicted by SPH 
simulations using both volume and shell wall models exhibit excellent agreement with the analytical solution 
throughout the entire simulation period. This consistency confirms the accuracy and reliability of the 
implemented Windkessel boundary condition in capturing physiologically realistic hemodynamic responses.

\subsection{Convergence test for deformable wall}

In the previous section, we conducted the convergence study for the fluid dynamics; in this section, 
that for the solid mechanics of the deformable wall for both volume model and shell model has been carried out.

\begin{figure}[htbp]
    \centering
    \includegraphics[width=7cm]{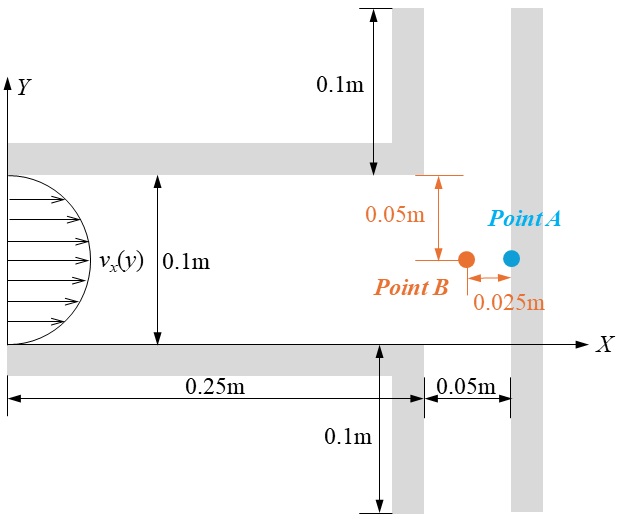}
    \caption{T-pipe flow: schematic illustration.}
    \label{fig: tpipe-schematic}
\end{figure}

The simple straight tube is replaced by the two-dimensional T-shaped pipe in this section to simulate the 
flow regimes at bifurcations. The geometry parameters are illustrated in Fig.\ref{fig: tpipe-schematic}. 
We set the fluid density to $\rho_f = 1000 \text{ kg}/\text{m}^3$ and the Reynolds number to $Re = 100$. 
The fluid viscosity is derived from the equation $\eta_f = \rho_f U_f d/Re$, 
with $U_f = 1.0 \text{ m/s}$ as the characteristic velocity and $d = 0.1 \text{ m}$ representing the inlet 
height. Deformable solids with thickness of $\text{TH} = 0.01 \text{ m}$ are modeled using parameters 
$\rho_s = 1200 \text{ kg}/\text{m}^3$, Young's modulus $E = 10 \text{ MPa}$, and Poisson's ratio of 0.45.

To simulate physiological conditions akin to blood flow, a pulsatile flow profile is implemented at the 
inlet, described by the following parabolic velocity distribution equation
\begin{equation}\label{eq: t-pipe-vinlet}
  v_x(y, t) = 1.5 v_{x,\text{avg}}\frac{4}{d^2} \left(\frac{d^2}{4}-(y-\frac{d}{2})^2 \right),
\end{equation}
with $v_{x,\text{avg}}$ varying as
\begin{equation}\label{eq: t-pipe-vavg}
v_{x,\text{avg}} =
\begin{cases}
0.5(1-\cos{\frac{\pi t}{T_{\text{ref}}}}) &{t < T_{\text{ref}}}\\
1.0 &{t \geq T_{\text{ref}}}\\
\end{cases}.
\end{equation}
The outlet pressure is set at zero.

\begin{figure}[htbp]
    \centering
    \includegraphics[width=9cm]{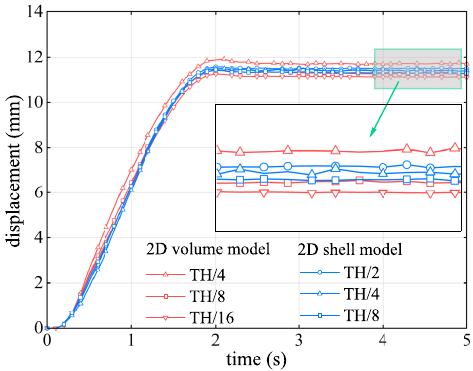}
    \caption{T-pipe flow: convergence study for displacement at Point A in the deformable wall.}
    \label{fig: 2d-tpipe-convergence}
\end{figure}

To evaluate the numerical convergence behavior of solid mechanics under the FSI framework, we monitor 
the displacement at Point A in the deformable wall, as shown in Fig.\ref{fig: 2d-tpipe-convergence}. 
For the volume model, particle spacings are from $dp^0 = \text{TH}/4$ due to the complete kernel support 
for near-wall fluid particles, while the shell model employs coarser resolutions beginning with 
$dp^0 = \text{TH}/2$ owing to its reduced dimensionality. The results demonstrate that the shell model 
achieves convergence at coarser resolutions compared to the volume model. Specifically, the shell model 
already yields stable displacement results at a resolution of $dp^0 = \text{TH}/4$ and even performs 
reasonably well at $dp^0 = \text{TH}/2$, while the volume model requires a finer resolution of 
$dp^0 = \text{TH}/8$ to attain similar accuracy. The computational efficiency of both approaches is 
quantitatively compared in Table \ref{table: tpipe-time}. All simulations were executed on an AMD Ryzen 
Threadripper PRO 5975WX 32-Core 3.60 GHz CPU. At the converged resolutions, the shell model with 
$dp^0 = \text{TH}/4$ speeds up more than 4 times compared to the volume model with $dp^0 = \text{TH}/8$.

\begin{table}[]
\centering
\caption{Time cost of T-pipe flow case.} \label{table: tpipe-time}
\begin{tabular}{cccclccc}
\hline
\multirow{3}{*}{particle spacing} & \multicolumn{3}{c}{volume model}                        &  & \multicolumn{3}{c}{shell model}                         \\ \cline{2-8} 
                                  & \multicolumn{2}{c}{particle number} & time cost (s)     &  & \multicolumn{2}{c}{particle number} & time cost (s)     \\ \cline{2-8} 
      & solid & fluid  &          &  & solid & fluid &         \\ \hline
TH/2  &       &        &          &  & 198   & 1600  & 56.283  \\
TH/4  & 1568  & 6400   & 138.175  &  & 398   & 6400  & 147.440 \\
TH/8  & 6272  & 25600  & 618.930  &  & 798   & 25600 & 521.438 \\
TH/16 & 25088 & 102400 & 5331.866 &  &       &       &         \\ \hline
\end{tabular}
\end{table}

\begin{figure}[htbp]
    \centering
    \includegraphics[width=14cm]{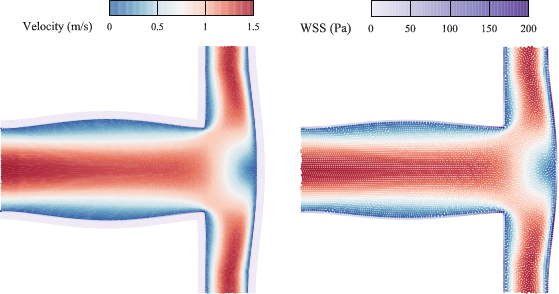}
    \caption{T-pipe flow: velocity and WSS distribution at the end of simulation (deformable wall). 
    Left: volume model with $dp^0 = \text{TH}/8$; right: shell model with $dp^0 = \text{TH}/4$.}
    \label{fig: 2d-tpipe-vipo-vel-wss-contour}
\end{figure}

\begin{figure}[htbp]
    \centering
    \includegraphics[width=9cm]{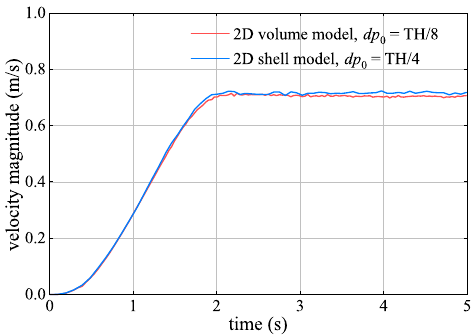}
    \caption{T-pipe flow: velocity magnitude at Point B.}
    \label{fig: 2d-tpipe-velocity-pointB}
\end{figure}

\begin{figure}[htbp]
    \centering
    \includegraphics[width=9cm]{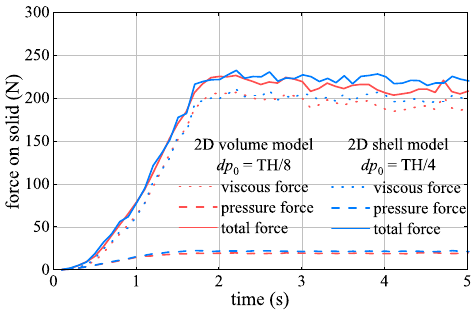}
    \caption{T-pipe flow: pressure and viscous forces on wall.}
    \label{fig: 2d-tpipe-force-on-solid}
\end{figure}

To demonstrate that the shell model, even at coarser spatial resolutions, can maintain high fidelity for fluid 
dynamics in FSI simulations, further comparative analyses for flow fields are performed between the volume 
model with particle spacing $dp^0 = \text{TH}/8$ and the shell model with $dp^0 = \text{TH}/4$. At the end of 
the simulation, where the flow reaches a steady state, Fig.\ref{fig: 2d-tpipe-vipo-vel-wss-contour} provides 
snapshots of the velocity and WSS distributions and a high degree of agreement is achieved between the two 
models in both flow and WSS fields. The temporal evolution of velocity magnitude at Point B, located along the 
central axis of the T-junction, is illustrated in Fig.\ref{fig: 2d-tpipe-velocity-pointB}. The shell model 
exhibits closely matching velocity trajectories over time with the volume model, further validating its 
effectiveness in capturing transient flow behavior. Additionally, the dynamic forces acting on the deformable 
wall, including viscous and pressure forces, are detailed in Fig.\ref{fig: 2d-tpipe-force-on-solid}. The shell 
and volume models display similar patterns of force evolution over time. These qualitative and quantitative 
comparisons confirm that, for fluid dynamics alone in this case, a spatial resolution of $dp^0 = \text{TH}/4$ 
is already sufficient to capture the key flow features with high fidelity. In contrast, the finer 
resolution $dp^0 = \text{TH}/8$ required by the volume model to ensure solid mechanical convergence results in 
redundant resolution for the fluid domain, leading to unnecessary computational overhead. 
Therefore, the shell model provides a more efficient and balanced approach for FSI simulations.

\subsection{Assessment of Wall Compliance}
\begin{figure}[htbp]
    \centering
    \includegraphics[width=6cm]{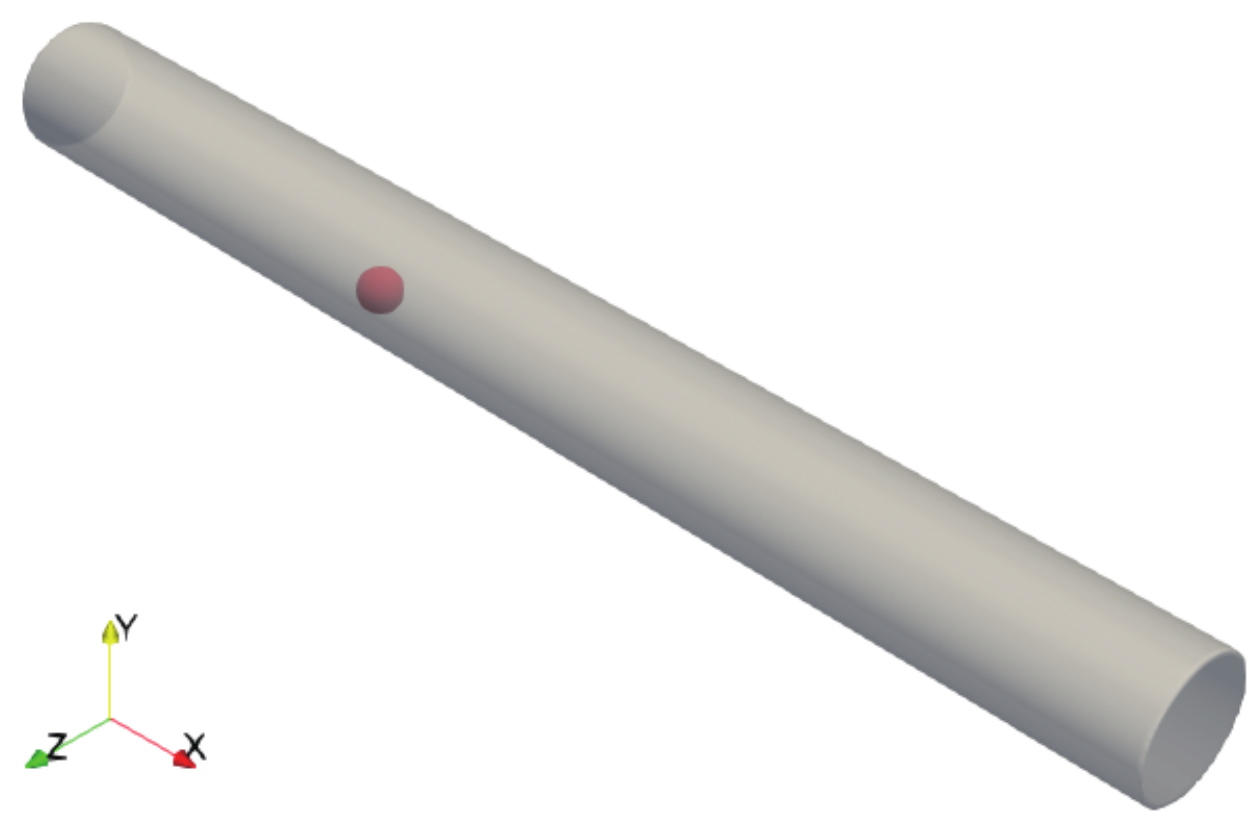}
    \caption{Stenosed cylinder: geometric illustration. The red sphere represents a localized stenosis 
    introduced to emulate vascular narrowing.}
    \label{fig: stenosed-cylinder-geo}
\end{figure}

To investigate the effects of arterial wall compliance, we conduct comparative simulations using both 
deformable and rigid wall configurations based on the shell model. The test geometry is a straight cylindrical 
vessel aligned along the positive $x$-axis, with a diameter of $D = 6$ mm and a length of $L = 60$ mm. 
To approximate a stenotic condition, a spherical obstruction as a rigid body with a radius of $r = 1$ mm is placed at the 
coordinate position (20, -1.5, 0) mm, as illustrated in Fig.\ref{fig: stenosed-cylinder-geo}. The fluid is 
modeled as an incompressible Newtonian fluid with density of $\rho_f = 1060 \text{ kg}/\text{m}^3$ and dynamic 
viscosity of $\eta_f = 0.004 \text{ Pa} \cdot \text{s}$. For the deformable configuration, the vessel wall is 
represented by a shell with a thickness of 0.6 mm, a density of $\rho_s = 1000 \text{ kg}/\text{m}^3$, 
Young's modulus $ E= 100\text{ MPa}$, and Poisson's ratio of 0.3. A pulsatile inflow velocity is prescribed 
according to Eq.\ref{eq: resistance-vinlet} and set to parabolic distribution, while a resistance-type outlet 
boundary condition with resistance $R = 5 \times 10^6 \text{ kg} \cdot \text{m}^{-4}\text{s}^{-1}$ is applied. 
The initial particle spacing is set to $dp^0 = 0.3$ mm for both the fluid and vessel wall domains, and refined 
to $0.5dp^0$ for the spherical stenosis to better resolve the geometry.

\begin{figure}[htbp]
    \centering
    \begin{subfigure}[b]{0.48\textwidth}
        \includegraphics[width=\textwidth]{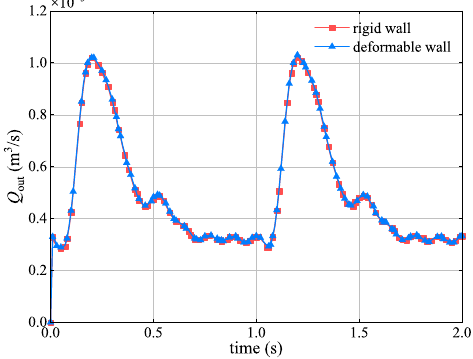}
        \caption{}
        \label{fig: stenosed-Q}
    \end{subfigure}
    \begin{subfigure}[b]{0.48\textwidth}
        \includegraphics[width=\textwidth]{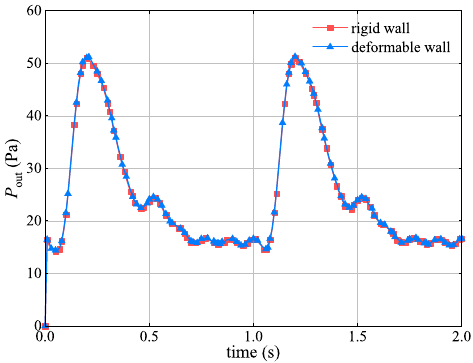}
        \caption{}
        \label{fig: stenosed-P}
    \end{subfigure}
    \caption{Stenosed cylinder: comparison of (a) outlet volume flow rate and (b) outlet pressure with rigid 
    and deformable configurations for the vessel wall.}
    \label{fig: stenosed-QandP}
\end{figure}

\begin{figure}[htbp]
    \centering
    \includegraphics[width=10cm]{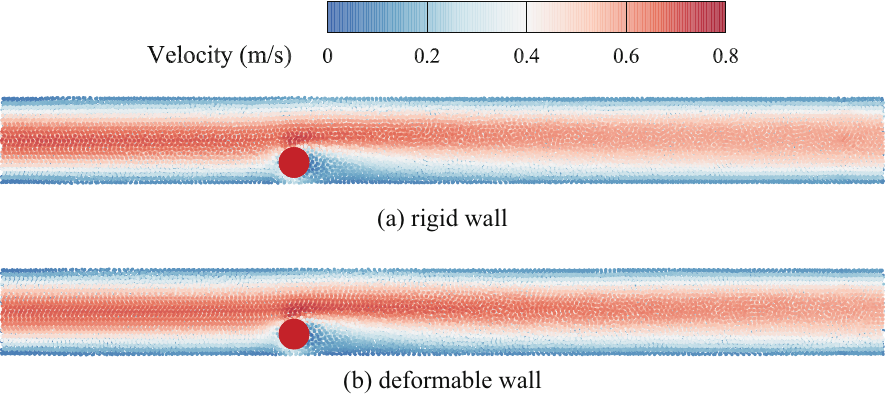}
    \caption{Stenosed cylinder: velocity distribution at the peak flow point during the second period.}
    \label{fig: stenosed-velocity-distribution}
\end{figure}

\begin{figure}[htbp]
    \centering
    \includegraphics[width=9cm]{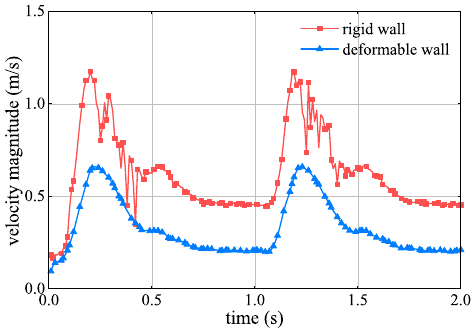}
    \caption{Stenosed cylinder: velocity magnitude at the center of midsection.}
    \label{fig: stenosed-midpoint-velocity}
\end{figure}

\begin{figure}[htbp]
    \centering
    \includegraphics[width=10cm]{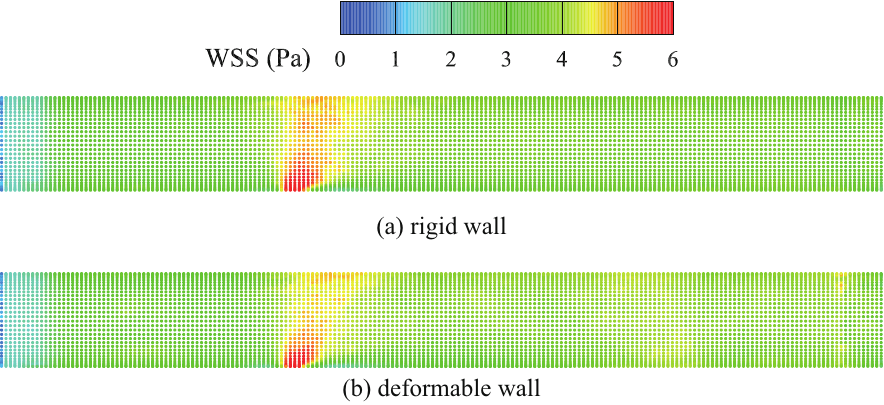}
    \caption{Stenosed cylinder: WSS distribution at the peak flow point during the second period.}
    \label{fig: stenosed-WSS-distribution}
\end{figure}

Fig.\ref{fig: stenosed-QandP} presents the outlet volume flow rate and pressure for both rigid and deformable 
wall configurations. Due to the weakly compressible nature of the fluid and the relatively small structural 
deformation, the results exhibit minimal differences between the two cases. Similarly, the velocity 
distributions at the peak flow point during the second cycle, shown in 
Fig.\ref{fig: stenosed-velocity-distribution}, reveal comparable patterns for both wall models. However, as 
shown in Fig.\ref{fig: stenosed-midpoint-velocity}, the deformable wall configuration demonstrates improved 
damping of wave oscillations, resulting in lower and smoother temporal variations of velocity at the midsection 
center. Notably, the WSS value, as one of the key parameter for hemodynamics, reveals significant discrepancies 
between the two cases as shown in Fig.\ref{fig: stenosed-WSS-distribution}, emphasizing the influence of wall 
compliance on local flow patterns.

\subsection{Patient-specific case I: carotid artery}

\begin{figure}[htbp]
    \centering
    \includegraphics[width=9cm]{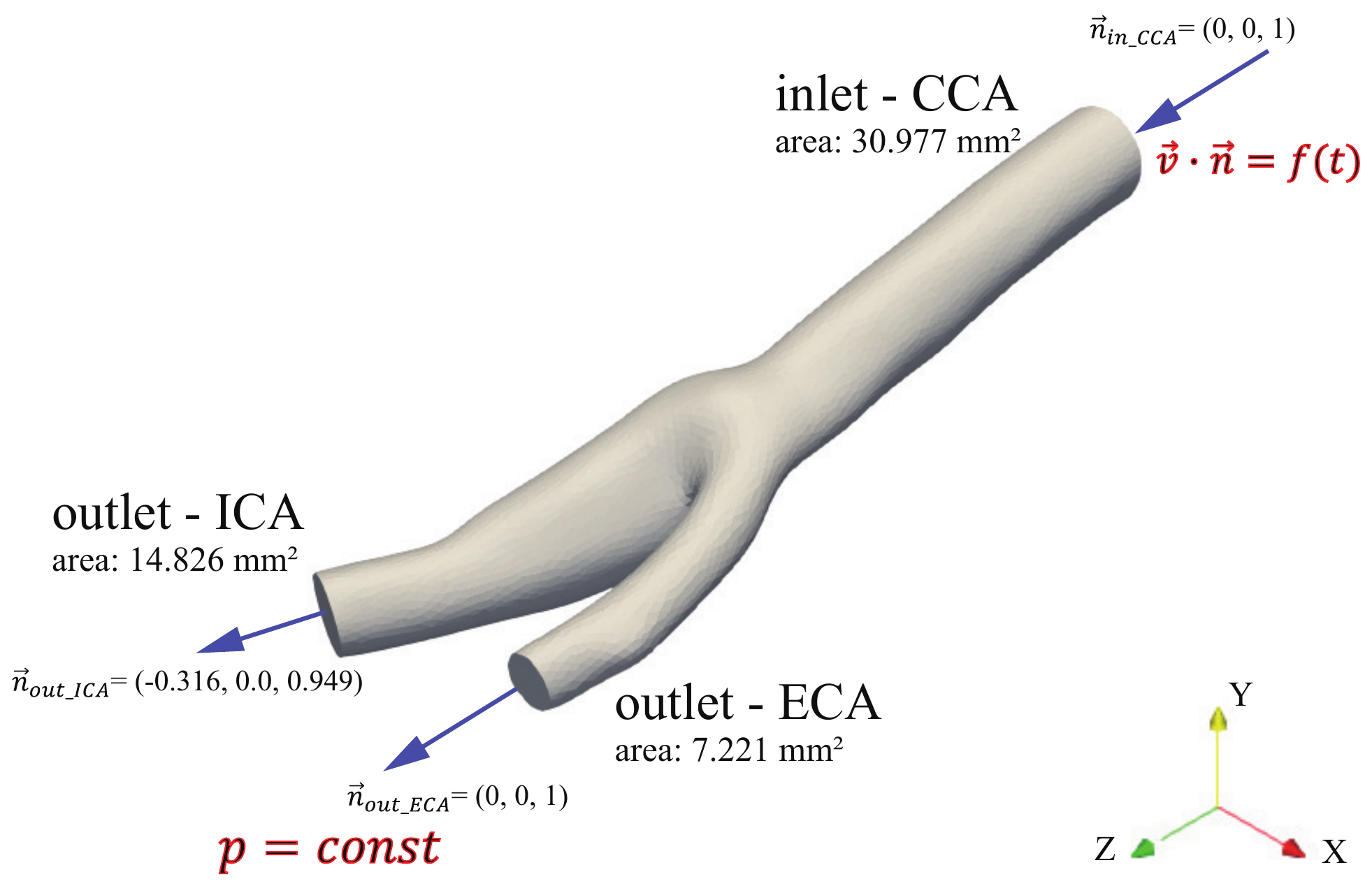}
    \caption{Carotid artery: geometric illustration.}
    \label{fig: carotid-fluid-geo}
\end{figure}

Following the validation in idealized geometries, we now apply the shell model to patient-specific cases. 
The first case we use carotid artery from the official benchmark case provided by ANSYS Fluent 
(\href{https://innovationspace.ansys.com/product/3d-bifurcating-artery/}{3D Bifurcating Artery}). 
The geometry of the carotid artery used in this study is illustrated in Fig.\ref{fig: carotid-fluid-geo}. 
Physiologically, blood enters the domain through the common carotid artery (CCA) and is distributed through 
two primary branches: the external carotid artery (ECA), which supplies blood to the muscles of face and neck, 
and the internal carotid artery (ICA), which delivers blood to the brain. In the present simulation, 
blood is modeled as a Newtonian fluid with a density of $\rho_f = 1060 \text{ kg}/\text{m}^3$ and a dynamic 
viscosity of $\eta_f = 0.0035 \text{ Pa} \cdot \text{s}$, following values reported in the literature \cite{lopes2019influence}.

At the inlet boundary (CCA), a time-dependent velocity function with a plug flow profile is prescribed to 
capture the pulsatile nature of blood circulation. The waveform characterizes distinct systolic and diastolic 
phases and is defined as
\begin{equation}\label{eq: carotid-inlet-vel}
v(t) =
\begin{cases}
0.5 \sin[4\pi(t+0.0160236)] &{0.5n < t \leq 0.5n + 0.218} \\
0.1 &{0.5n + 0.218 < t \leq 0.5(n+1)}, \\
\end{cases}
\end{equation}
where $ n = 0, 1, 2...$ denotes the cycle number. The flow direction is oriented along the normal vector of the 
inlet boundary surface. At the outlet boundaries (ICA and ECA), a constant pressure of $100 \text{ mmHg}$ is 
imposed, consistent with physiological arterial conditions. However, in the WCSPH formulation used in this 
study, the flow field is governed by pressure differences rather than absolute pressure values. 
Therefore, to maintain numerical stability, the outlet pressure in this case is normalized to zero in the SPH simulations, 
and the inlet pressure is adjusted accordingly by the governing equations to preserve the intended pressure gradient. 
During post-processing, the resulting pressure field is uniformly shifted by $100 \text{ mmHg}$ to restore the 
original reference pressure level, allowing for meaningful comparison with clinical and literature-reported values.

To validate the physical fidelity of the SPH models in simulating fluid dynamics and hemodynamic behavior, 
the vessel wall is initially modeled as rigid, omitting wall deformation effects. Following the grid 
independence study, the particle spacing in the fluid domain is set to $dp_f = 0.2 \text{ mm}$, resulting in 
approximately 224,000 particles. For the solid domain, a finer resolution is adopted with a particle spacing 
of $dp_s = 0.5dp_f$. Consequently, the volume model comprises approximately 840,000 solid particles, 
while the shell model uses around 121,000 particles.

\begin{figure}[htbp]
    \centering
    \includegraphics[width=12cm]{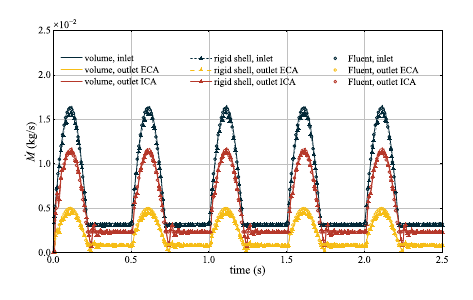}
    \caption{Hemodynamics in carotid artery (rigid wall): comparison of mass flow rate at three boundaries 
    among SPH results with volume and shell models as well as FVM result.}
    \label{fig: carotid-mass-flow-rate}
\end{figure}

Fig.\ref{fig: carotid-mass-flow-rate} presents the time histories of the mass flow rate at the inlet and two 
outlets, comparing results from the SPH volume and rigid shell model, as well as the FVM implemented in ANSYS 
Fluent with 142,833 cells. All models successfully capture the characteristic pulsatile waveform across the 
cardiac cycle, demonstrating strong agreement in both waveform shape and amplitude. Notably, the SPH results 
exhibit slight oscillations at the onset of diastole, which are attributed to the weakly compressible formulation 
of the SPH method. These fluctuations are physically reasonable and consistent with prior studies employing 
WCSPH, reflecting transient acoustic effects during rapid pressure relaxation.

\begin{figure}
    \centering
    \begin{subfigure}[b]{0.5\textwidth}
        \includegraphics[width=\textwidth]{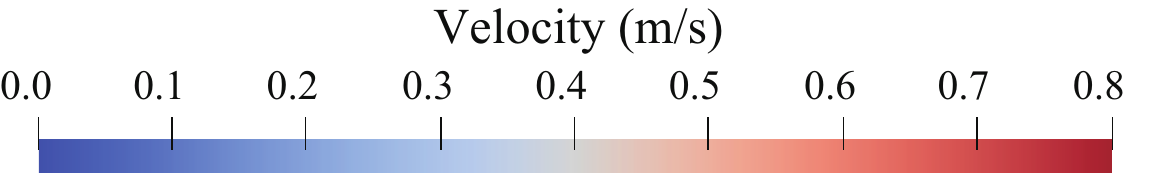}
    \end{subfigure}
    \begin{subfigure}[b]{\textwidth}
        \centering
        \begin{subfigure}[b]{0.32\textwidth}
            \caption*{\text{volume model}}
            \includegraphics[width=\textwidth]{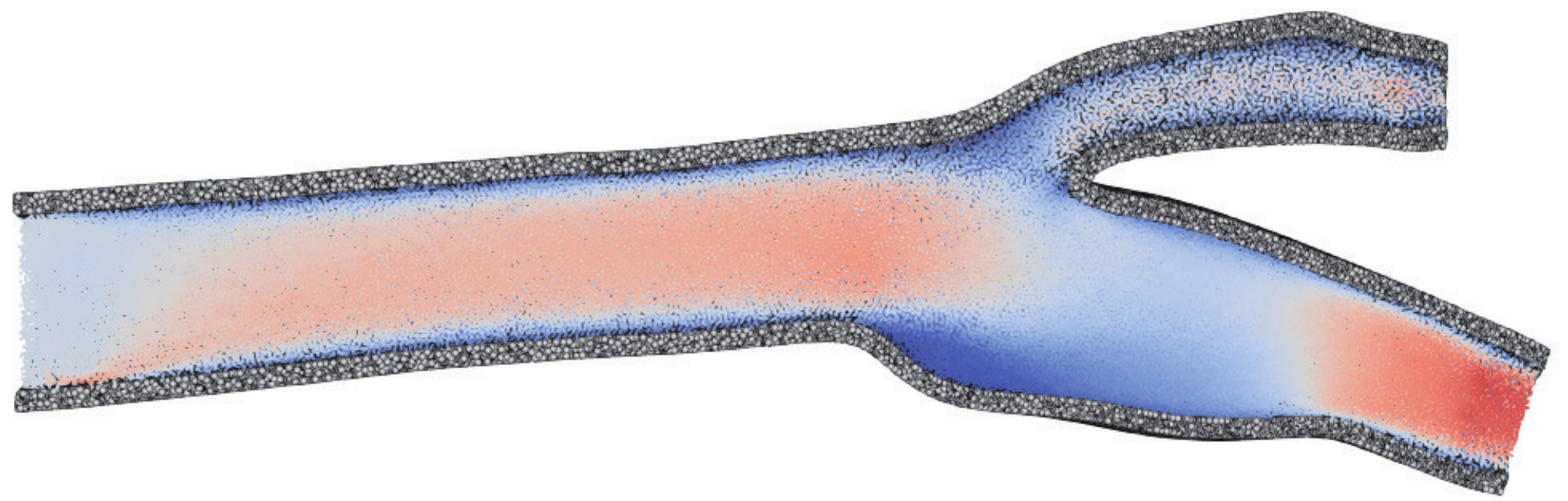}
        \end{subfigure}
        \begin{subfigure}[b]{0.32\textwidth}
            \caption*{\text{shell model}}
            \includegraphics[width=\textwidth]{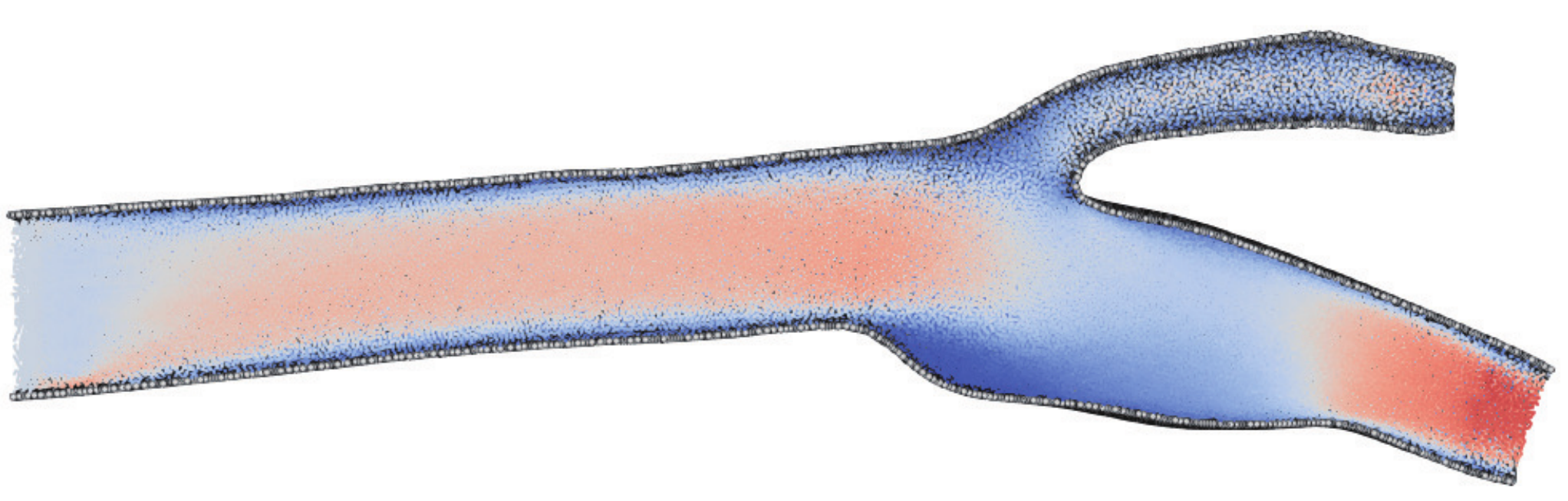}
        \end{subfigure}
        \begin{subfigure}[b]{0.32\textwidth}
            \caption*{\text{FVM}}
            \includegraphics[width=\textwidth]{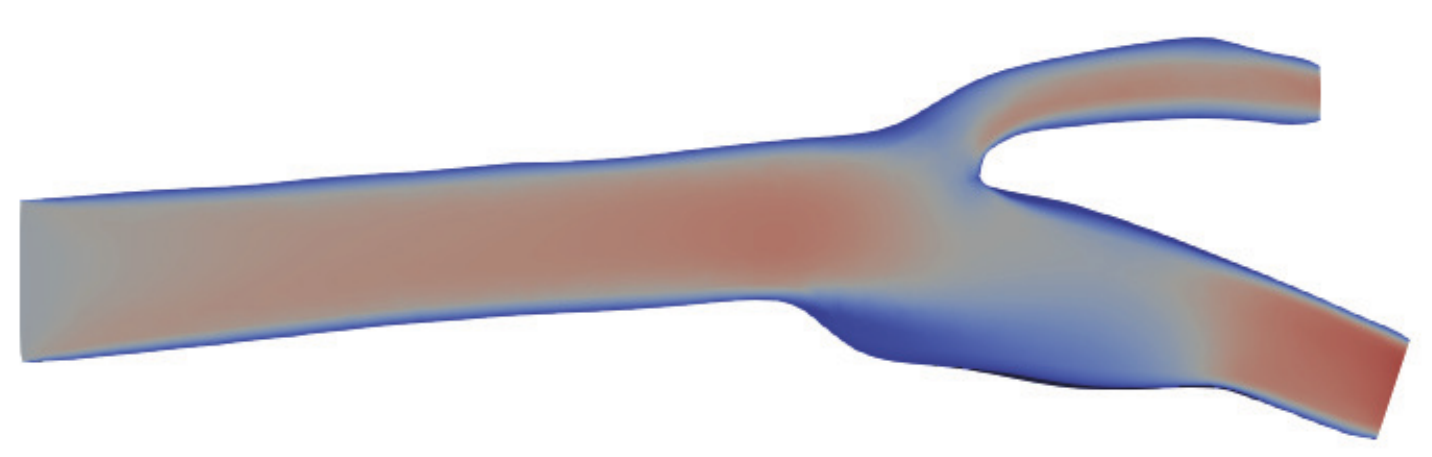}
        \end{subfigure}
        \caption*{\text{(a)} $t$ = 2.05s}
    \end{subfigure}
    \begin{subfigure}[b]{\textwidth}
        \centering
        \begin{subfigure}[b]{0.32\textwidth}
            \includegraphics[width=\textwidth]{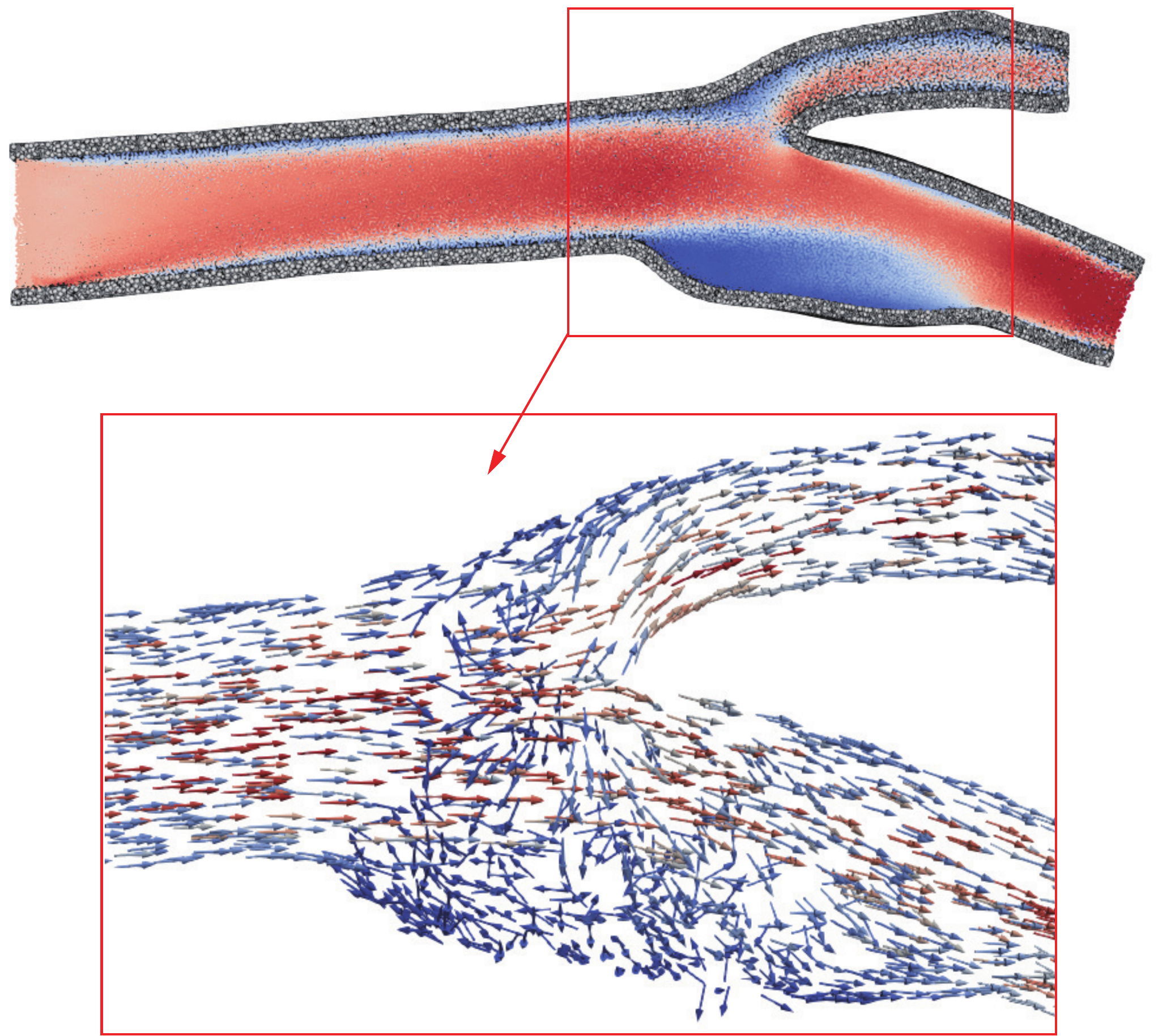}
        \end{subfigure}
        \begin{subfigure}[b]{0.32\textwidth}
            \includegraphics[width=\textwidth]{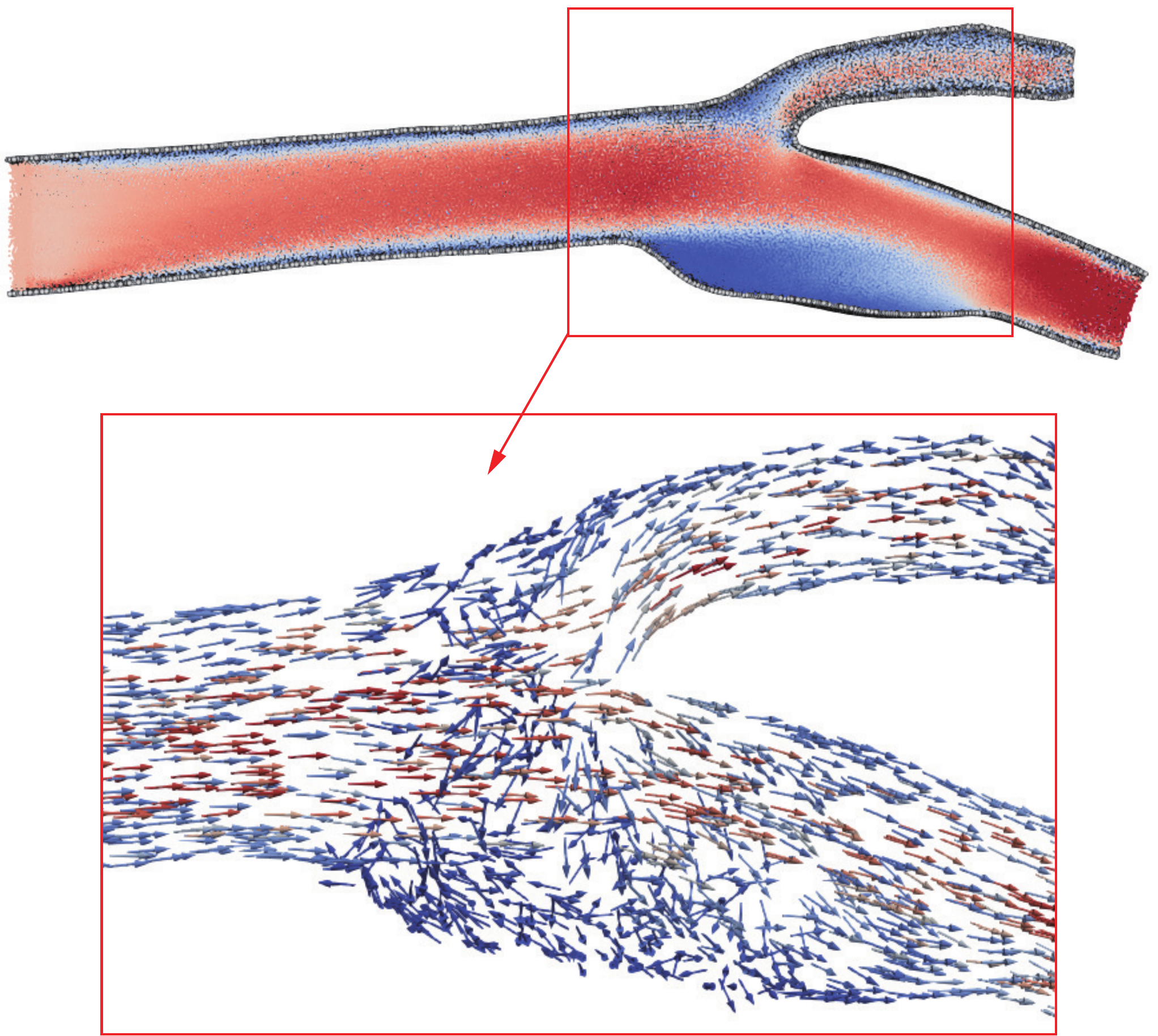}
        \end{subfigure}
        \begin{subfigure}[b]{0.32\textwidth}
            \includegraphics[width=\textwidth]{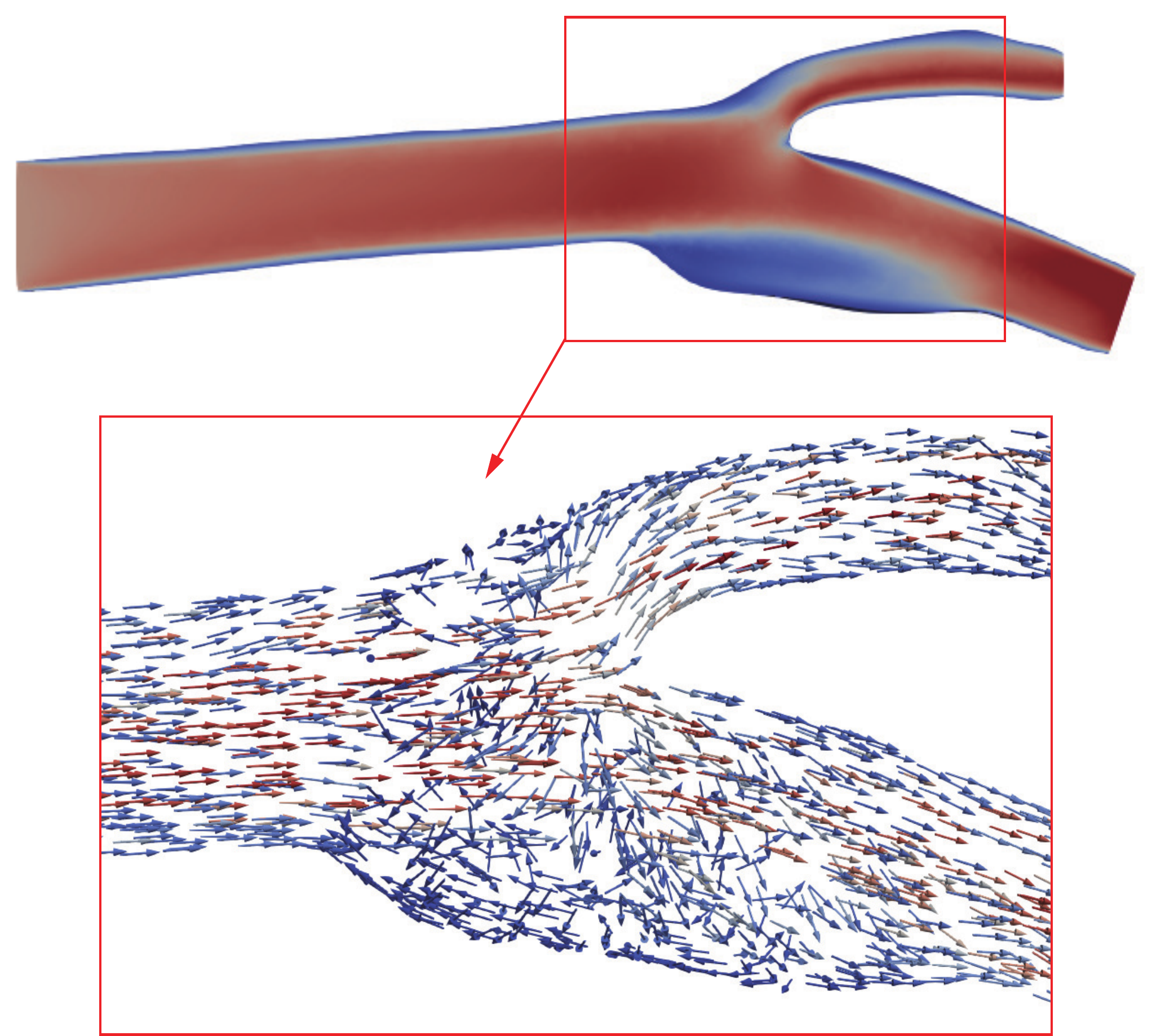}
        \end{subfigure}
        \caption*{\text{(b)} $t$ = 2.1s}
    \end{subfigure}
    \begin{subfigure}[b]{\textwidth}
        \centering
        \begin{subfigure}[b]{0.32\textwidth}
            \includegraphics[width=\textwidth]{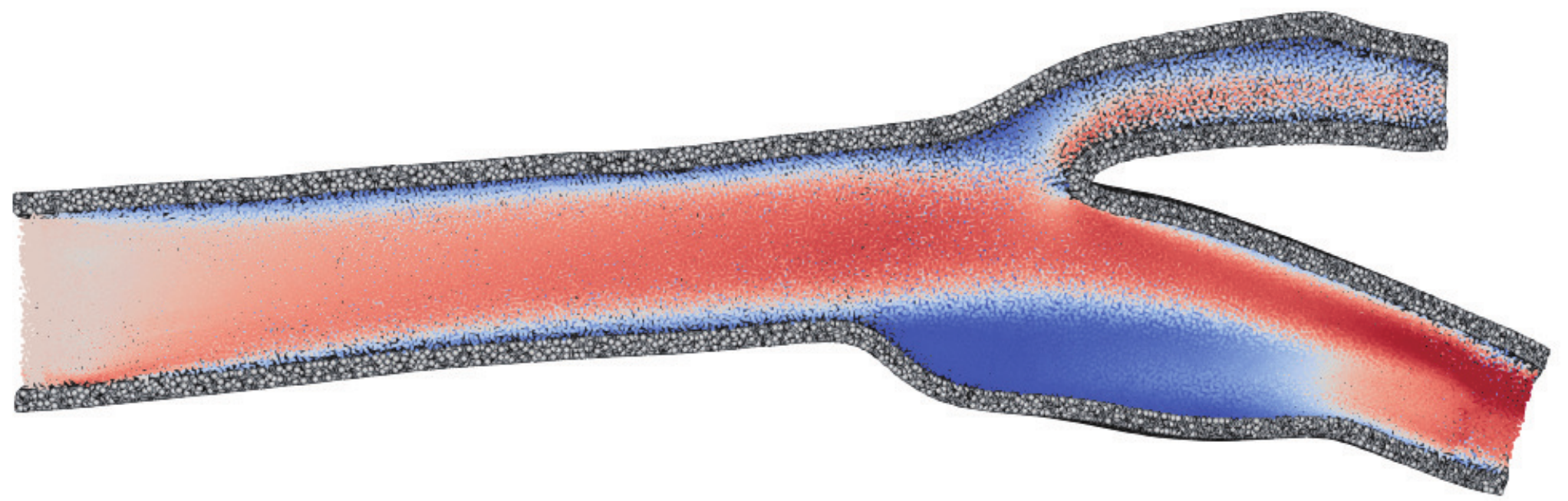}
        \end{subfigure}
        \begin{subfigure}[b]{0.32\textwidth}
            \includegraphics[width=\textwidth]{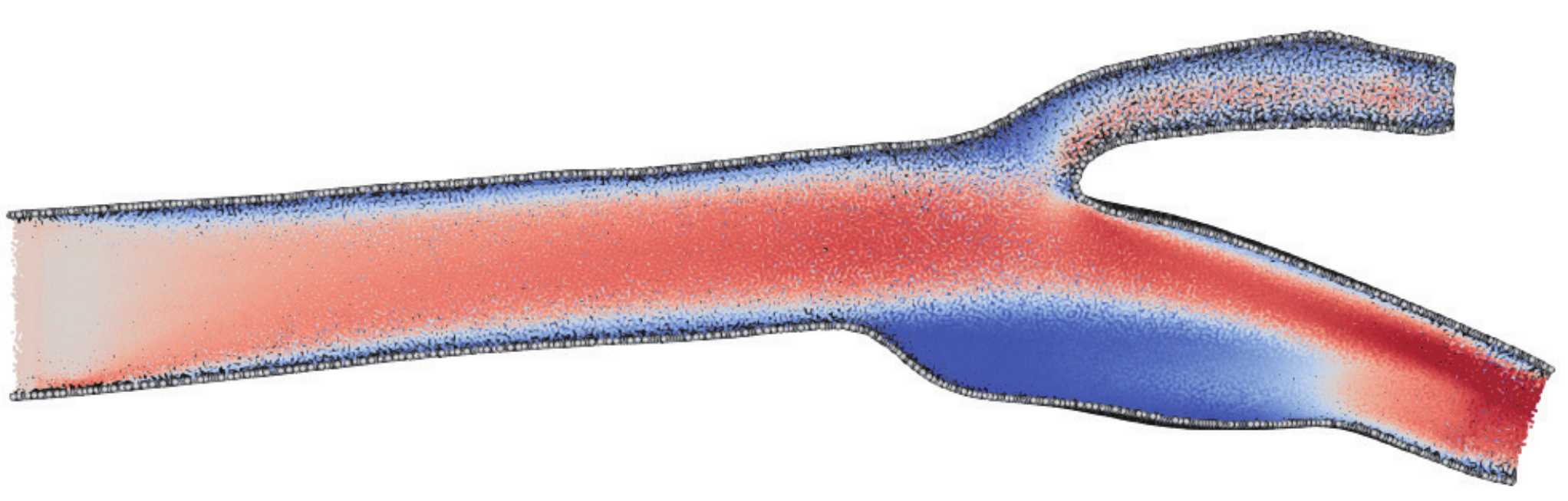}
        \end{subfigure}
        \begin{subfigure}[b]{0.32\textwidth}
            \includegraphics[width=\textwidth]{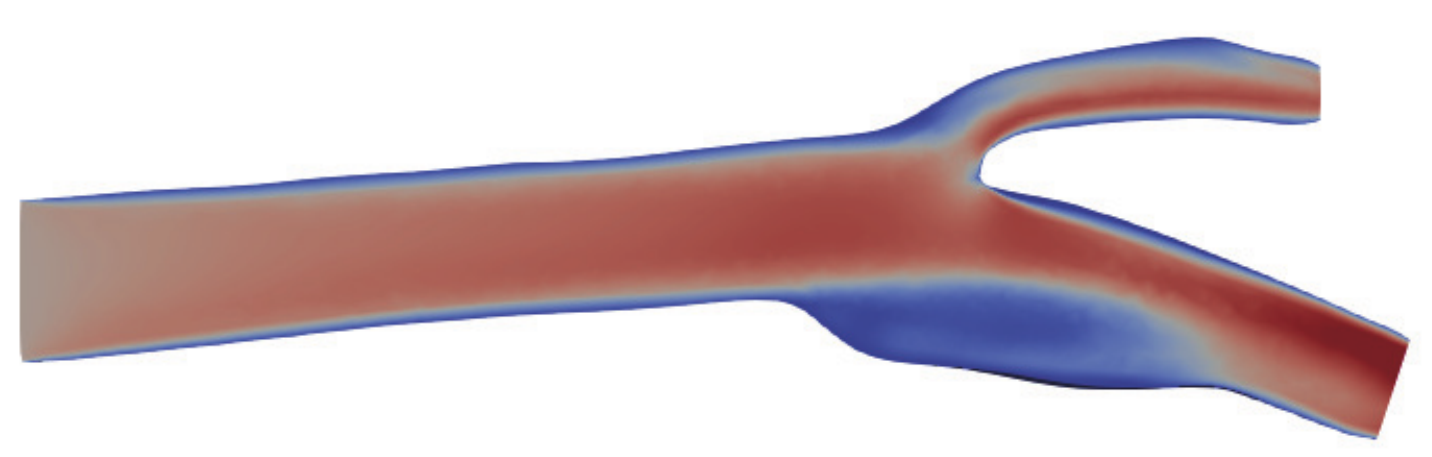}
        \end{subfigure}
        \caption*{\text{(c)} $t$ = 2.15s}
    \end{subfigure}
    \begin{subfigure}[b]{\textwidth}
        \centering
        \begin{subfigure}[b]{0.32\textwidth}
            \includegraphics[width=\textwidth]{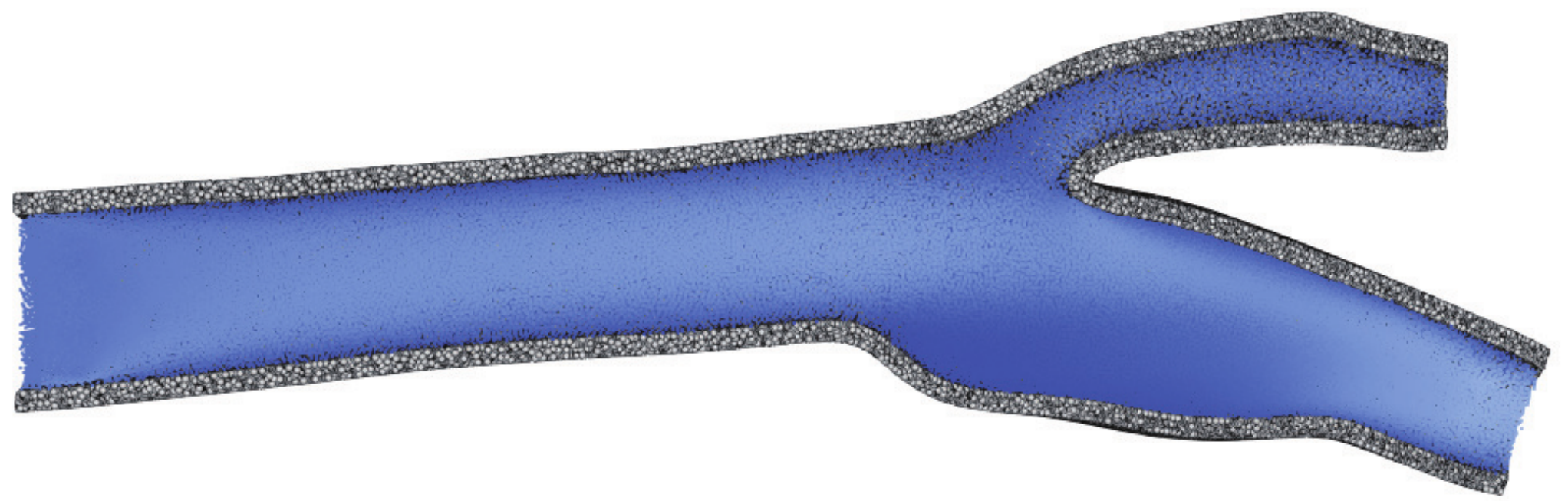}
        \end{subfigure}
        \begin{subfigure}[b]{0.32\textwidth}
            \includegraphics[width=\textwidth]{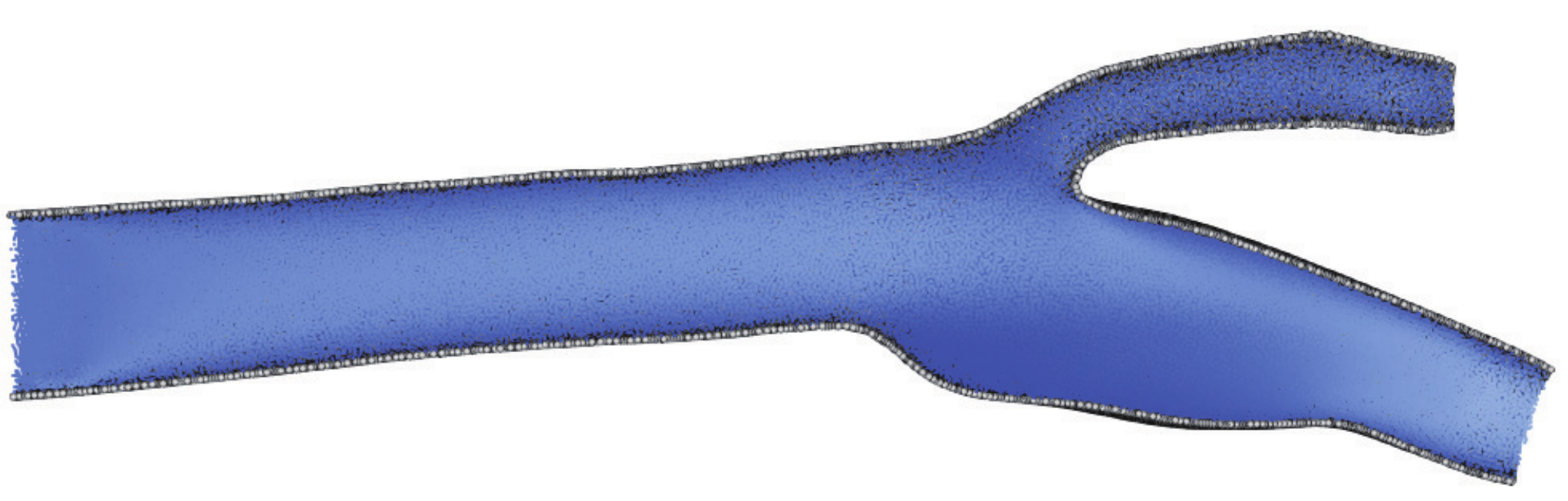}
        \end{subfigure}
        \begin{subfigure}[b]{0.32\textwidth}
            \includegraphics[width=\textwidth]{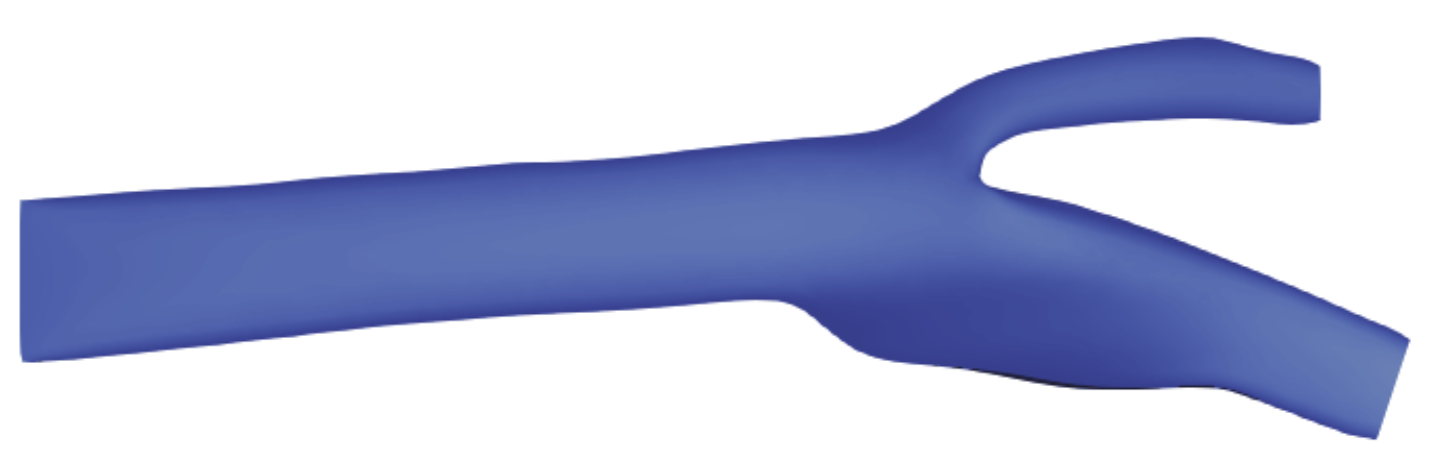}
        \end{subfigure}
        \caption*{\text{(d)} $t$ = 2.4s}
    \end{subfigure}
    \caption{Hemodynamics in carotid artery (rigid wall): velocity distributions at four time instants in the 
    fifth cardiac cycle. Left: SPH result with volume-based wall model; middle: SPH result with shell-based 
    wall model; right: FVM reference solution.}
    \label{fig: carotid-VIPO-velocity}
\end{figure}

\begin{figure}
    \centering
    \begin{subfigure}[b]{0.5\textwidth}
        \includegraphics[width=\textwidth]{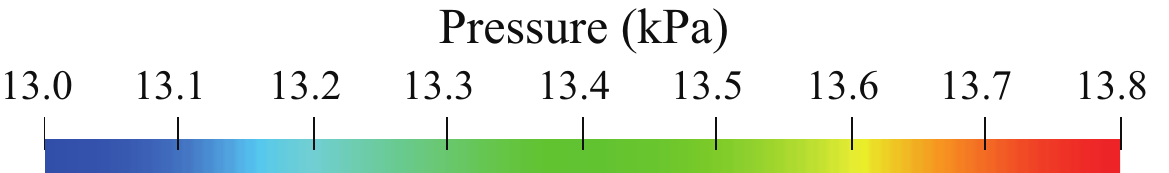}
    \end{subfigure}
    \begin{subfigure}[b]{\textwidth}
        \centering
        \begin{subfigure}[b]{0.32\textwidth}
            \caption*{\text{volume model}}
            \includegraphics[width=\textwidth]{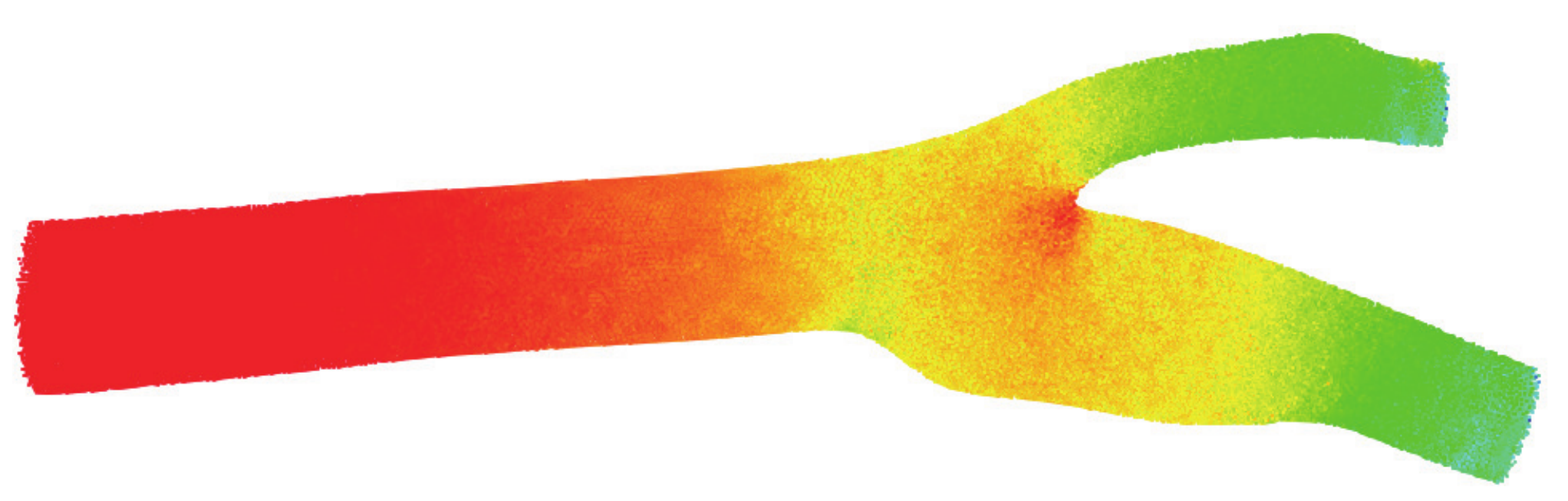}
        \end{subfigure}
        \begin{subfigure}[b]{0.32\textwidth}
            \caption*{\text{shell model}}
            \includegraphics[width=\textwidth]{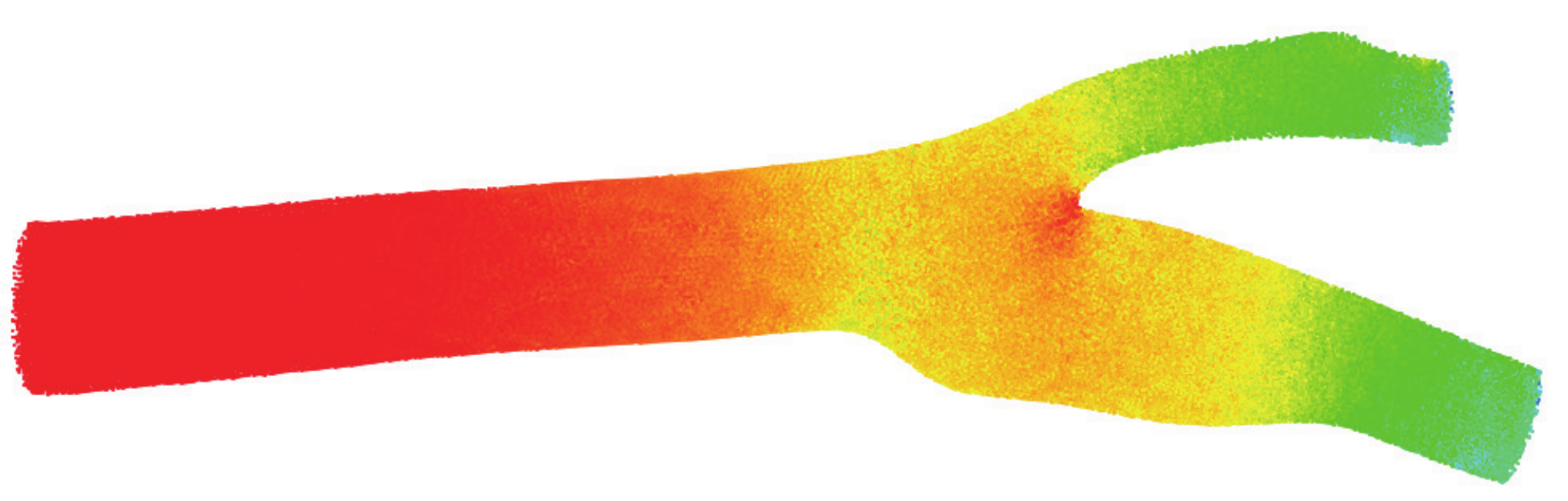}
        \end{subfigure}
        \begin{subfigure}[b]{0.32\textwidth}
            \caption*{\text{FVM}}
            \includegraphics[width=\textwidth]{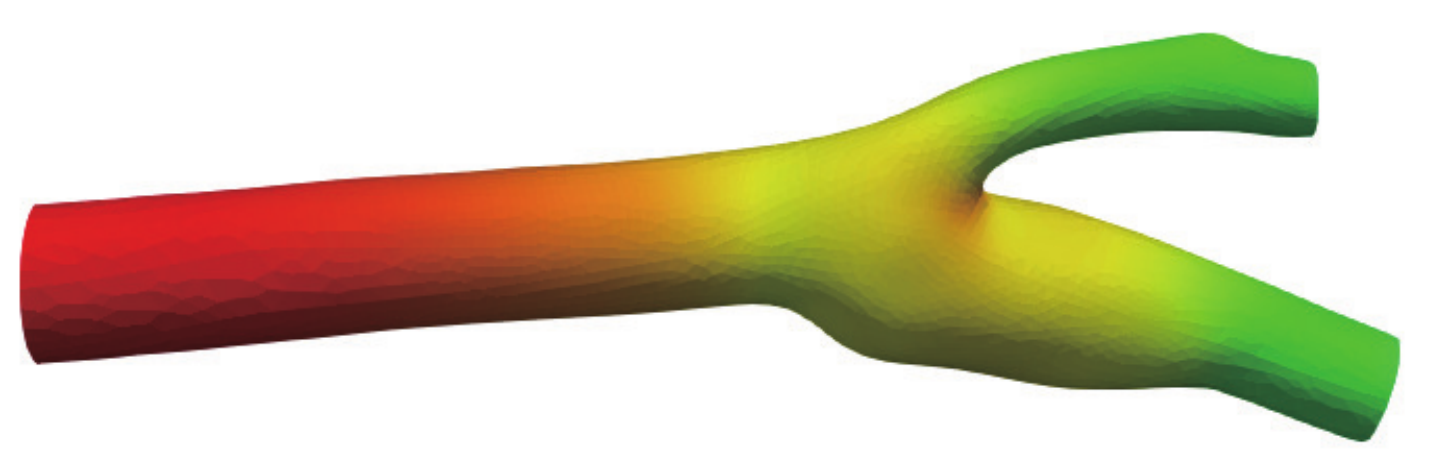}
        \end{subfigure}
        \caption*{\text{(a)} $t$ = 2.05s}
    \end{subfigure}
    \begin{subfigure}[b]{\textwidth}
        \centering
        \begin{subfigure}[b]{0.32\textwidth}
            \includegraphics[width=\textwidth]{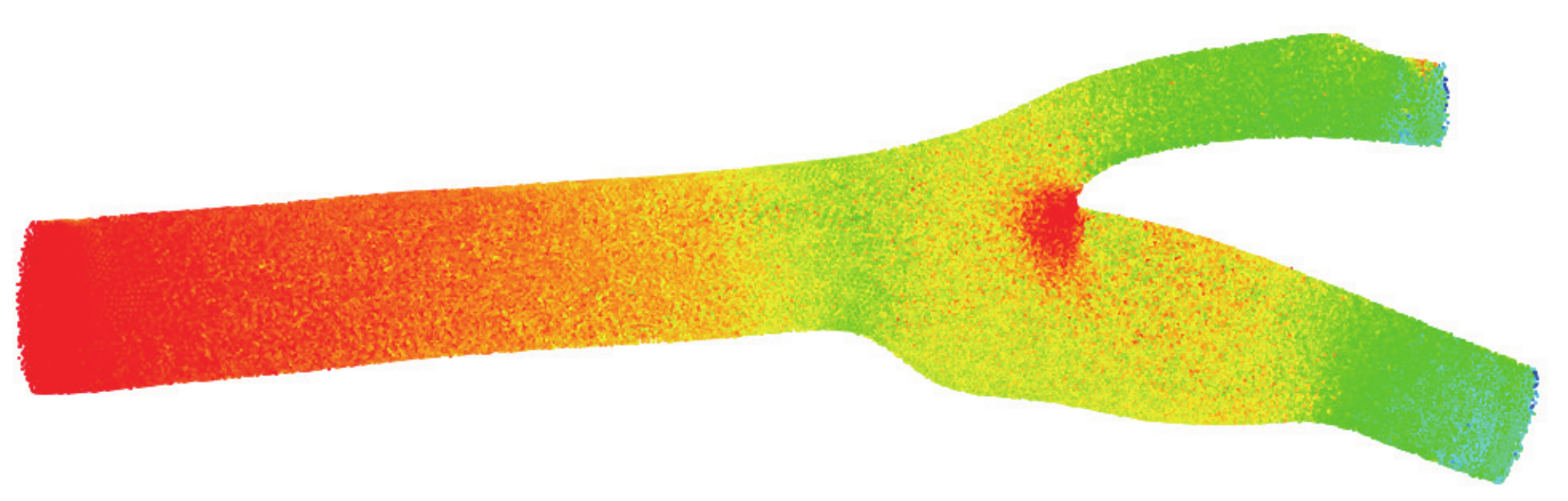}
        \end{subfigure}
        \begin{subfigure}[b]{0.32\textwidth}
            \includegraphics[width=\textwidth]{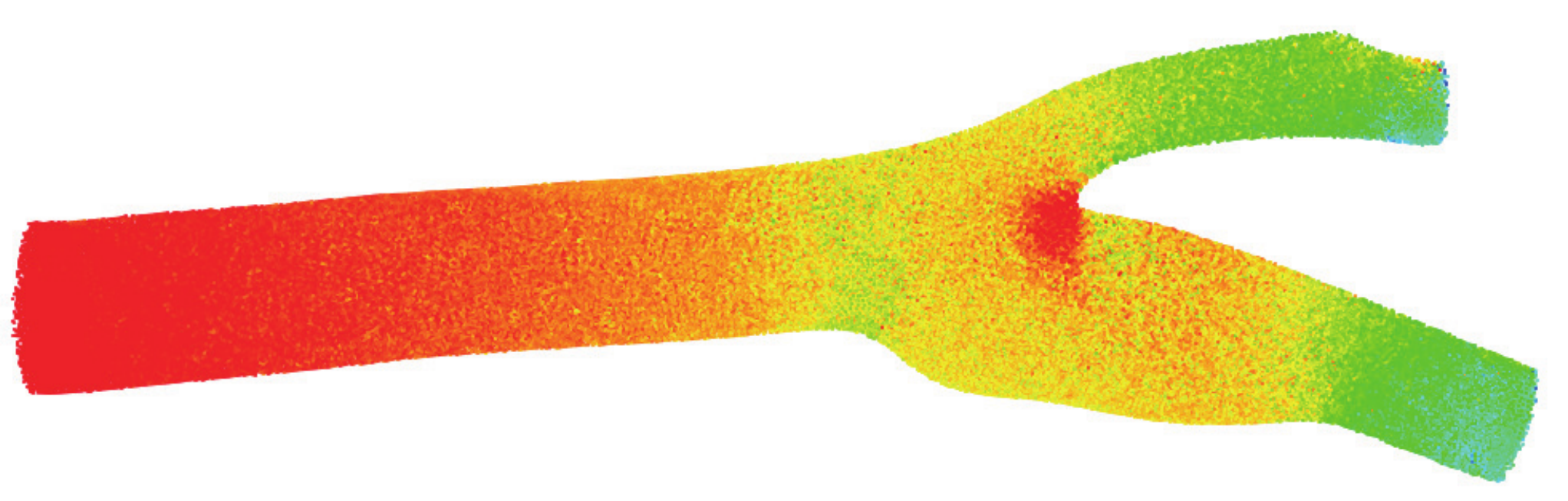}
        \end{subfigure}
        \begin{subfigure}[b]{0.32\textwidth}
            \includegraphics[width=\textwidth]{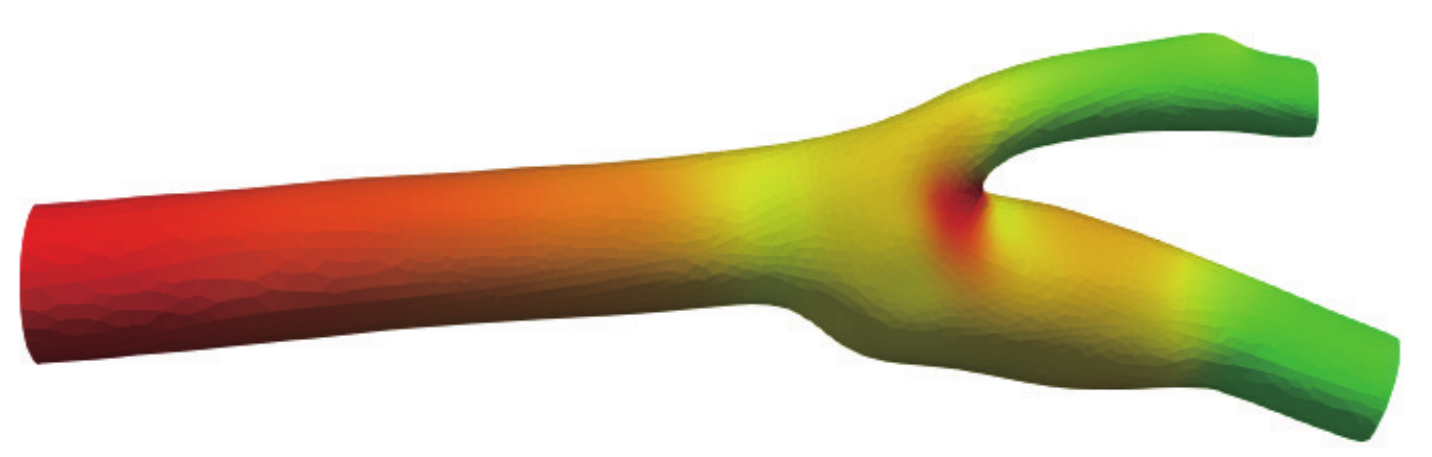}
        \end{subfigure}
        \caption*{\text{(b)} $t$ = 2.1s}
    \end{subfigure}
    \begin{subfigure}[b]{\textwidth}
        \centering
        \begin{subfigure}[b]{0.32\textwidth}
            \includegraphics[width=\textwidth]{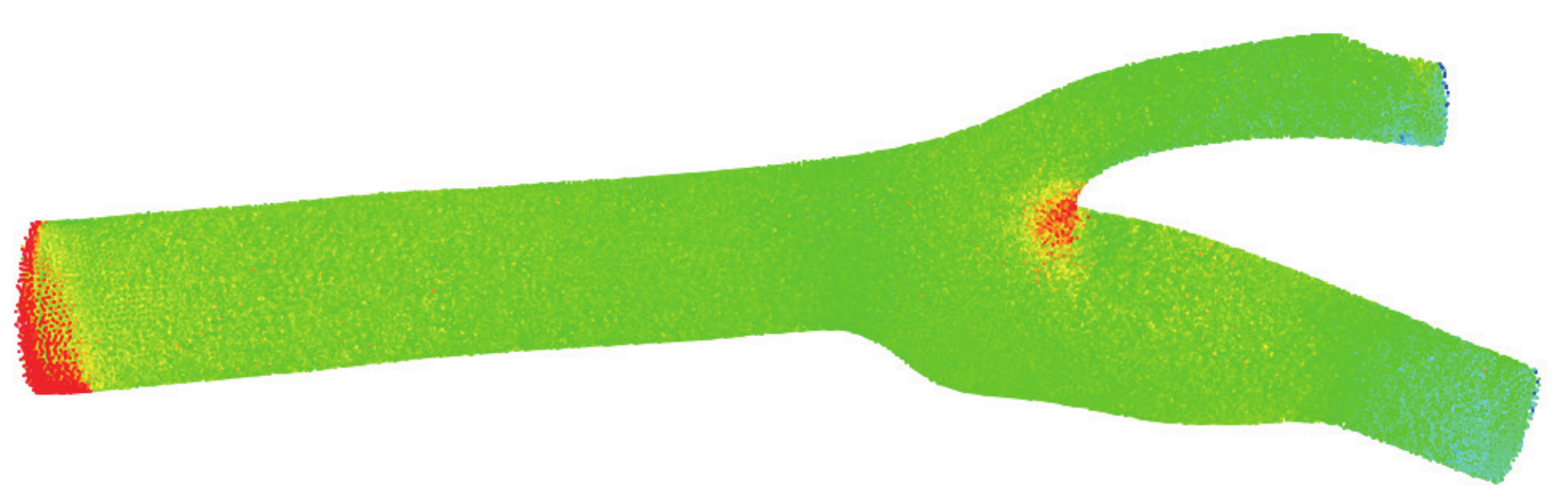}
        \end{subfigure}
        \begin{subfigure}[b]{0.32\textwidth}
            \includegraphics[width=\textwidth]{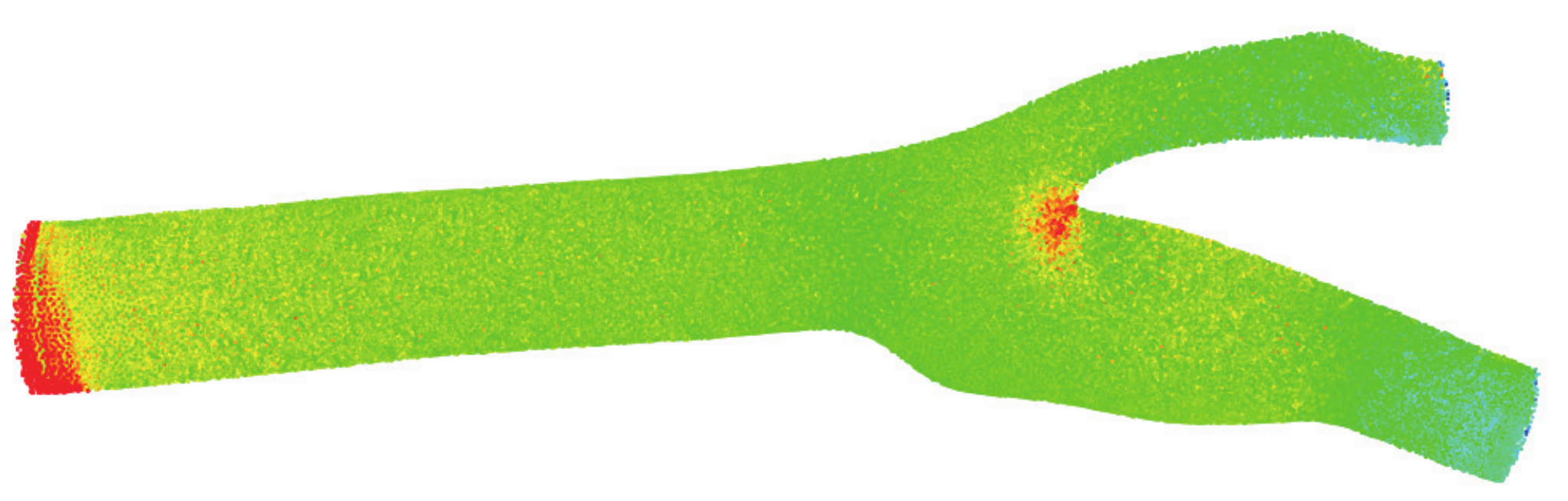}
        \end{subfigure}
        \begin{subfigure}[b]{0.32\textwidth}
            \includegraphics[width=\textwidth]{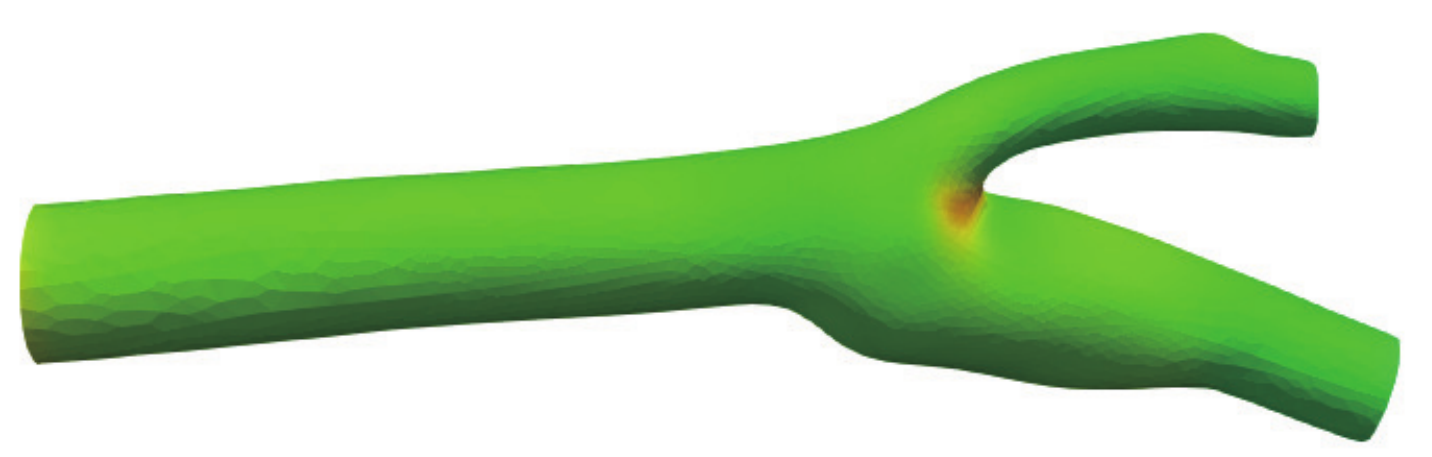}
        \end{subfigure}
        \caption*{\text{(c)} $t$ = 2.15s}
    \end{subfigure}
    \begin{subfigure}[b]{\textwidth}
        \centering
        \begin{subfigure}[b]{0.32\textwidth}
            \includegraphics[width=\textwidth]{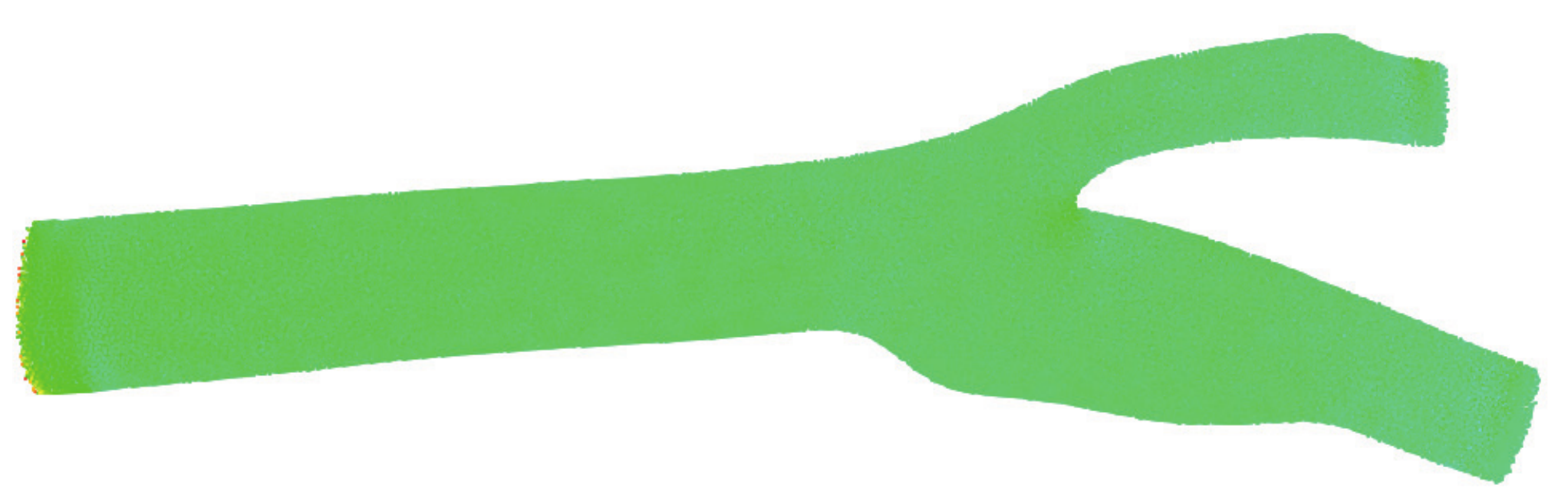}
        \end{subfigure}
        \begin{subfigure}[b]{0.32\textwidth}
            \includegraphics[width=\textwidth]{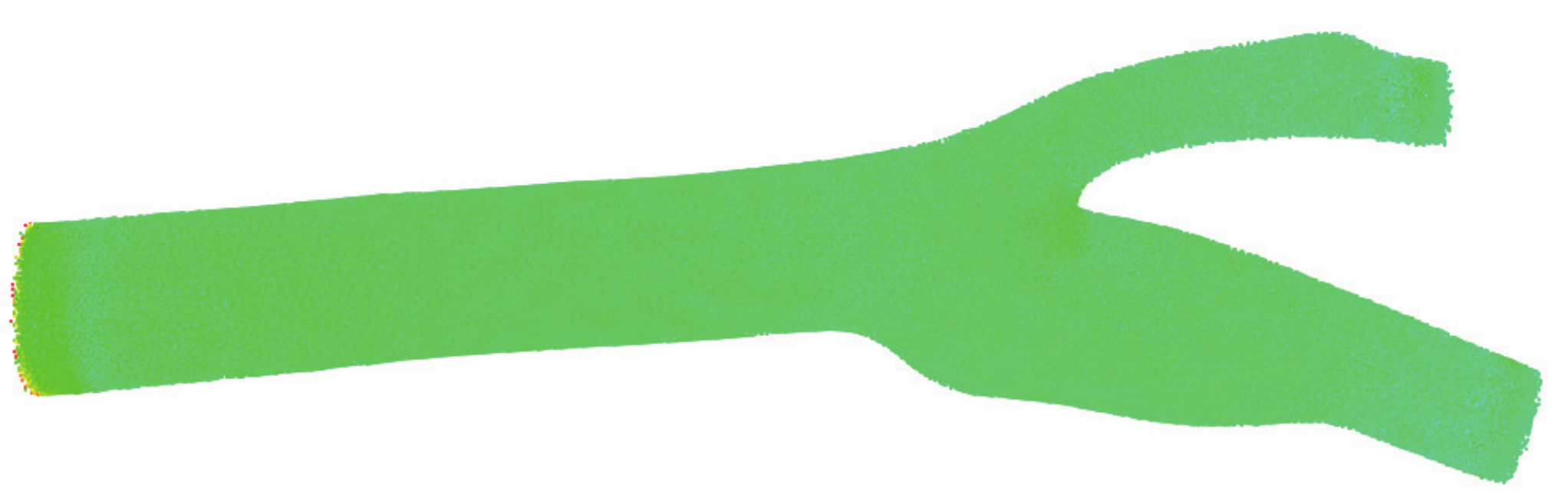}
        \end{subfigure}
        \begin{subfigure}[b]{0.32\textwidth}
            \includegraphics[width=\textwidth]{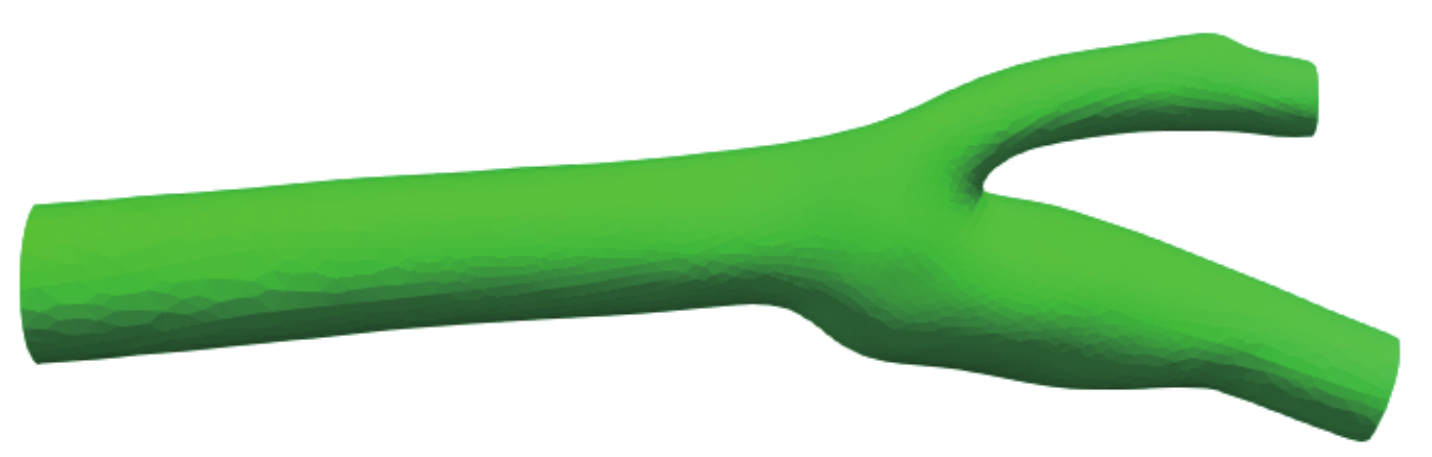}
        \end{subfigure}
        \caption*{\text{(d)} $t$ = 2.4s}
    \end{subfigure}
    \caption{Hemodynamics in carotid artery (rigid wall): pressure distributions at four time instants in the 
    fifth cardiac cycle. Left: SPH result with volume-based wall model; middle: SPH result with shell-based 
    wall model; right: FVM reference solution.}
    \label{fig: carotid-VIPO-pressure}
\end{figure}

\begin{figure}
    \centering
    \begin{subfigure}[b]{0.5\textwidth}
        \includegraphics[width=\textwidth]{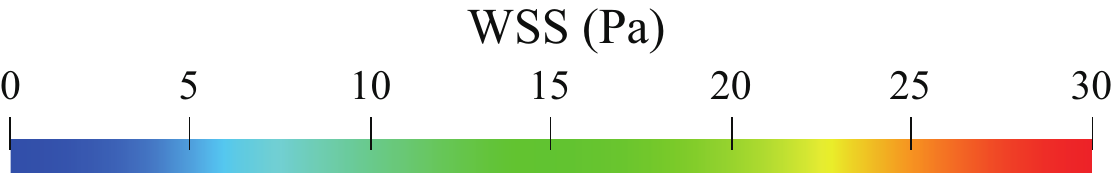}
    \end{subfigure}
    \begin{subfigure}[b]{\textwidth}
        \centering
        \begin{subfigure}[b]{0.32\textwidth}
            \caption*{\text{volume model}}
            \includegraphics[width=\textwidth]{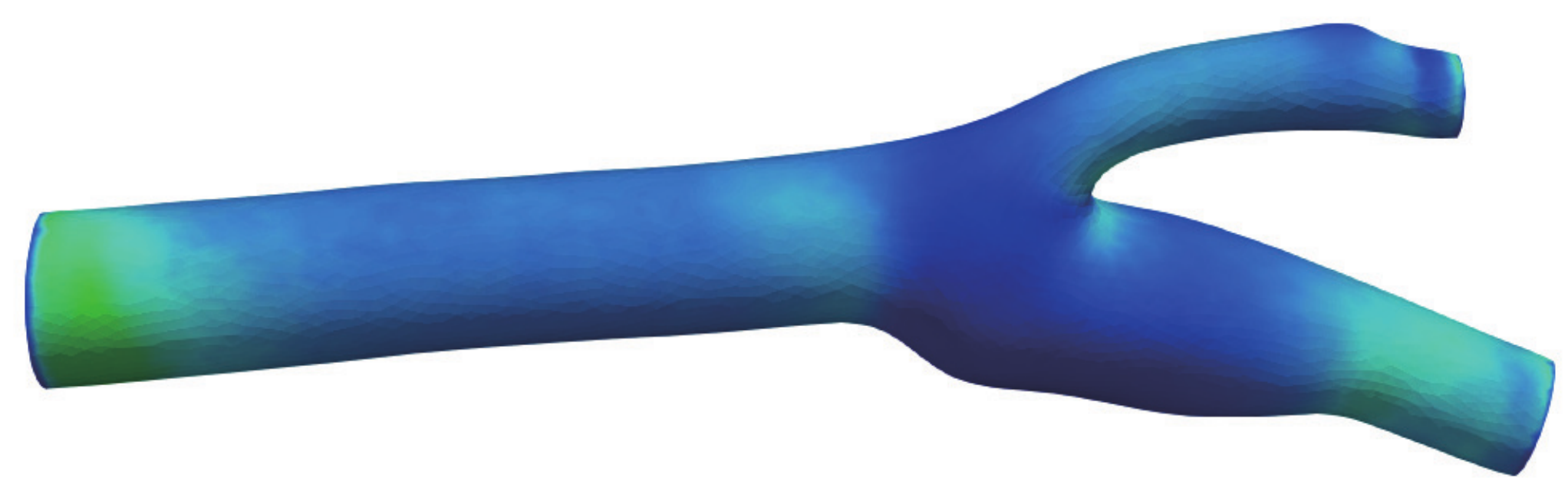}
        \end{subfigure}
        \begin{subfigure}[b]{0.32\textwidth}
            \caption*{\text{shell model}}
            \includegraphics[width=\textwidth]{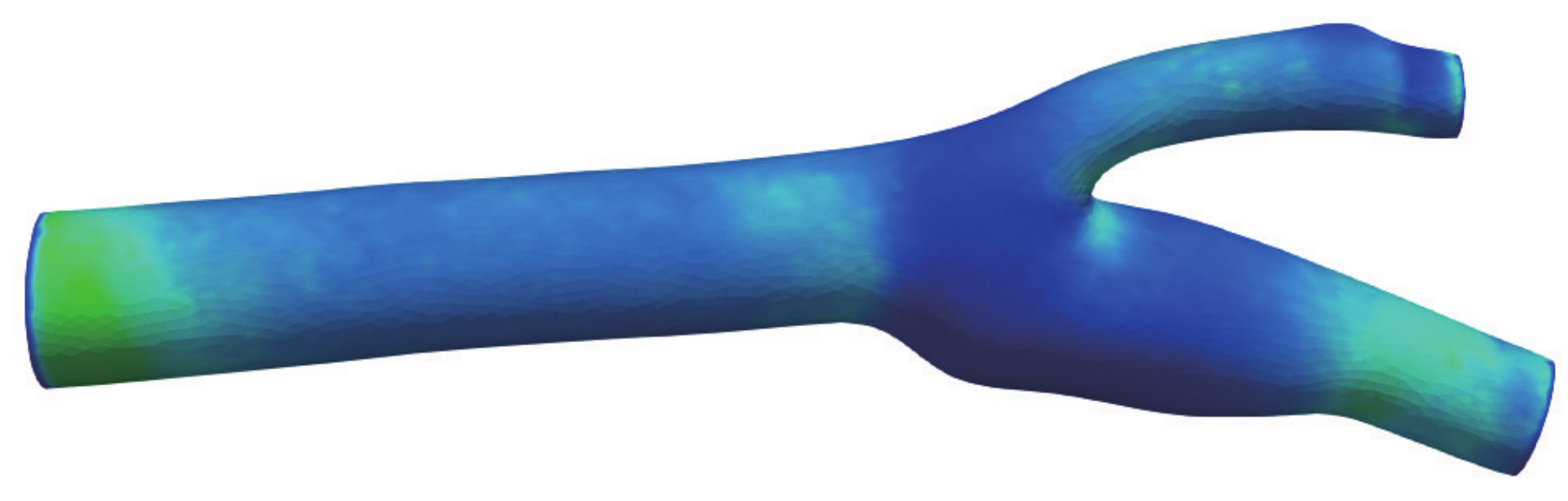}
        \end{subfigure}
        \begin{subfigure}[b]{0.32\textwidth}
            \caption*{\text{FVM}}
            \includegraphics[width=\textwidth]{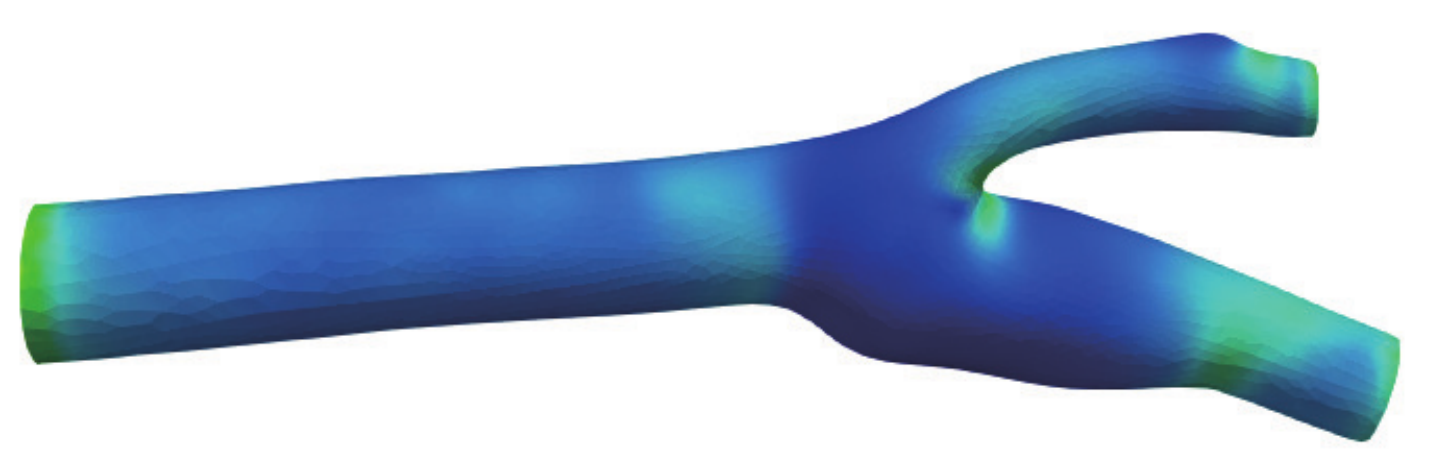}
        \end{subfigure}
        \caption*{\text{(a)} $t$ = 2.05s}
    \end{subfigure}
    \begin{subfigure}[b]{\textwidth}
        \centering
        \begin{subfigure}[b]{0.32\textwidth}
            \includegraphics[width=\textwidth]{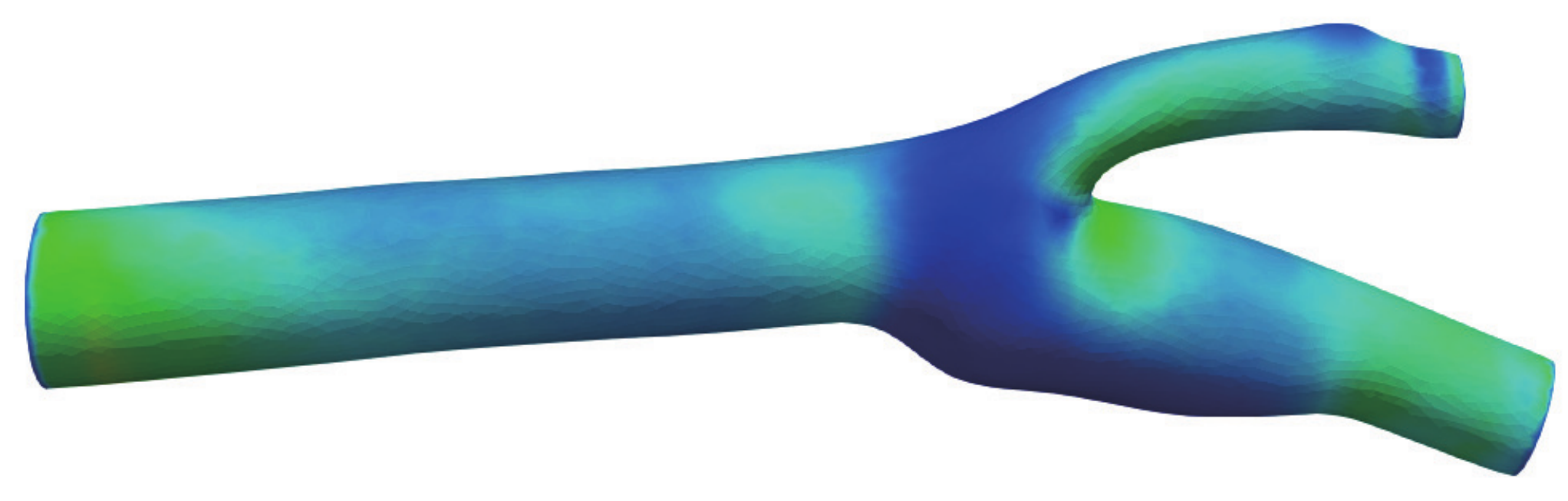}
        \end{subfigure}
        \begin{subfigure}[b]{0.3\textwidth}
            \includegraphics[width=\textwidth]{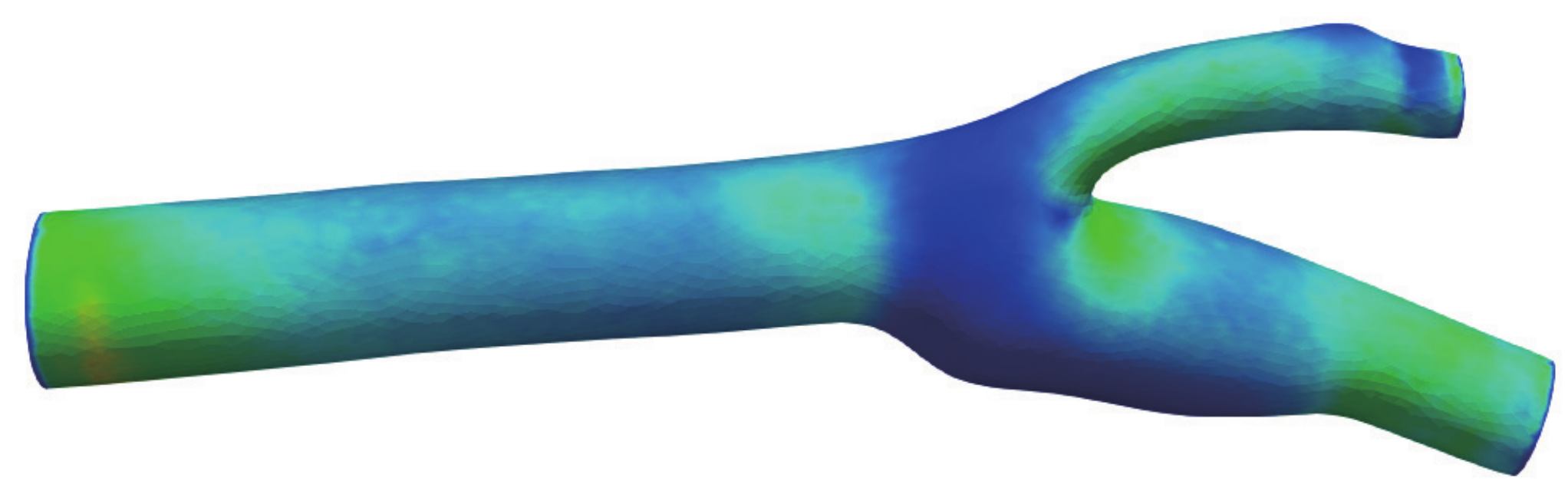}
        \end{subfigure}
        \begin{subfigure}[b]{0.32\textwidth}
            \includegraphics[width=\textwidth]{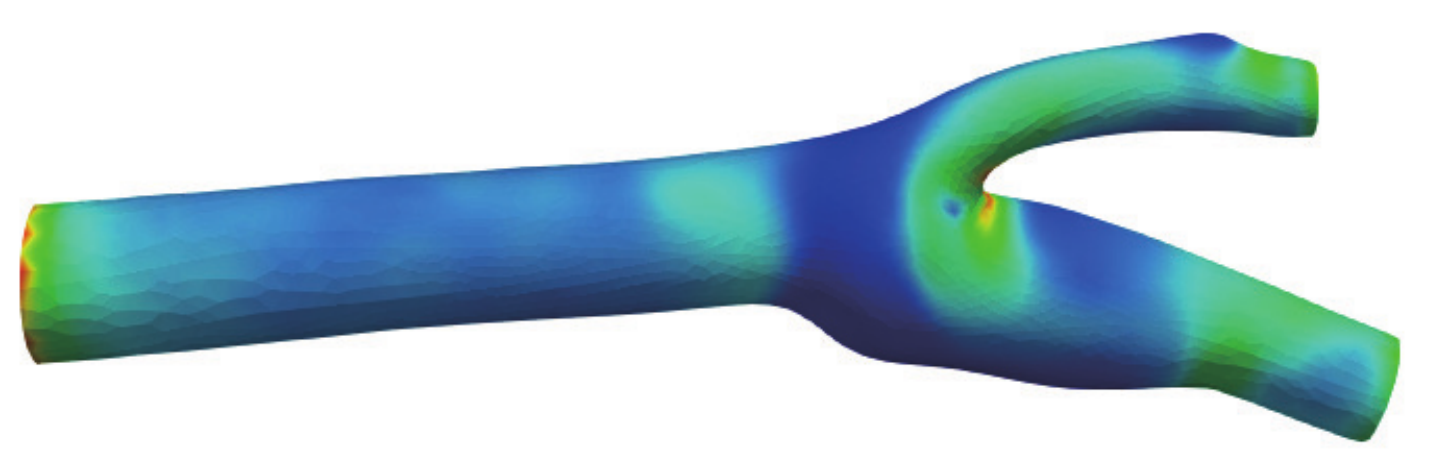}
        \end{subfigure}
        \caption*{\text{(b)} $t$ = 2.1s}
    \end{subfigure}
    \begin{subfigure}[b]{\textwidth}
        \centering
        \begin{subfigure}[b]{0.32\textwidth}
            \includegraphics[width=\textwidth]{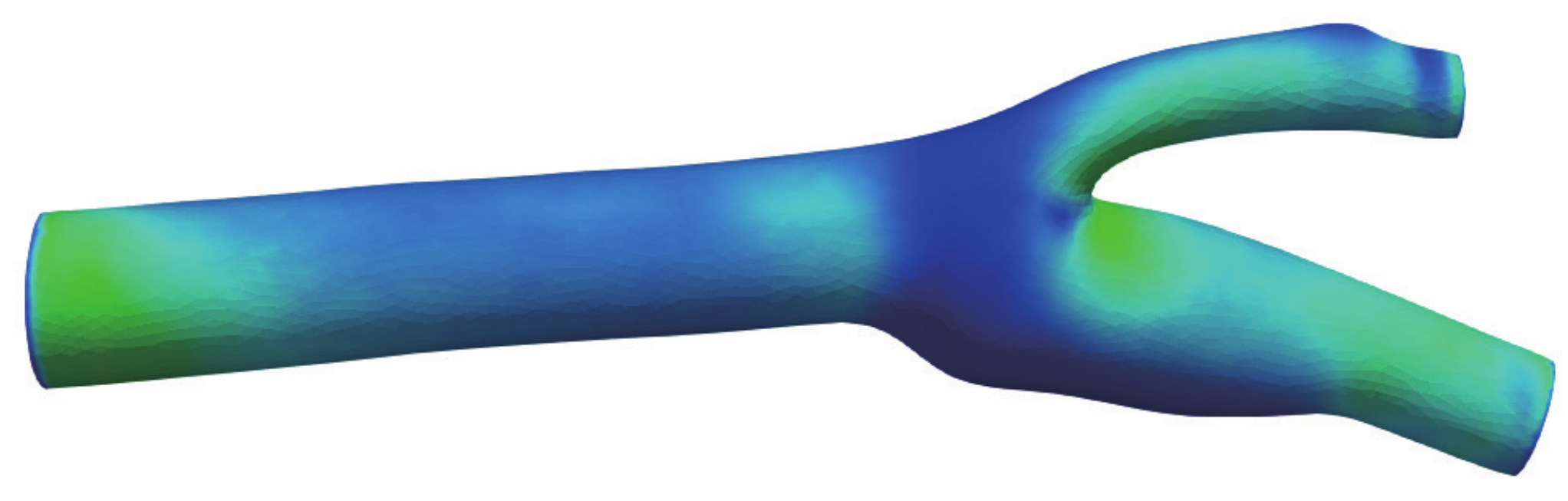}
        \end{subfigure}
        \begin{subfigure}[b]{0.32\textwidth}
            \includegraphics[width=\textwidth]{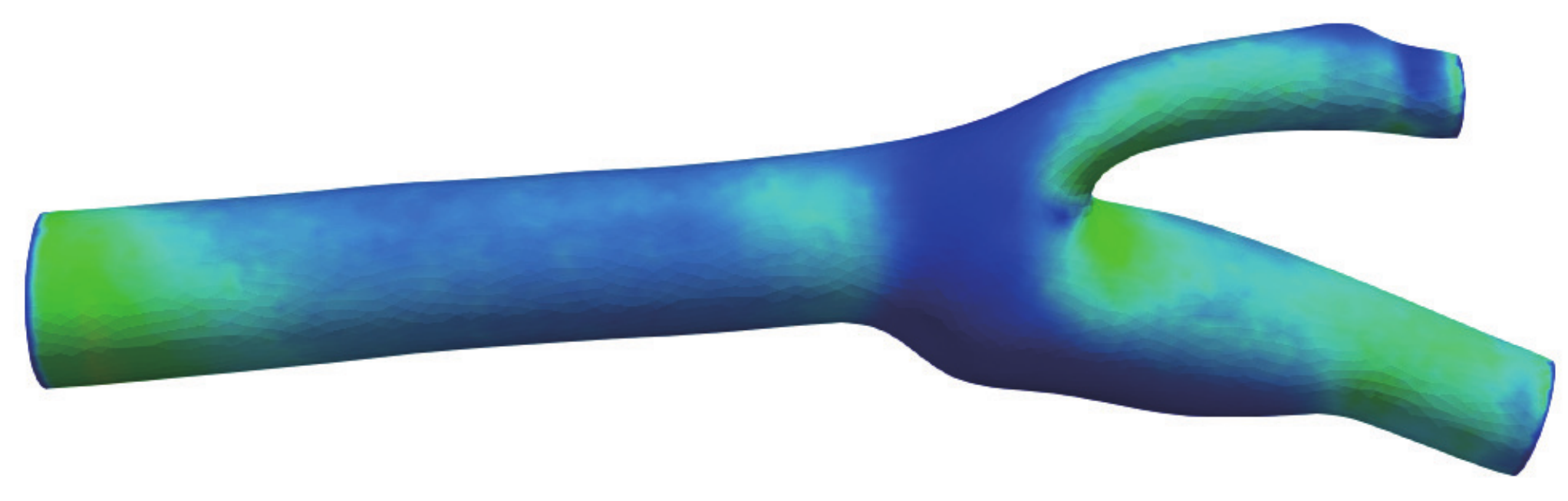}
        \end{subfigure}
        \begin{subfigure}[b]{0.32\textwidth}
            \includegraphics[width=\textwidth]{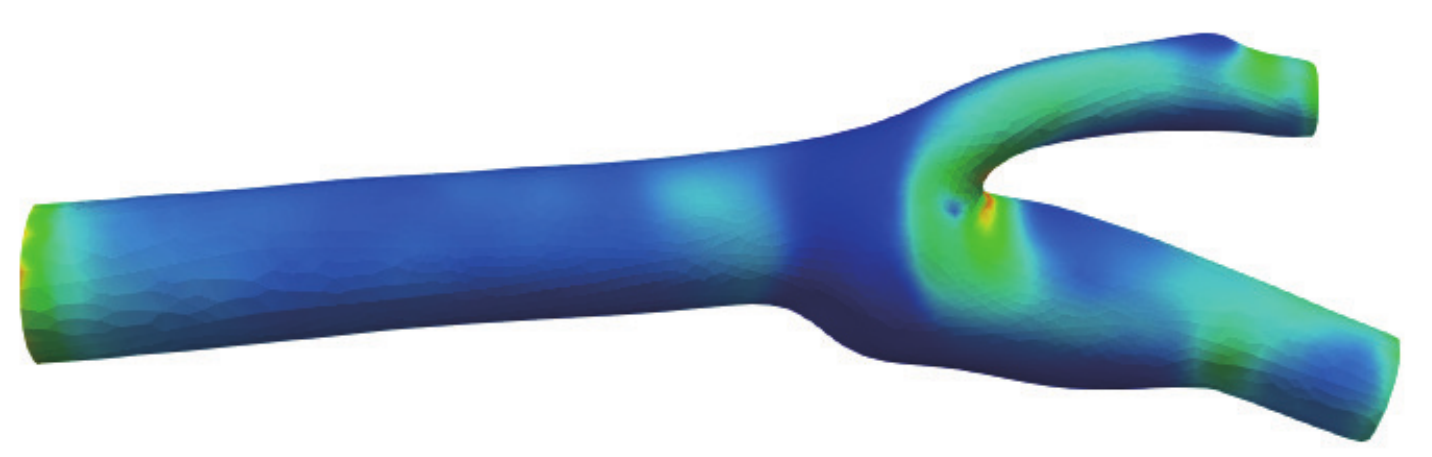}
        \end{subfigure}
        \caption*{\text{(c)} $t$ = 2.15s}
    \end{subfigure}
    \begin{subfigure}[b]{\textwidth}
        \centering
        \begin{subfigure}[b]{0.32\textwidth}
            \includegraphics[width=\textwidth]{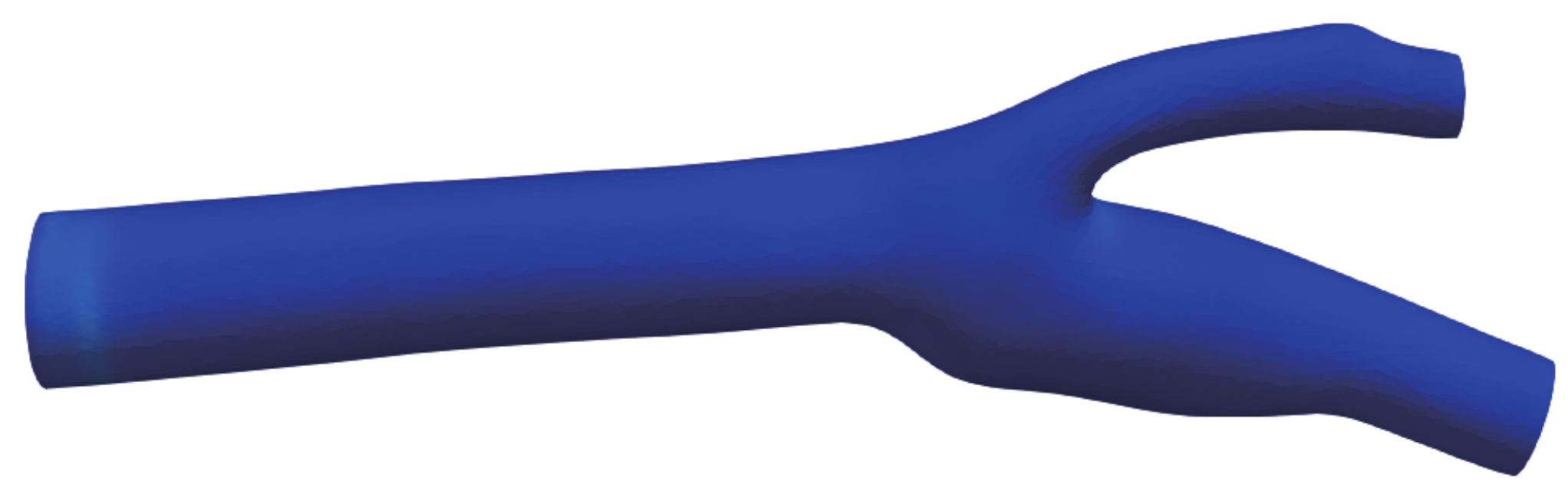}
        \end{subfigure}
        \begin{subfigure}[b]{0.32\textwidth}
            \includegraphics[width=\textwidth]{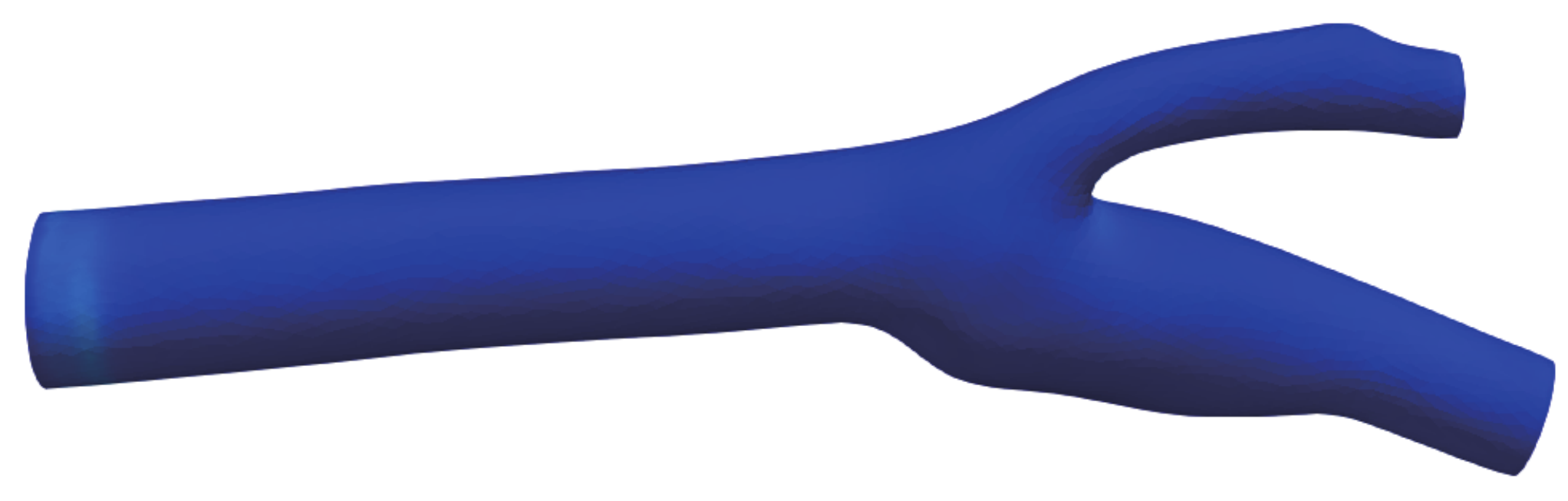}
        \end{subfigure}
        \begin{subfigure}[b]{0.32\textwidth}
            \includegraphics[width=\textwidth]{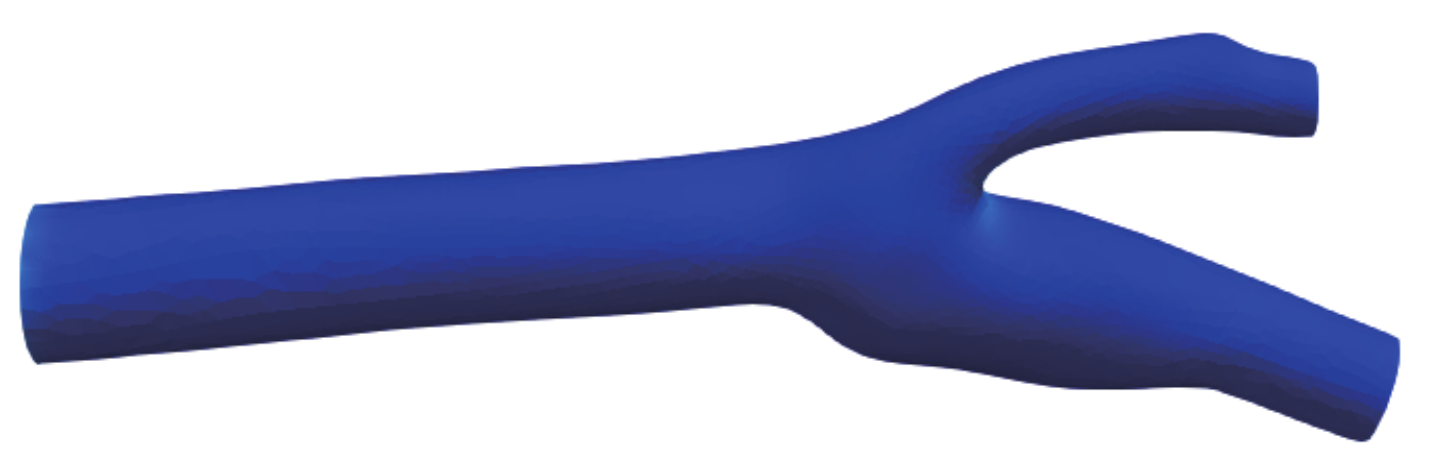}
        \end{subfigure}
        \caption*{\text{(d)} $t$ = 2.4s}
    \end{subfigure}
    \caption{Hemodynamics in carotid artery (rigid wall): WSS distributions at four time instants in the fifth 
    cardiac cycle. Left: SPH result with volume-based wall model; middle: SPH result with shell-based wall 
    model; right: FVM reference solution.}
    \label{fig: carotid-VIPO-WSS}
\end{figure}

Fig.\ref{fig: carotid-VIPO-velocity} and Fig.\ref{fig: carotid-VIPO-pressure} illustrate the temporal evolution 
of velocity and pressure fields throughout a representative cardiac cycle in the carotid bifurcation. 
Snapshots are taken at four characteristic time instants during the fifth cycle, corresponding to early 
systole, peak systolic flow, post-systole, and stable phase of diastole. At $t = 2.05\text{s}$, the flow begins to 
accelerate, marking the onset of systole. The velocity field is well-developed along the CCA, and the pressure 
field presents the highest value at the inlet and begins to taper smoothly downstream with a mild gradient near 
the bifurcation. At $t = 2.1\text{s}$, the inflow reaches its maximum velocity. All of these three numerical models 
successfully capture the formation of the flow separation zones near the bifurcation, and severe localized flow 
reversal appears near the carotid bulb. Concurrently, the pressure gradient becomes notably steeper at the 
bifurcation and into the downstream branches. At $t = 2.15\text{s}$, flow deceleration is evident. Simultaneously, 
pressure levels start to decline and exhibit a more spatially uniform distribution throughout the vascular 
domain. A noticeable pressure increase of SPH results is observed in the inlet buffer zone due to the imposed 
inflow condition; however, this does not affect the downstream hemodynamic field. 
At $t = 2.4\text{s}$, during the diastolic resting phase, the velocity magnitude decreases significantly, 
consistent with the expected low-pressure regime. Fig.\ref{fig: carotid-VIPO-WSS} illustrates the temporal 
evolution of wall shear stress distributions at different moments. The WSS values, originally computed at the 
wall particles within the SPH simulations, are interpolated onto the STL surface of the original fluid geometry 
using ParaView’s post-processing tools. This surface mapping enhances visual clarity and facilitates direct 
comparison with FVM results from ANSYS Fluent. These results indicate that the SPH method is capable of 
accurately capturing the near-wall velocity gradients necessary for reliable hemodynamic shear stress 
prediction. Notably, high-shear regions are observed near the bifurcation apex and along the inner walls of 
the ICA and ECA during systole. This further reflects the flow deviation toward the inner curvature of the 
branches, where shear stress intensifies, in contrast to the outer walls, where recirculation or flow 
separation leads to lower shear stress magnitudes.

\begin{figure}
    \centering
    \begin{subfigure}[b]{\textwidth}
        \centering
        \begin{subfigure}[b]{0.4\textwidth}
            \includegraphics[width=\textwidth]{Figure/carotid_vipo_velocity_legend.pdf}
        \end{subfigure}
        \begin{subfigure}[b]{0.4\textwidth}
            \includegraphics[width=\textwidth]{Figure/carotid_vipo_wss_legend.pdf}
        \end{subfigure}
    \end{subfigure}
    \begin{subfigure}[b]{\textwidth}
        \centering
        \begin{subfigure}[b]{0.4\textwidth}
            \includegraphics[width=\textwidth]{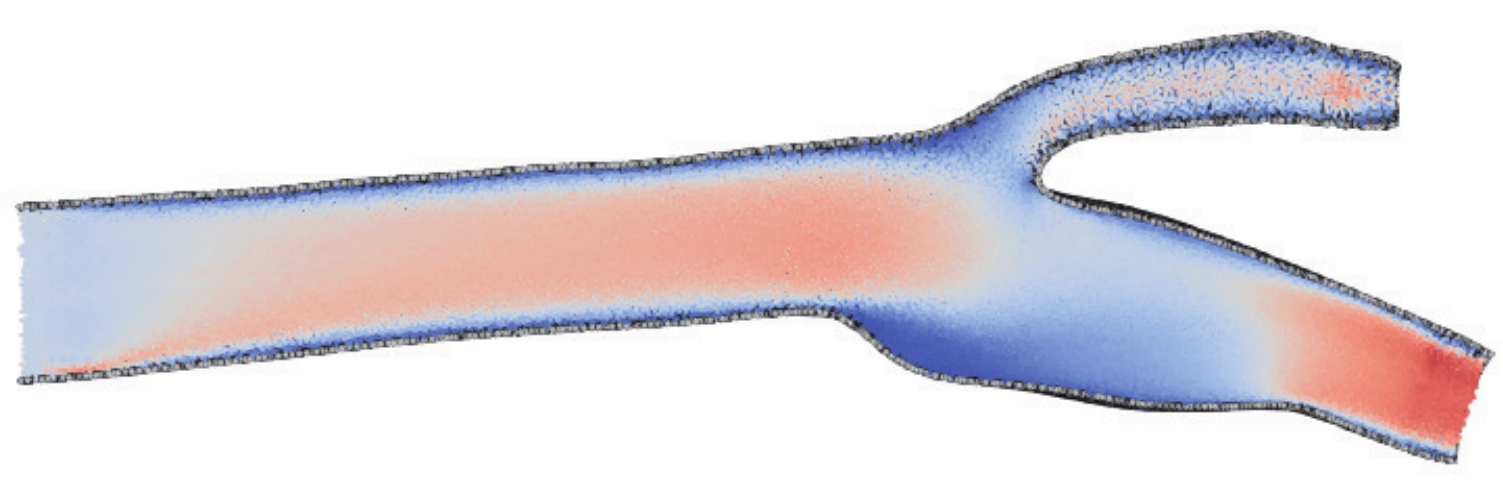}
        \end{subfigure}
        \begin{subfigure}[b]{0.4\textwidth}
            \includegraphics[width=\textwidth]{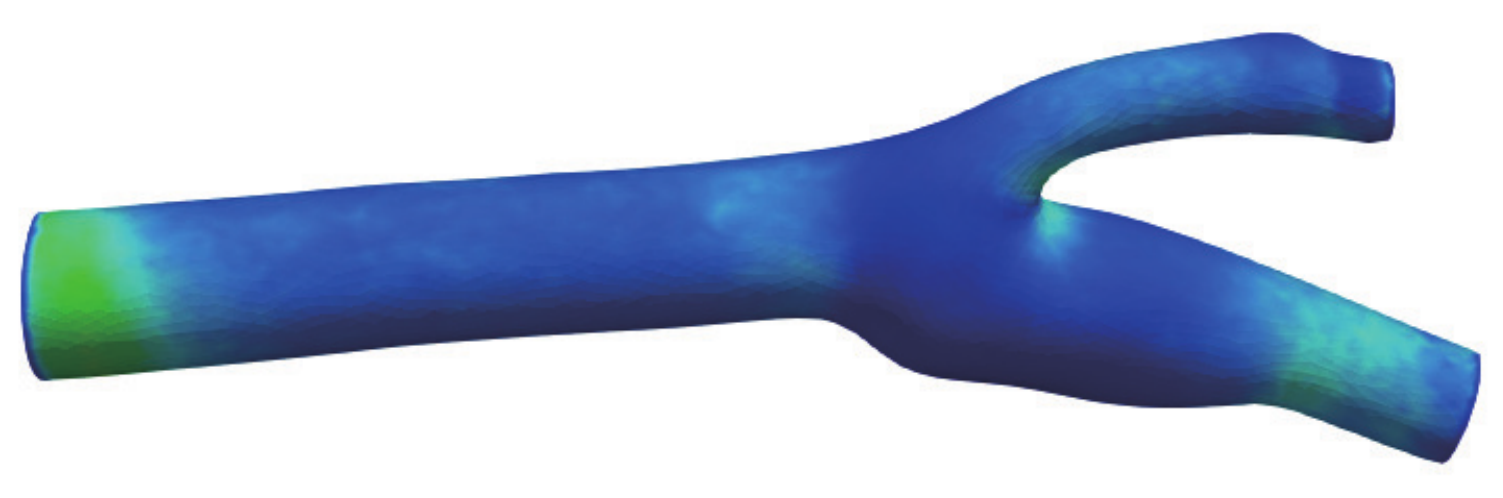}
        \end{subfigure}
        \caption*{\text{(a)} $t$ = 2.05s}
    \end{subfigure}
    \begin{subfigure}[b]{\textwidth}
        \centering
        \begin{subfigure}[b]{0.4\textwidth}
            \includegraphics[width=\textwidth]{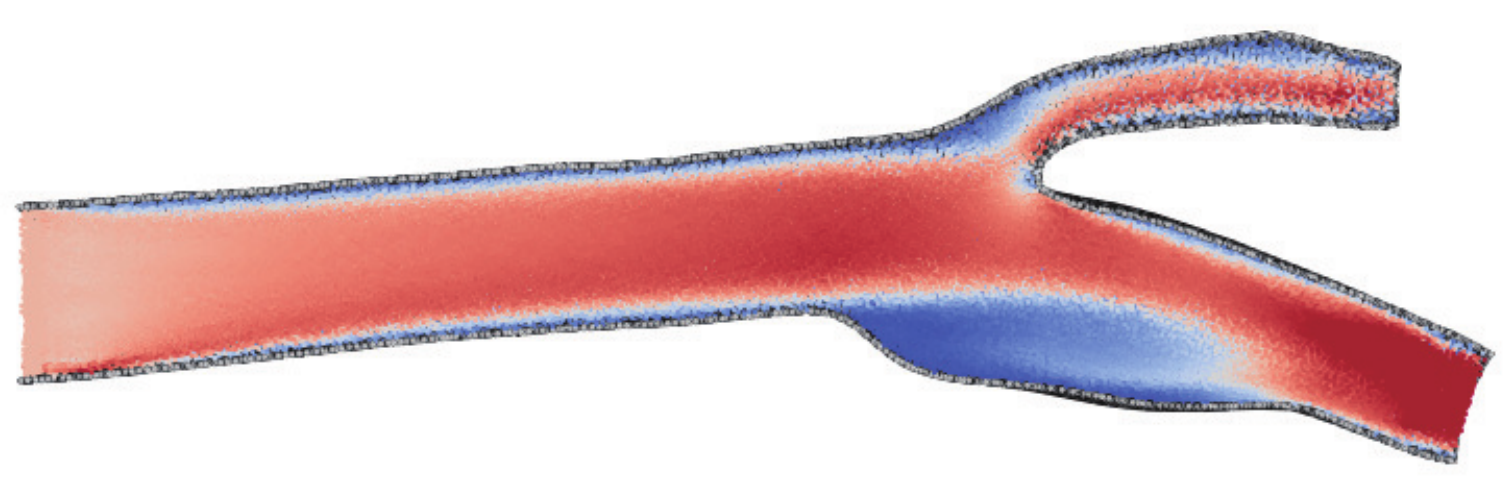}
        \end{subfigure}
        \begin{subfigure}[b]{0.4\textwidth}
            \includegraphics[width=\textwidth]{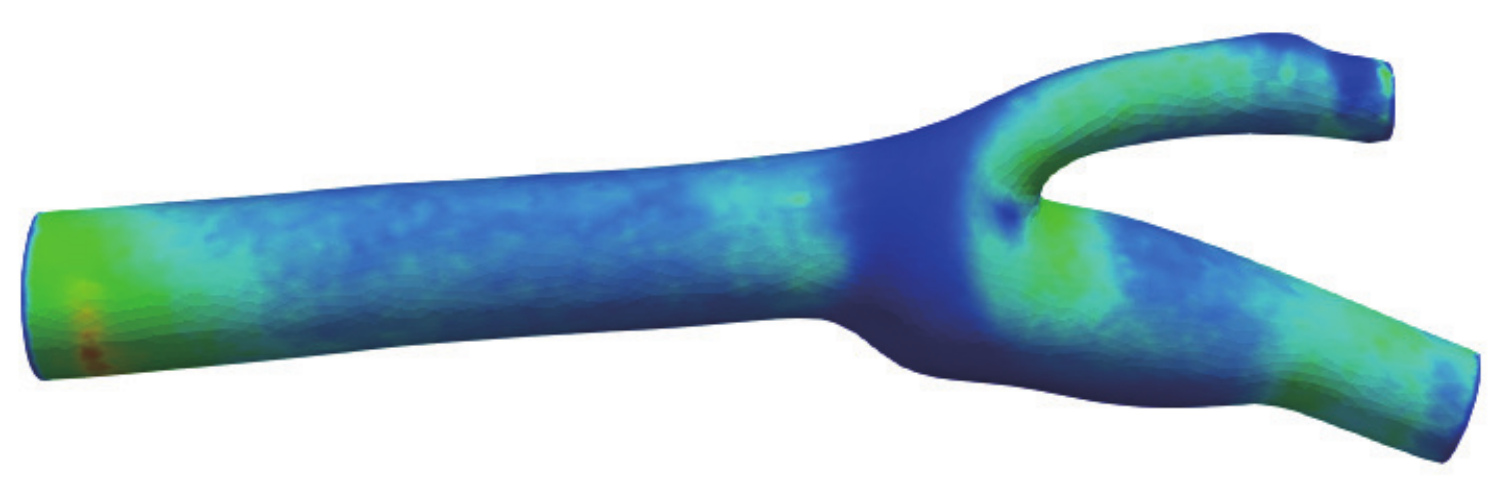}
        \end{subfigure}
        \caption*{\text{(b)} $t$ = 2.1s}
    \end{subfigure}
    \begin{subfigure}[b]{\textwidth}
        \centering
        \begin{subfigure}[b]{0.4\textwidth}
            \includegraphics[width=\textwidth]{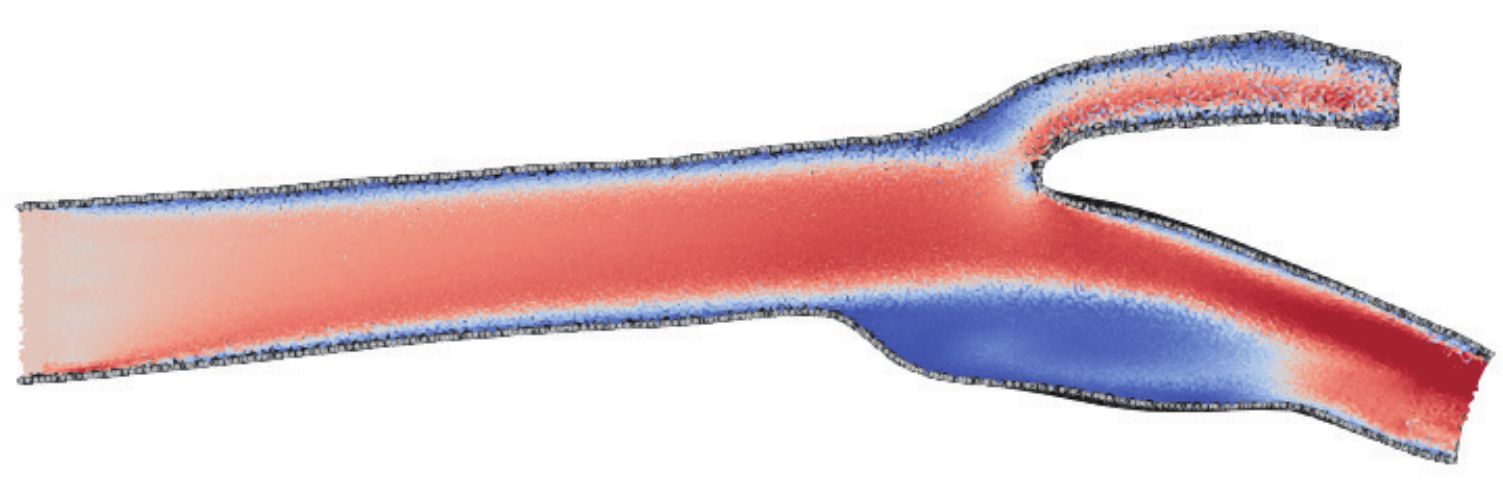}
        \end{subfigure}
        \begin{subfigure}[b]{0.4\textwidth}
            \includegraphics[width=\textwidth]{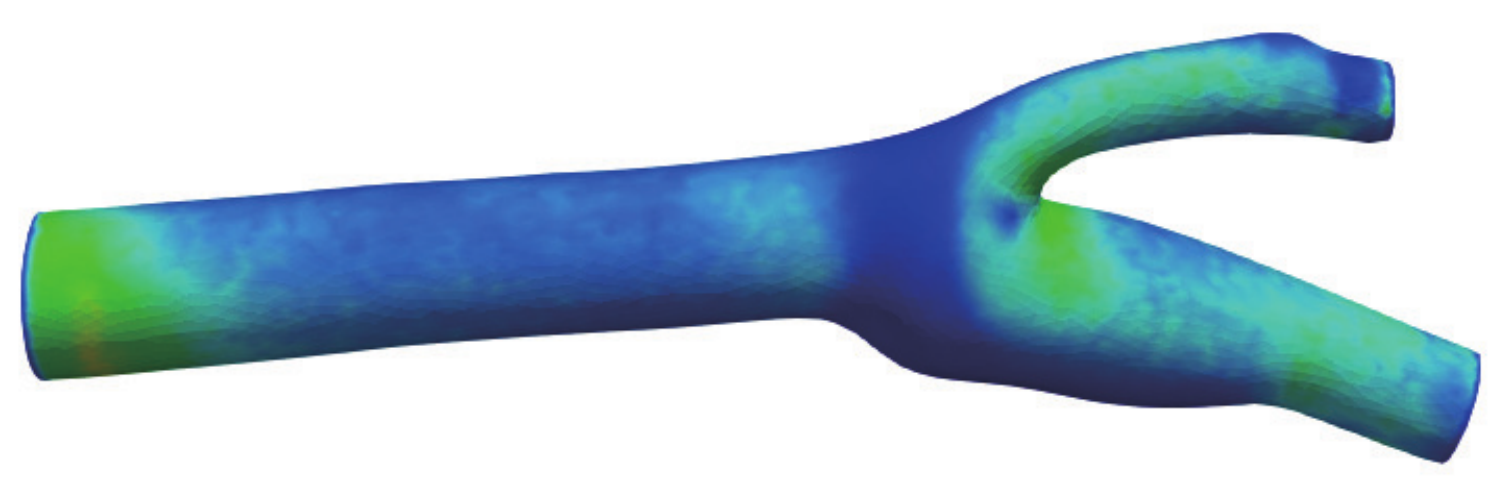}
        \end{subfigure}
        \caption*{\text{(c} $t$ = 2.15s}
    \end{subfigure}
        \begin{subfigure}[b]{\textwidth}
        \centering
        \begin{subfigure}[b]{0.4\textwidth}
            \includegraphics[width=\textwidth]{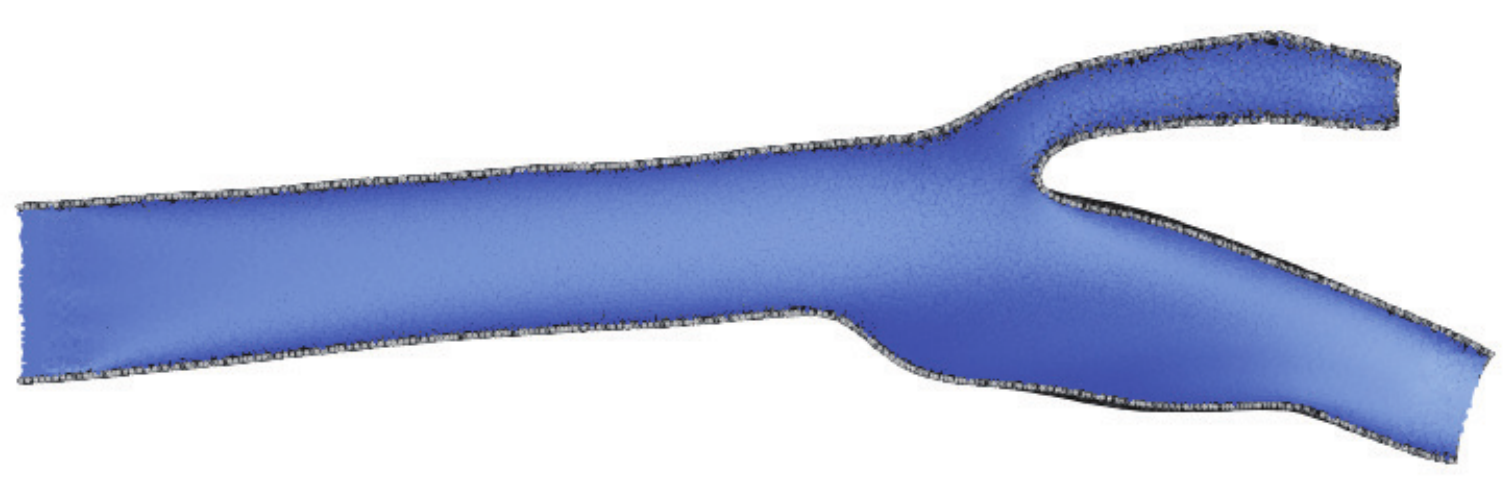}
        \end{subfigure}
        \begin{subfigure}[b]{0.4\textwidth}
            \includegraphics[width=\textwidth]{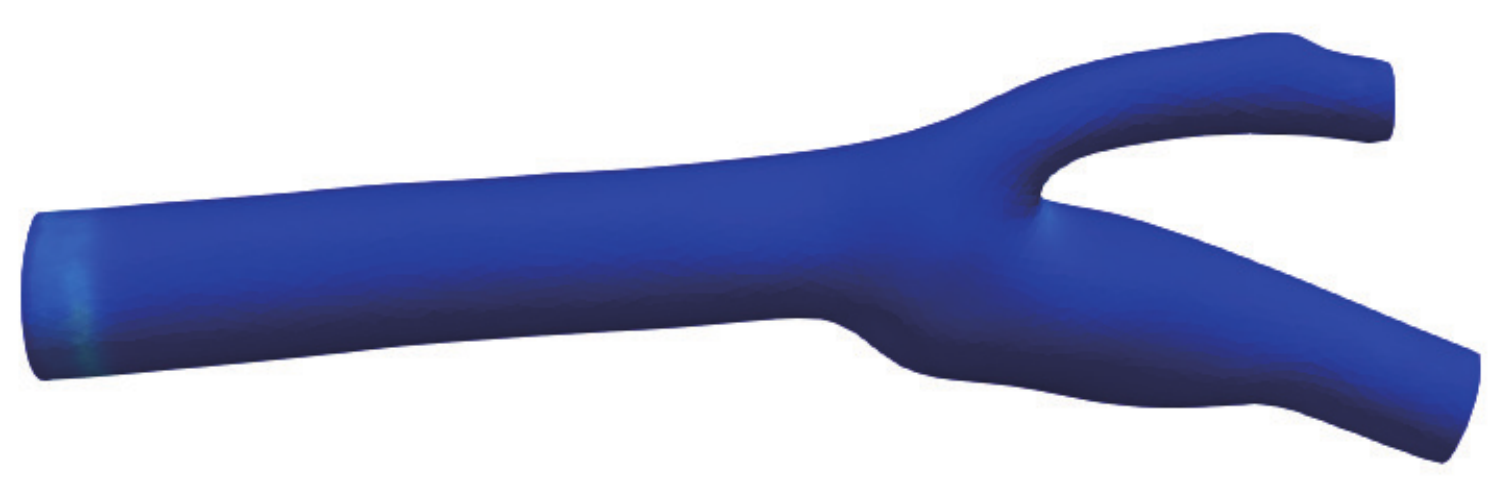}
        \end{subfigure}
        \caption*{\text{(d)} $t$ = 2.4s}
    \end{subfigure}
    \caption{Hemodynamics in carotid artery (deformable wall with shell model): velocity (left) and WSS 
    (right) distributions at four time instants during the fifth cardiac cycle.}
    \label{fig: carotid-VIPO-shell}
\end{figure}

\begin{figure}
    \centering
    \begin{subfigure}[b]{0.5\textwidth}
        \includegraphics[width=\textwidth]{Figure/carotid_vipo_velocity_legend.pdf}
    \end{subfigure}
    \begin{subfigure}[b]{\textwidth}
        \centering
        \begin{subfigure}[b]{0.49\textwidth}
            \includegraphics[width=\textwidth]{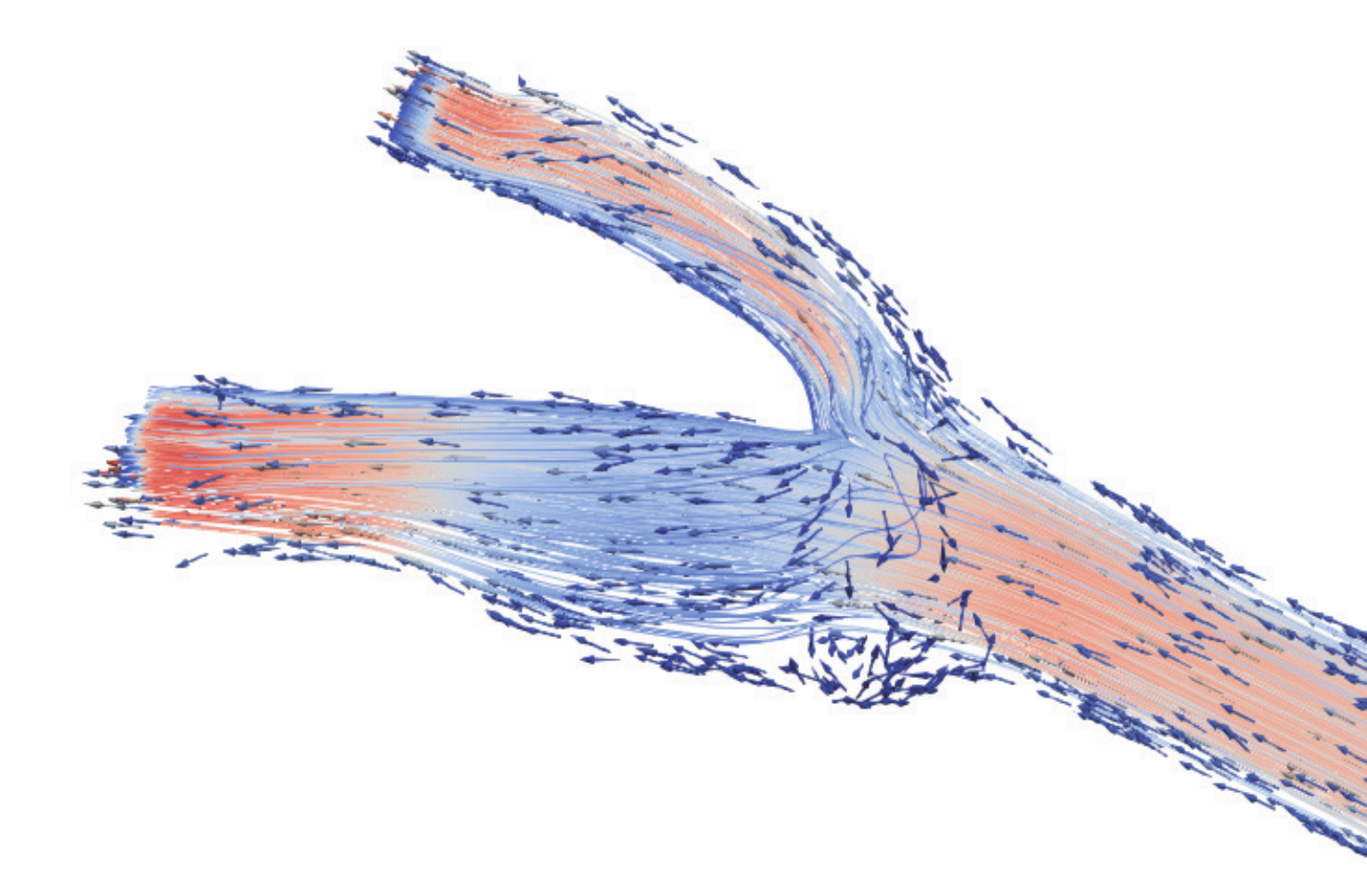}
            \caption{$t$ = 2.05s}
        \end{subfigure}
        \begin{subfigure}[b]{0.49\textwidth}
            \includegraphics[width=\textwidth]{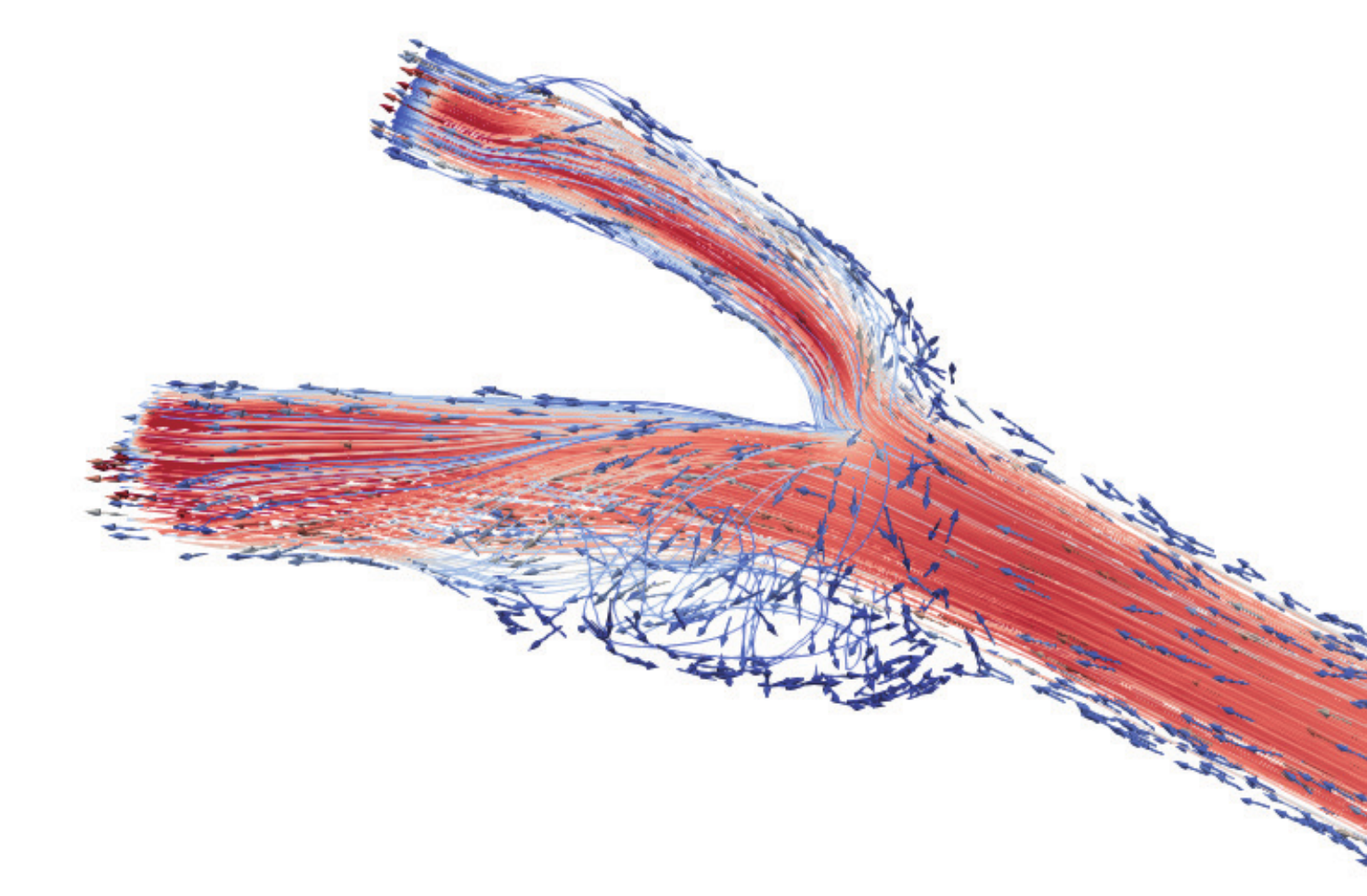}
            \caption{$t$ = 2.1s}
        \end{subfigure}
    \end{subfigure}
    \begin{subfigure}[b]{\textwidth}
        \centering
        \begin{subfigure}[b]{0.49\textwidth}
            \includegraphics[width=\textwidth]{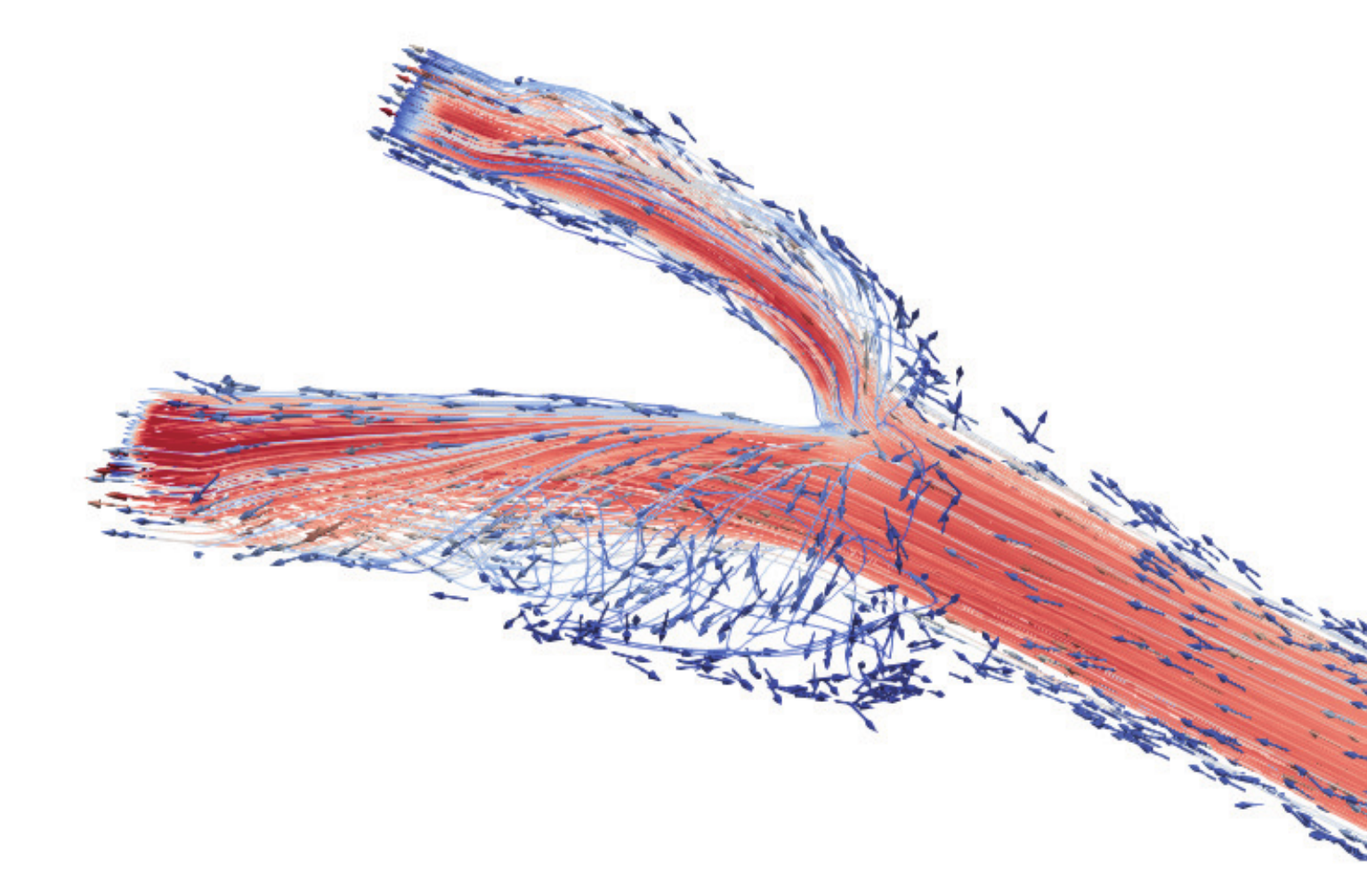}
            \caption{$t$ = 2.15s}
        \end{subfigure}
        \begin{subfigure}[b]{0.49\textwidth}
            \includegraphics[width=\textwidth]{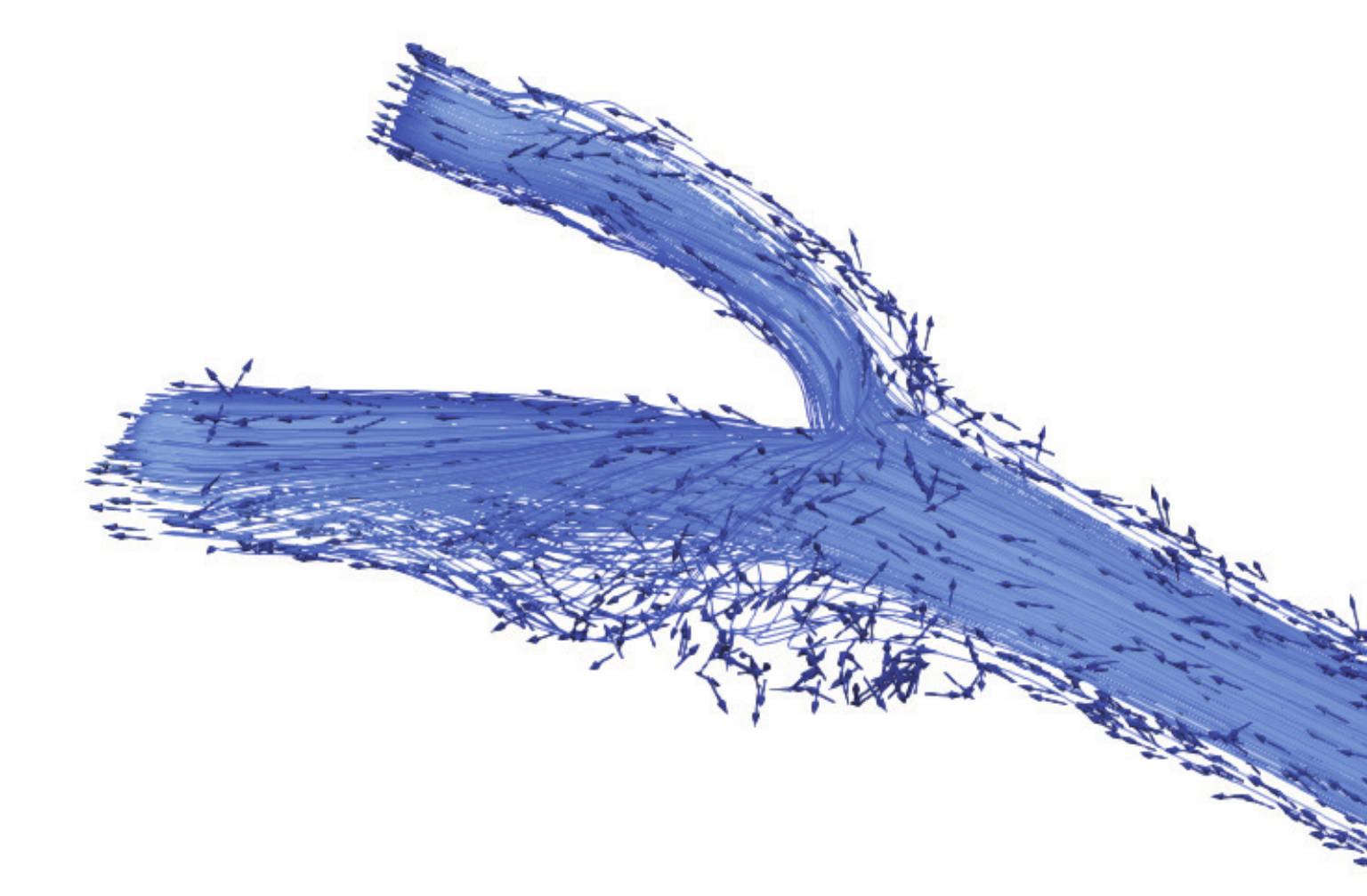}
            \caption{$t$ = 2.4s}
        \end{subfigure}
    \end{subfigure}
    \caption{Hemodynamics in carotid artery (deformable wall with shell model): streamline and velocity 
    vector at four time instants during the fifth cardiac cycle.}
    \label{fig: carotid-VIPO-shell-vel-vector}
\end{figure}

Following the configuration in Ref.\cite{lopes2019influence}, the deformable artery wall with thickness of 0.6 
mm \cite{paul2012measurement} is modeled as a linear elastic and isotropic material with a density 
$\rho_s = 1120 \text{ kg}/\text{m}^3$, Young's modulus $ E= 1.106\text{ MPa}$ and Poisson's ratio is 0.45. 
In the present SPH framework, the arterial wall is represented using a single layer of shell particles. 
Fig.\ref{fig: carotid-VIPO-shell} presents the instantaneous velocity and WSS contours, 
while Fig.\ref{fig: carotid-VIPO-shell-vel-vector} illustrates the corresponding velocity vector fields at four 
representative time instants during the fifth cardiac cycle. We did not adopt the volume model for the 
deformable wall, as the vessel's thinness would require a prohibitively high number of solid particles to 
ensure accuracy, which in turn demands a significantly denser fluid resolution to maintain numerical stability. 
Additionally, we did not employ ANSYS Fluent for this case due to the lack of built-in support for FSI; 
it typically requires coupling with other modules in ANSYS Workbench, which considerably complicates the 
simulation workflow.

\begin{figure}
    \centering
    \begin{subfigure}[b]{0.4\textwidth}
        \includegraphics[width=\textwidth]{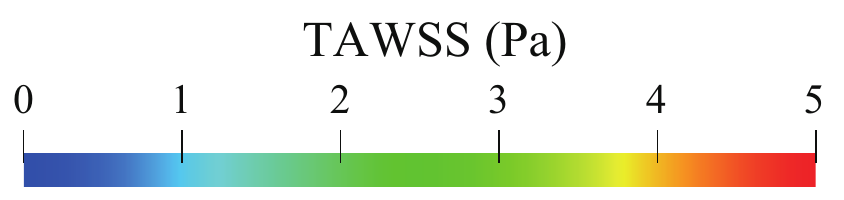}
    \end{subfigure}
    \begin{subfigure}[b]{\textwidth}
        \centering
        \begin{subfigure}[b]{0.35\textwidth}
            \includegraphics[width=\textwidth]{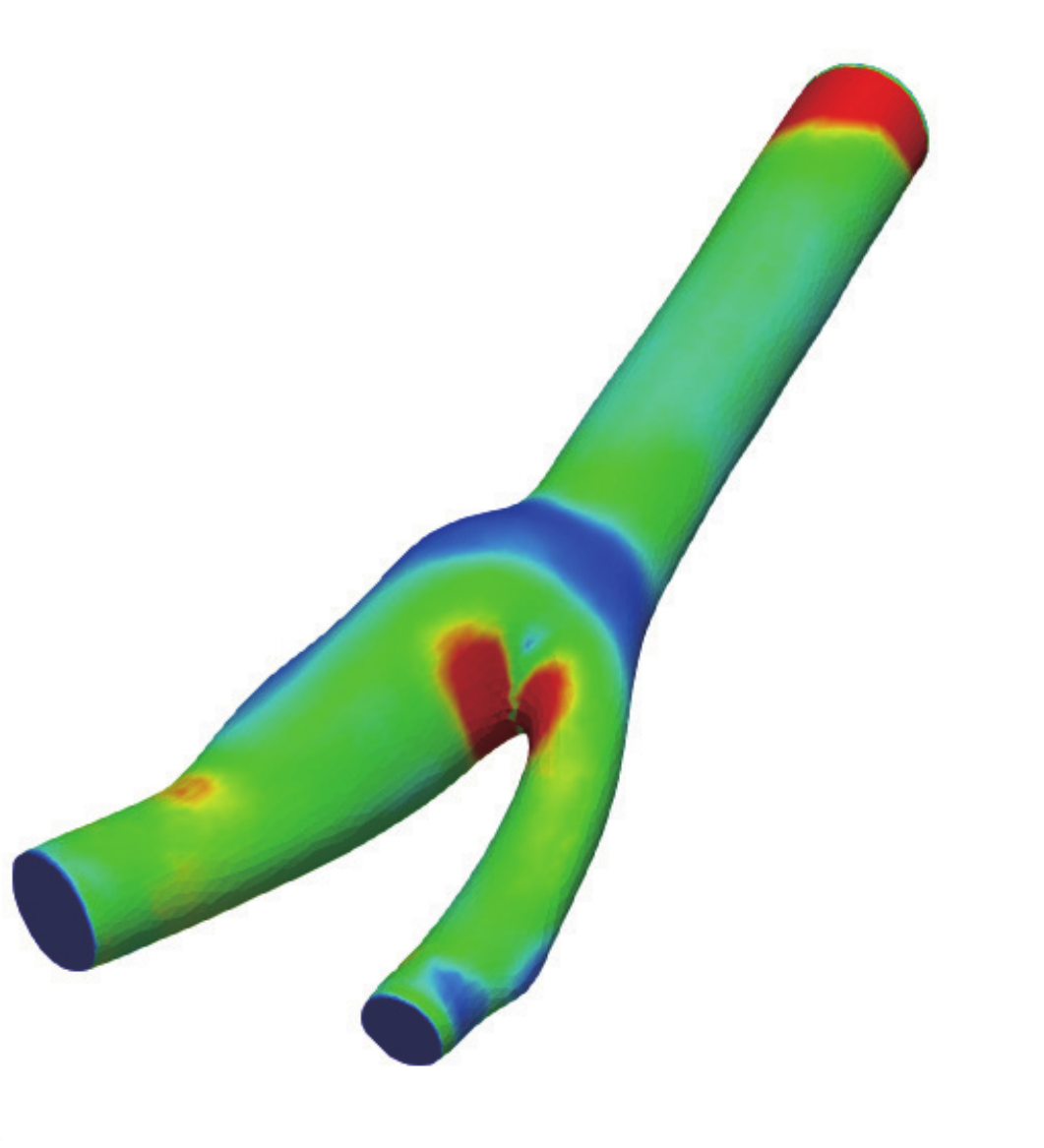}
        \end{subfigure}
        \begin{subfigure}[b]{0.35\textwidth}
            \includegraphics[width=\textwidth]{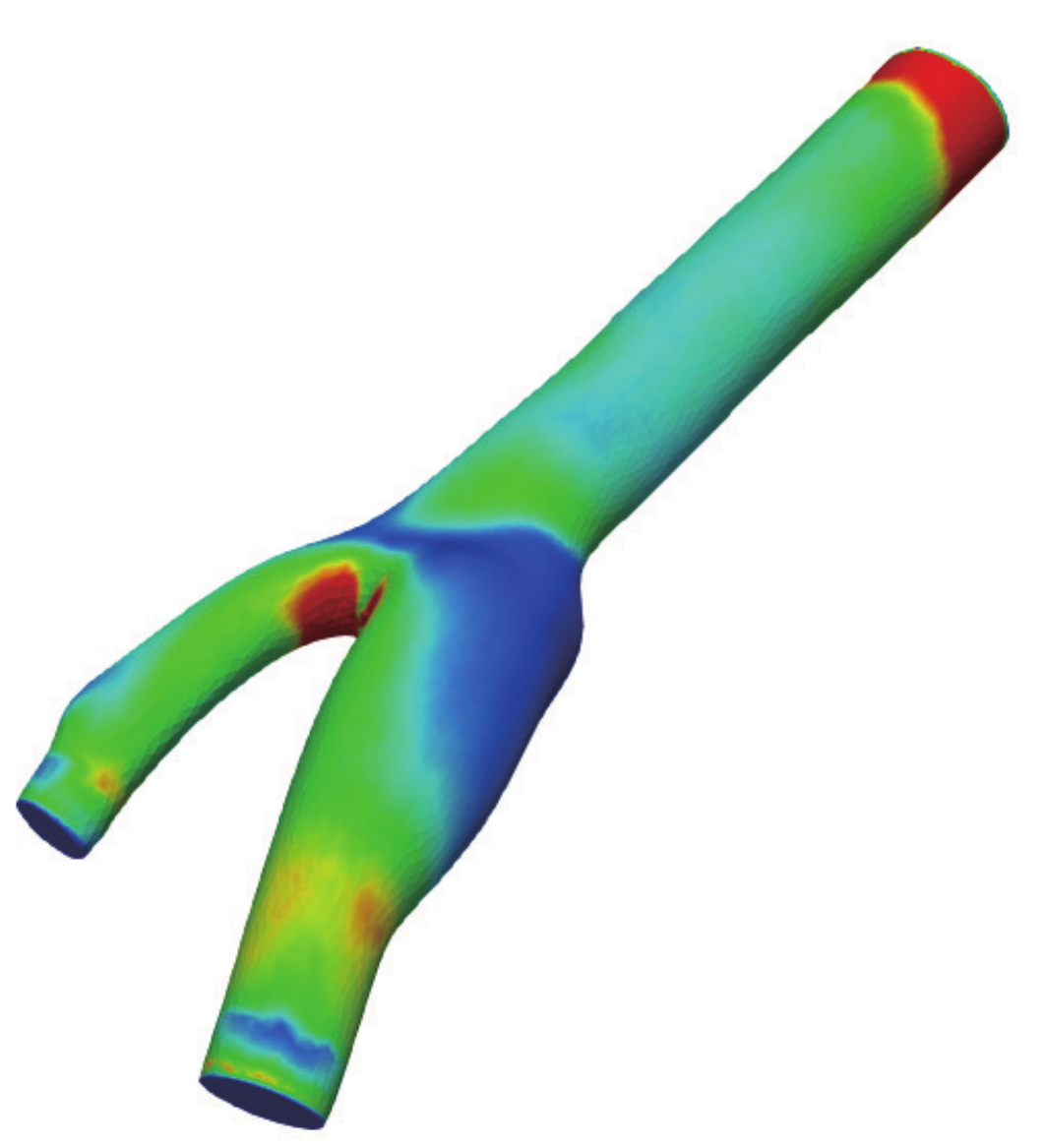}
        \end{subfigure}
        \caption*{\text{(a) TAWSS}}
    \end{subfigure}
    \begin{subfigure}[b]{0.4\textwidth}
        \includegraphics[width=\textwidth]{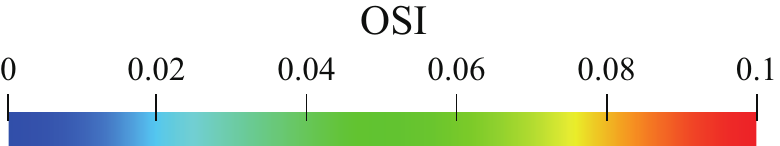}
    \end{subfigure}
    \begin{subfigure}[b]{\textwidth}
        \centering
        \begin{subfigure}[b]{0.35\textwidth}
            \includegraphics[width=\textwidth]{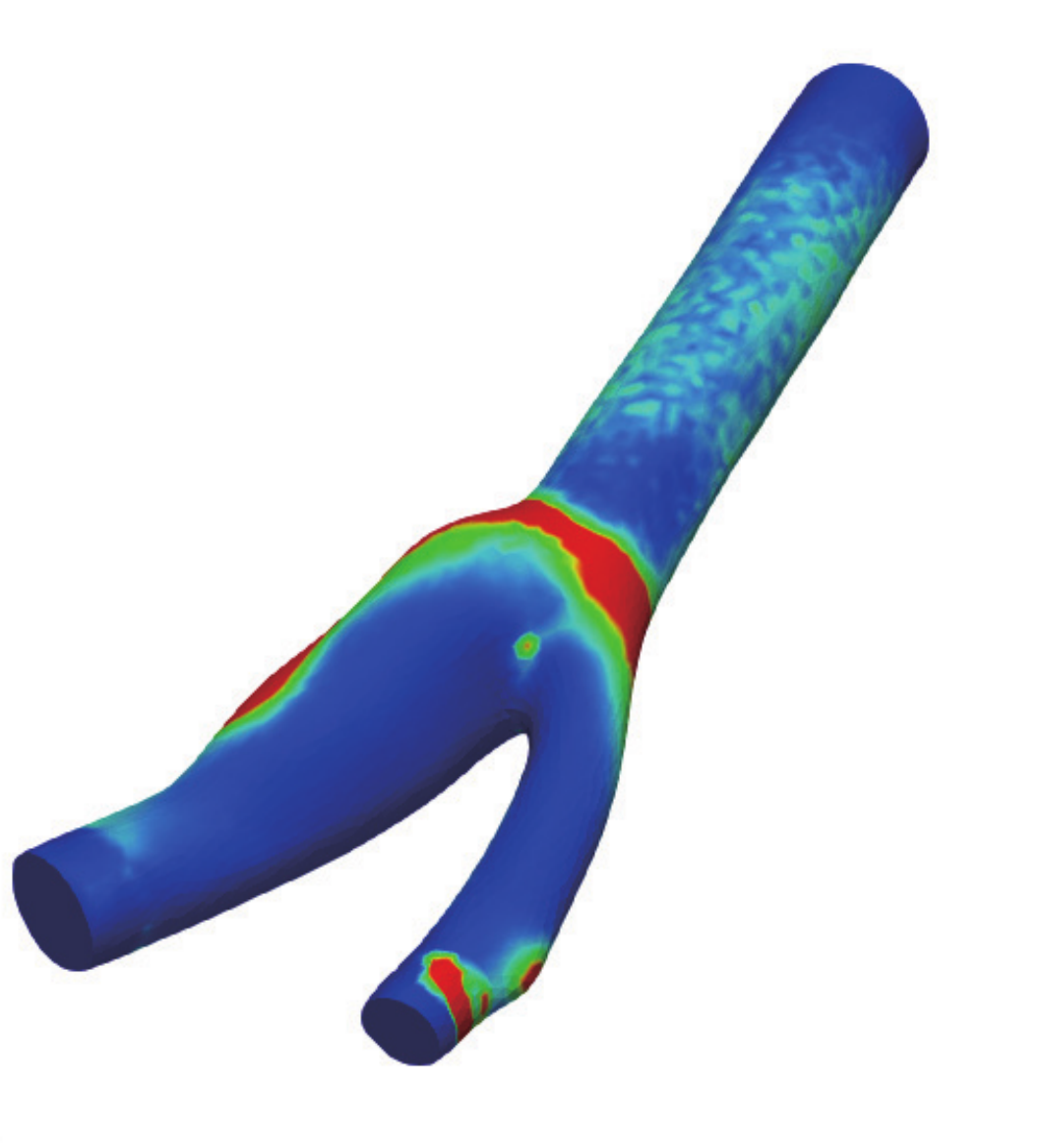}
        \end{subfigure}
        \begin{subfigure}[b]{0.35\textwidth}
            \includegraphics[width=\textwidth]{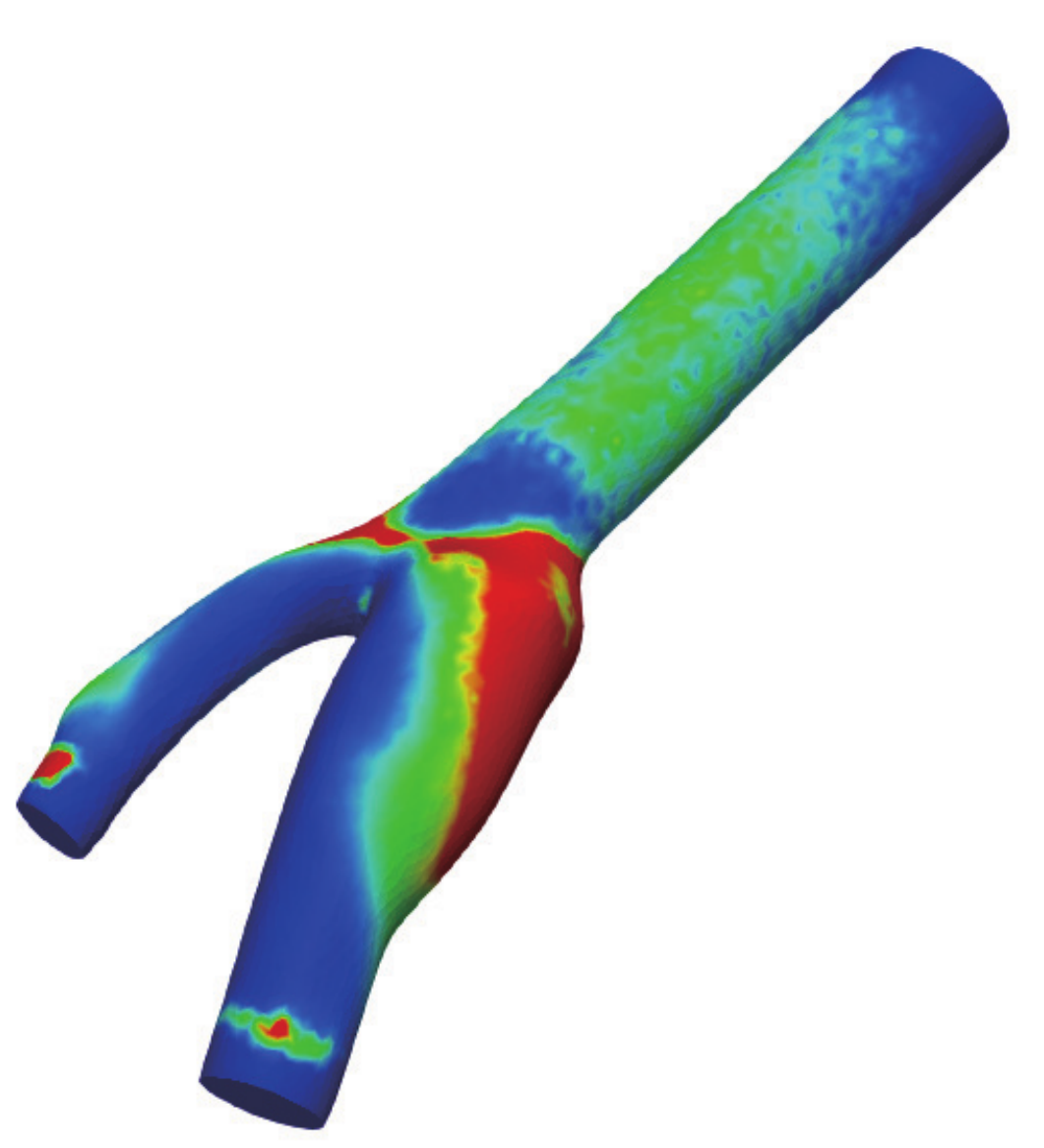}
        \end{subfigure}
        \caption*{\text{(b) OSI}}
    \end{subfigure}
    \caption{Hemodynamics in carotid artery (deformable wall with shell model): (a) TAWSS and (b) OSI 
    distributions of the fifth cardiac cycle.}
    \label{fig: carotid-VIPO-shell-TAWSS-OSI}
\end{figure}

In Fig.\ref{fig: carotid-VIPO-shell-TAWSS-OSI} the spatial distributions of TAWSS and OSI over the fifth cardiac cycle 
are visualized, both of which are widely used to assess hemodynamic risk factors. 
TAWSS represents the temporal average of the WSS over a full cardiac cycle \cite{schoenborn2022fluid}, defined as
\begin{equation}\label{eq: TAWSS}
  \mathrm{TAWSS} = \frac{1}{T}\int_{0}^{T} \left\vert \tau_{wall} \right\vert \mathrm{d}t,
\end{equation}
where $T$ is the duration of the cardiac cycle. OSI measures the directional variability of WSS during a 
cardiac cycle. High OSI values are often linked to regions with a high likelihood of atherosclerotic lesion 
formation \cite{schoenborn2022fluid}. OSI is calculated as
\begin{equation}\label{eq: OSI}
  \mathrm{OSI} = \frac{1}{2}(1-\frac{\left\vert \int_{0}^{T} \tau_{wall} \mathrm{d}t \right\vert}{\int_{0}^{T} \left\vert \tau_{wall} \right\vert \mathrm{d}t}).
\end{equation}
Elevated TAWSS is observed near the bifurcation apex, where the parent artery divides and redirects flow into 
the internal and external branches. This region experiences strong deceleration and velocity gradients, 
leading to locally intensified shear. Conversely, regions of low TAWSS and elevated OSI are predominantly found 
in the carotid bulb, where the flow undergoes recirculation and complex secondary motion. These disturbed flow 
patterns promote significant temporal variation in shear direction, as reflected by high OSI, and are known to 
be associated with increased risk of thrombus formation and atherogenesis due to the pro-inflammatory and pro-coagulant 
endothelial responses under such hemodynamic environments. These results 
confirm that the SPH shell model can effectively capture the FSI mechanisms and their 
impact on key hemodynamic indicators in anatomically realistic arterial geometries.

\begin{figure}
    \centering
    \begin{subfigure}[b]{\textwidth}
        \centering
        \begin{subfigure}[b]{0.4\textwidth}
            \includegraphics[width=\textwidth]{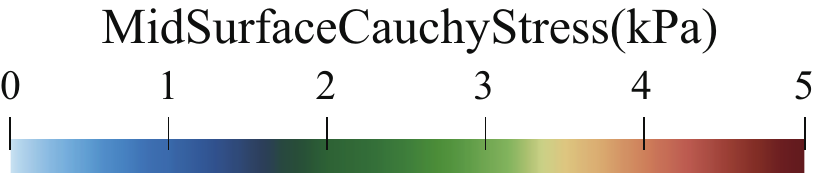}
        \end{subfigure}
        \begin{subfigure}[b]{0.4\textwidth}
            \includegraphics[width=\textwidth]{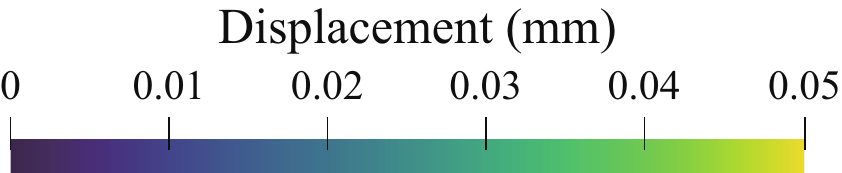}
        \end{subfigure}
    \end{subfigure}
    \begin{subfigure}[b]{\textwidth}
        \centering
        \begin{subfigure}[b]{0.4\textwidth}
            \includegraphics[width=\textwidth]{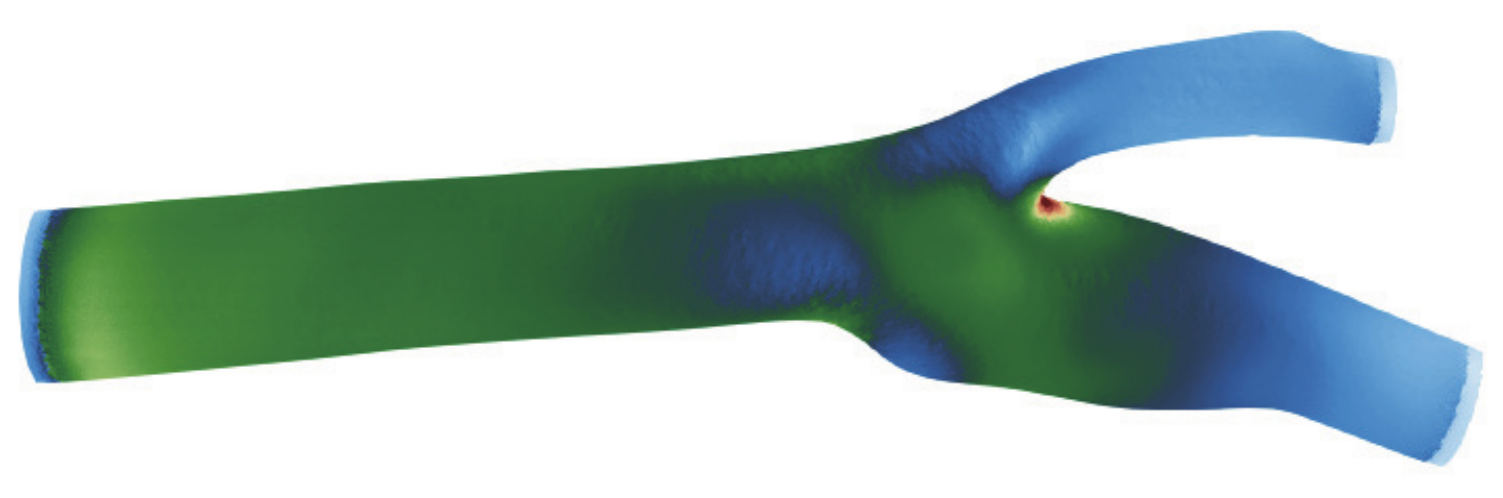}
        \end{subfigure}
        \begin{subfigure}[b]{0.4\textwidth}
            \includegraphics[width=\textwidth]{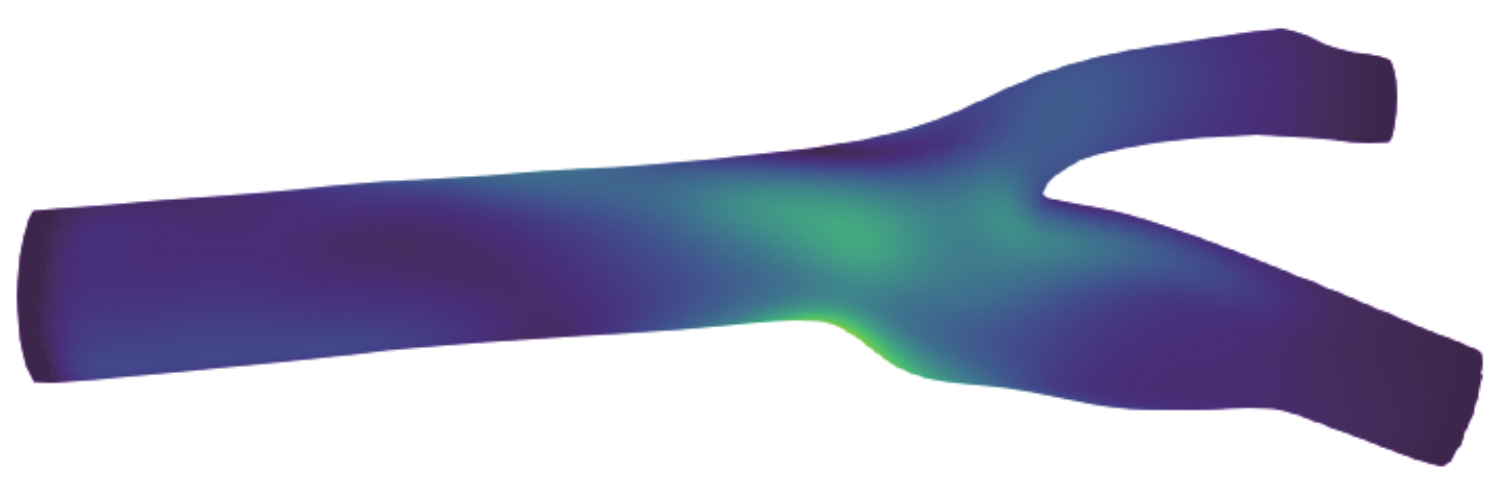}
        \end{subfigure}
        \caption*{\text{(a)} $t$ = 2.05s}
    \end{subfigure}
    \begin{subfigure}[b]{\textwidth}
        \centering
        \begin{subfigure}[b]{0.4\textwidth}
            \includegraphics[width=\textwidth]{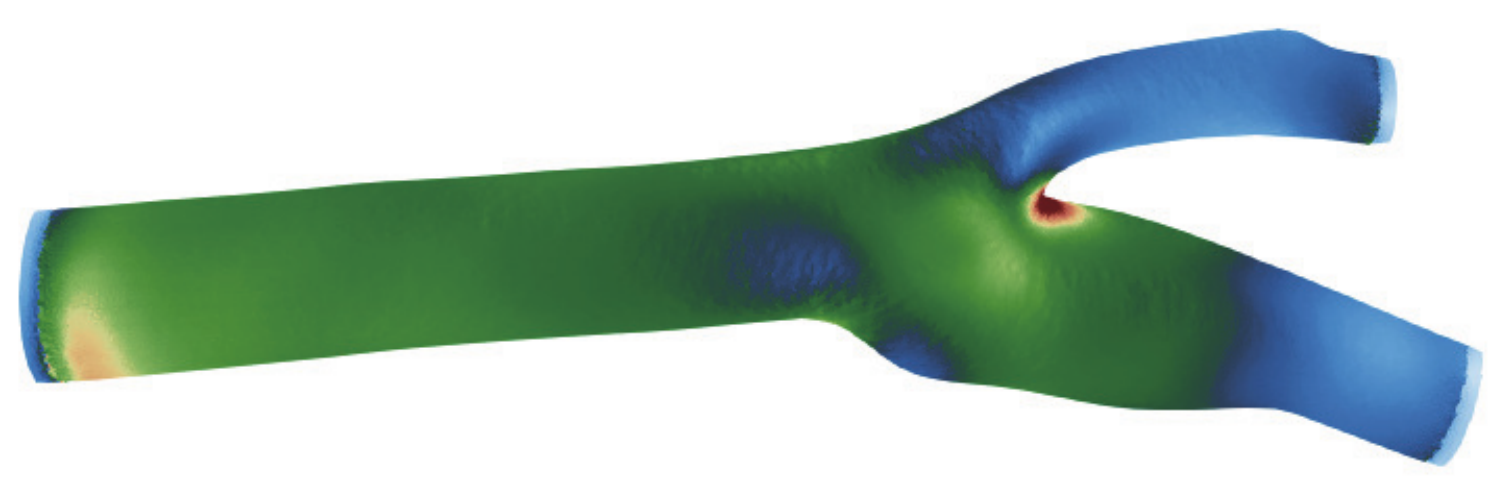}
        \end{subfigure}
        \begin{subfigure}[b]{0.4\textwidth}
            \includegraphics[width=\textwidth]{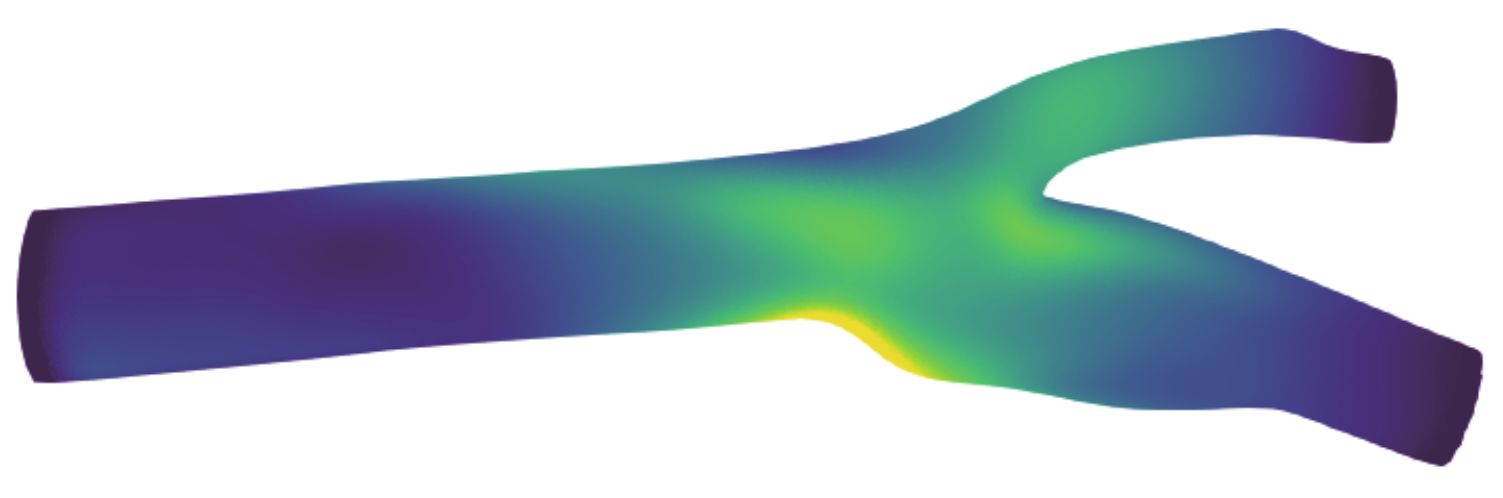}
        \end{subfigure}
        \caption*{\text{(b)} $t$ = 2.05s}
    \end{subfigure}
    \begin{subfigure}[b]{\textwidth}
        \centering
        \begin{subfigure}[b]{0.4\textwidth}
            \includegraphics[width=\textwidth]{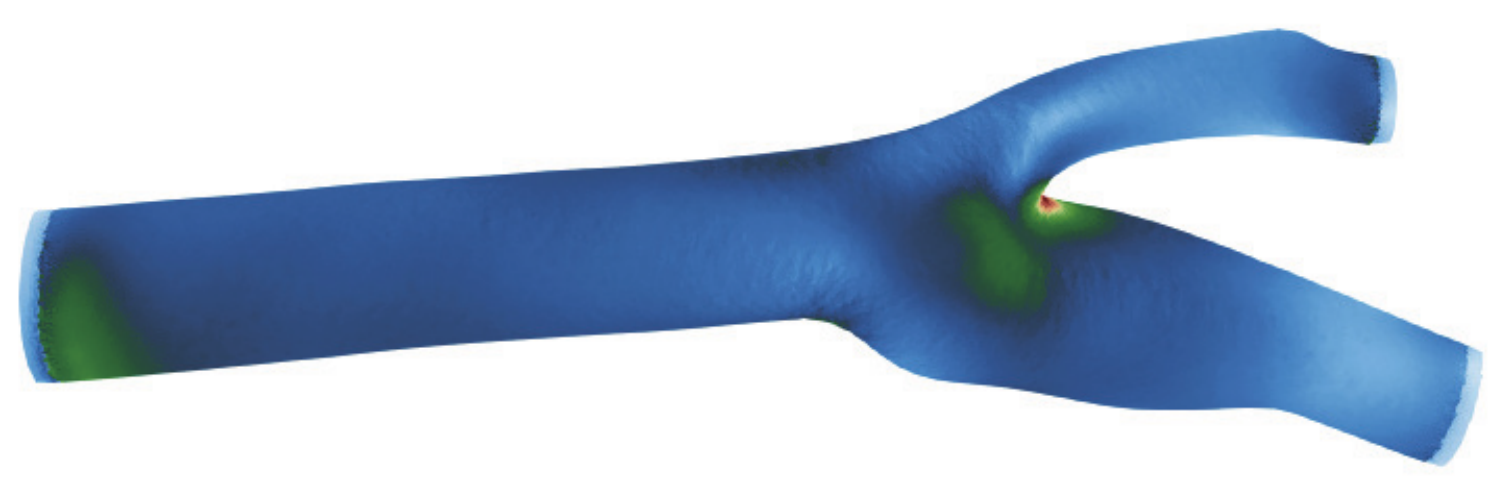}
        \end{subfigure}
        \begin{subfigure}[b]{0.4\textwidth}
            \includegraphics[width=\textwidth]{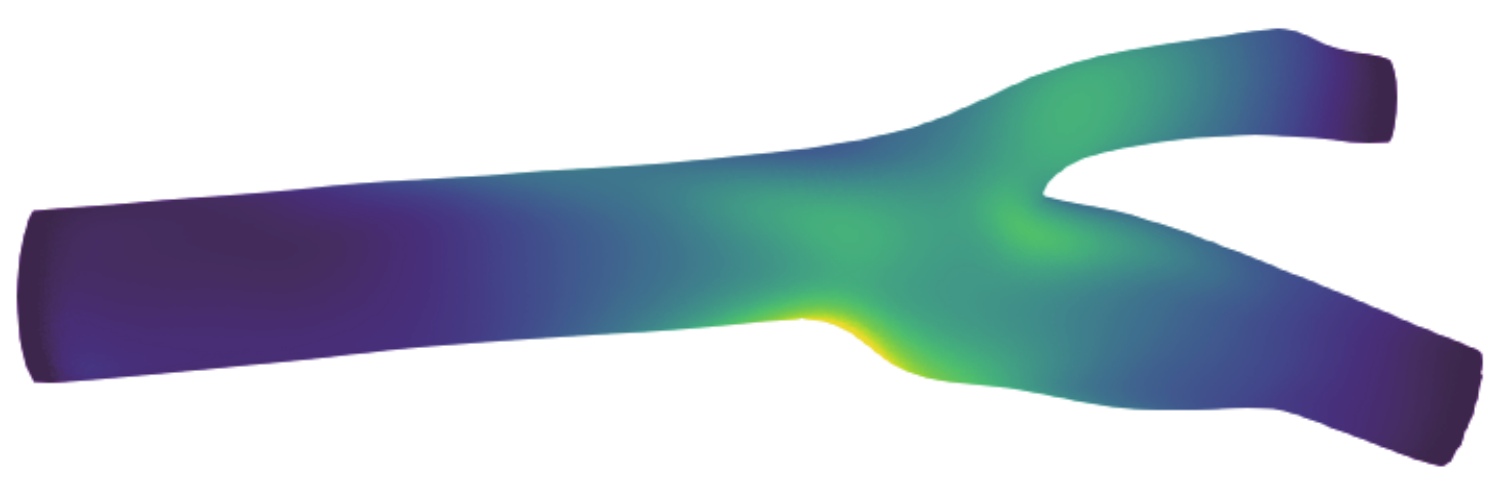}
        \end{subfigure}
        \caption*{\text{(c)} $t$ = 2.15s}
    \end{subfigure}
        \begin{subfigure}[b]{\textwidth}
        \centering
        \begin{subfigure}[b]{0.4\textwidth}
            \includegraphics[width=\textwidth]{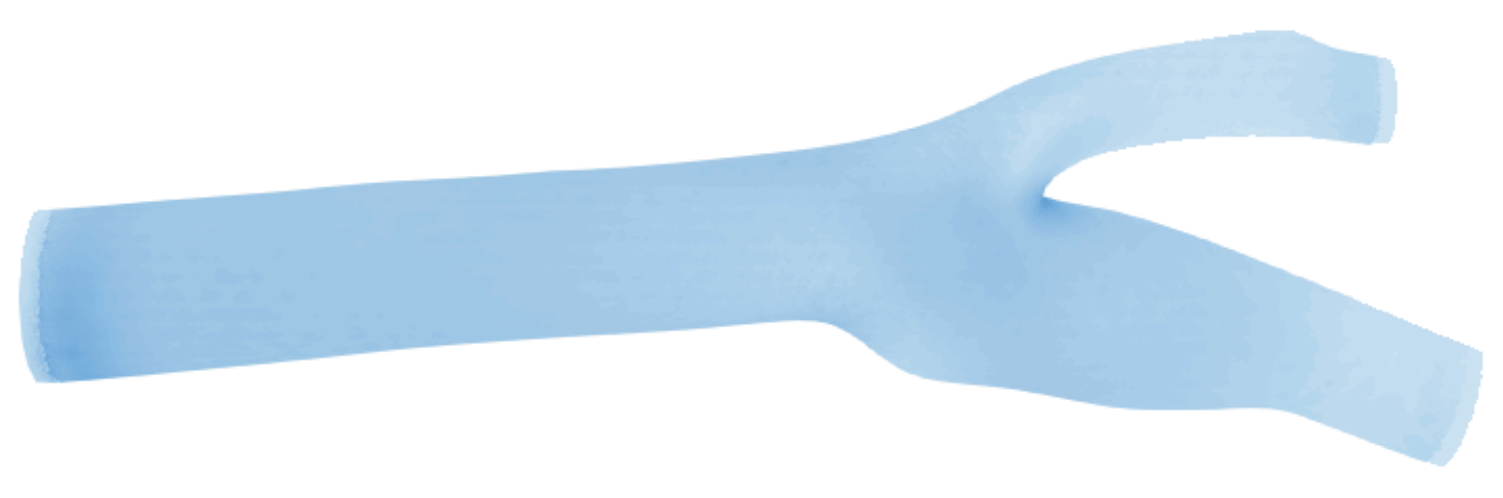}
        \end{subfigure}
        \begin{subfigure}[b]{0.4\textwidth}
            \includegraphics[width=\textwidth]{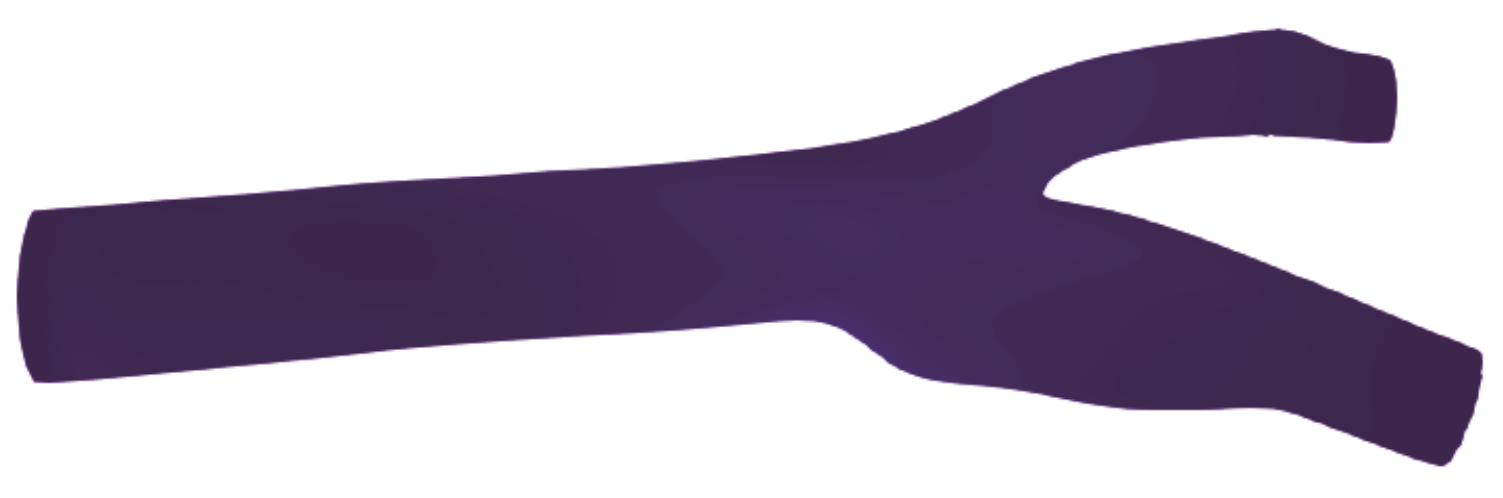}
        \end{subfigure}
        \caption*{\text{(d)} $t$ = 2.4s}
    \end{subfigure}
    \caption{Structural response of carotid artery wall (deformable wall with shell model): mid-surface Cauchy 
    stress (left) and displacement (right) distributions at four time instants in the fifth cardiac cycle.}
    \label{fig: carotid-VIPO-shell-stress}
\end{figure}

Fig.\ref{fig: carotid-VIPO-shell-stress} presents the distributions of mid-surface Cauchy stress and 
displacement of the shell. These results reflect the structural response of the arterial wall under pulsatile 
blood flow. Peak stress values are observed near the bifurcation apex and the flow-divider region, particularly 
during systole, where strong wall shear and pressure gradients coincide. These stress peaks indicate zones of 
dominant fluid-structure interaction and mechanical loading. In addition, the maximum shell deformation occurs 
in the carotid bulb and at the outer curvature of the bifurcation. The magnitude and spatial extent of 
deformation are in line with expected physiological wall compliance and are temporally synchronized with the 
systolic peak.

\subsection{Patient-specific case II: aorta}

\begin{figure}[htbp]
    \centering
    \includegraphics[width=12cm]{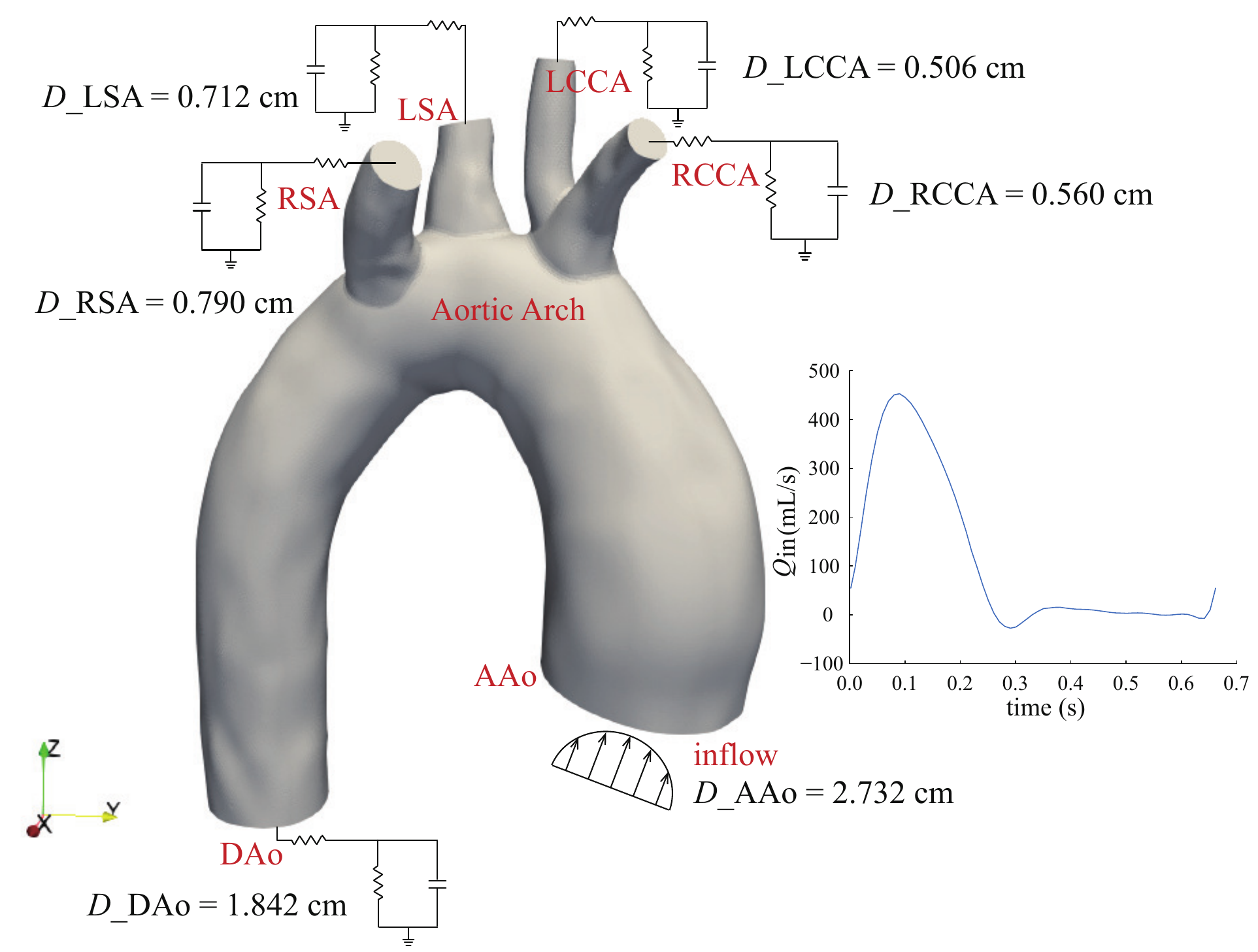}
    \caption{Patient-specific aorta case: illustration of geometry and boundary conditions.}
    \label{fig: aorta-geometry}
\end{figure}

In this section, the proposed fluid-shell interaction method is applied to a patient-specific aortic model. 
The anatomical geometry and boundary conditions are derived from a publicly available dataset in the Vascular 
Model Repository (ID: 0024\_H\_AO\_H), as illustrated in Fig.\ref{fig: aorta-geometry}. The ascending aorta 
(AAo) is prescribed as the inlet, where a pulsatile velocity profile with a cardiac period of $T = 0.66\text{s}$ is 
imposed. This inflow condition follows a parabolic velocity distribution whose temporal variation is described 
analytically by Eq.\ref{eq: aorta-vinlet}, and the corresponding volumetric flow rate is visualized in 
Fig.\ref{fig: aorta-geometry}. Five distal branches are left common carotid artery (LCCA), right common 
carotid artery (RCCA), left subclavian artery (LSA), right subclavian artery (RSA) and descending aorta (DAo), 
and all of them are treated as outlets. A three-element Windkessel model is employed at each outlet to represent 
the downstream vascular impedance, with the specific parameters listed in Table \ref{table: aorta-windkessel}.

\begin{equation}\label{eq: aorta-vinlet}
  v_{x, \text{avg}} = 5.0487+\sum_{i=1}^{8}[a_i\cos(\omega it) + b_i \sin(\omega it)],
\end{equation}
where the coefficients are
\begin{align*}\label{eq: aorta-coefficients}
  a &= [4.5287,\ -4.3509,\ -5.8551,\ -1.5063,\ 1.2800,\ 0.9012,\ 0.0855,\ -0.0480],\\
  b &= [-8.0420,\ -6.2637,\ 0.7465,\ 3.5239,\ 1.6283,\ -0.1306,\ -0.2738,\ -0.0449],\\
  \omega &= 2 \pi.
\end{align*}

\begin{table}[]
\centering
\caption{Parameters of Windkessel model for the patient-specific aorta.} \label{table: aorta-windkessel}
\begin{tabular}{cccc}
\hline
 & $R_p$ ($\text{kg} \cdot \text{m}^{-4}\text{s}^{-1}$) & $C$ ($\text{m}^4\text{s}^2 \cdot \text{kg}^{-1}$) & $R_d$ ($\text{kg} \cdot \text{m}^{-4}\text{s}^{-1}$) \\ \hline
LCCA & 7.13 $\times 10^7$ & 8.26 $\times 10^{-10}$ & 1.20 $\times 10^9$ \\
RCCA & 7.13 $\times 10^7$ & 8.26 $\times 10^{-10}$ & 1.20 $\times 10^9$ \\
LSA  & 6.02 $\times 10^7$ & 9.79 $\times 10^{-10}$ & 1.01 $\times 10^9$ \\
RSA  & 6.89 $\times 10^7$ & 8.55 $\times 10^{-10}$ & 1.16 $\times 10^9$ \\
DAo  & 9.80 $\times 10^6$ & 6.02 $\times 10^{-9}$  & 1.65 $\times 10^8$ \\ \hline
\end{tabular}
\end{table}

The blood is modeled as a weakly compressible Newtonian fluid with a density of 
$\rho_f = 1060 \text{ kg}/\text{m}^3$ and a dynamic viscosity of $\eta_f = 0.0035 \text{ Pa} \cdot \text{s}$. 
The material properties of the aorta wall are adopted from Ref.\cite{lu2024gpu}, with a solid density of 
$\rho_s = 1000 \text{ kg}/\text{m}^3$, Young's modulus $ E= 0.75\text{ MPa}$ and Poisson's ratio of 0.49. 
The wall is represented using a shell model with a uniform thickness of 0.25 cm \cite{liu2015evolution}. 
For this simulation, the initial particle spacing for both fluid and solid domains is set to $dp^0 = 0.06$ cm. 
The total number of particles used is 327,874 for the fluid domain at the beginning of the simulation 
(which will change as a result of particle injection and deletion) and 34,433 for the shell structure. 
The corresponding wall-clock time for the simulation is 44,493 seconds on a 32-core CPU, which is comparable 
to the computational time reported for a similar aorta case using the ALE method on a 388-core CPU in Ref.\cite{lu2024gpu}.

\begin{figure}[htbp]
    \centering
    \begin{subfigure}[b]{1.0\textwidth}
        \includegraphics[width=\textwidth]{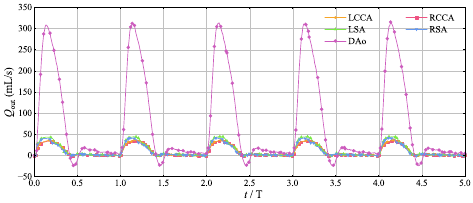}
        \caption{}
    \end{subfigure}
    \begin{subfigure}[b]{1.0\textwidth}
        \includegraphics[width=\textwidth]{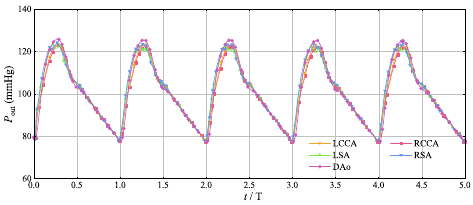}
        \caption{}
    \end{subfigure}
    \caption{Patient-specific aorta case: (a) volume flow rate and (b) pressure in five cardiac cycles at the outlets.}
    \label{fig: aorta-outlet-PandQ-5cycles}
\end{figure}

\begin{figure}[htbp]
    \centering
    \includegraphics[width=14cm]{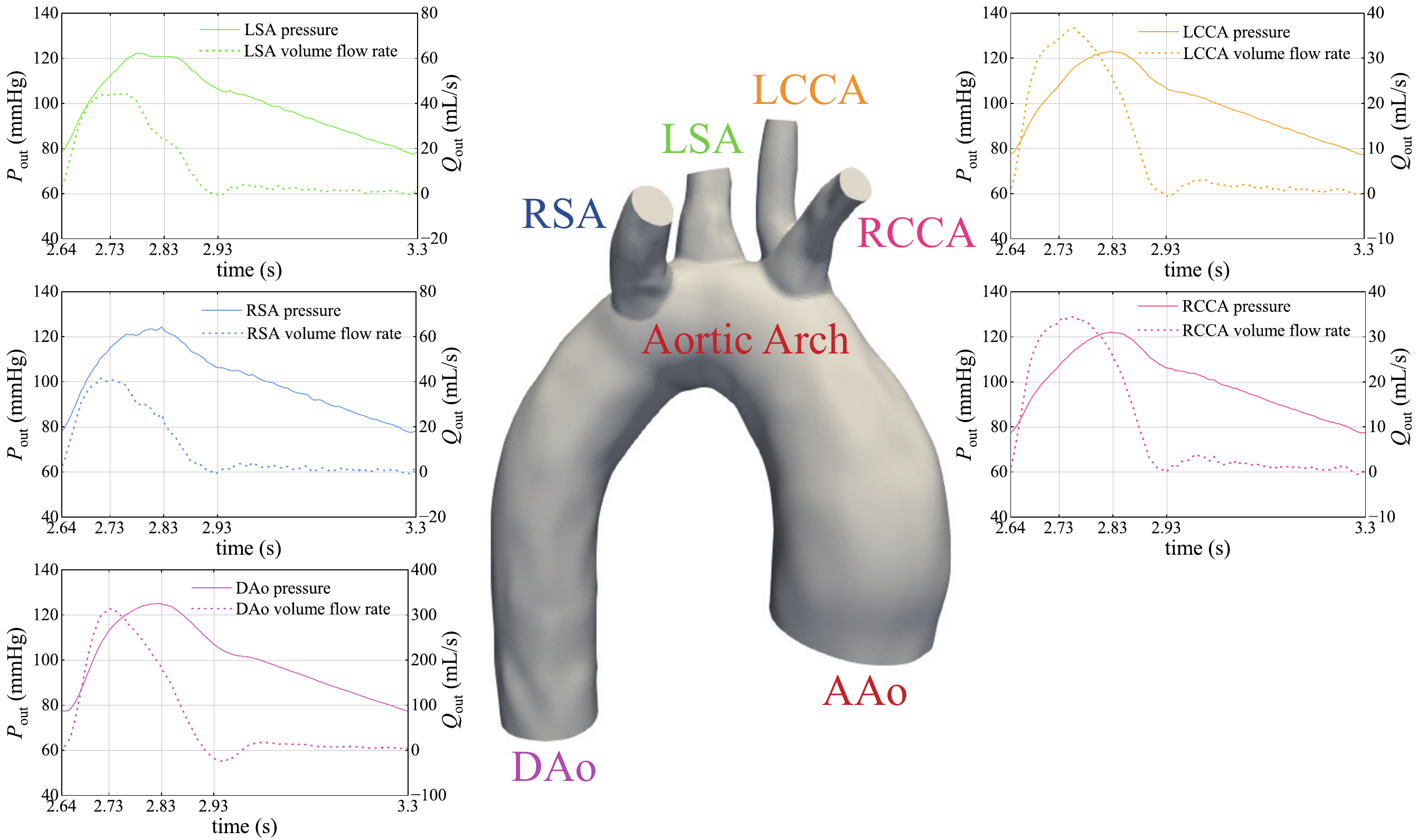}
    \caption{Patient-specific aorta case: volume flow rate and pressure during the fifth cardiac cycle at the outlets.}
    \label{fig: aorta-outlet-PandQ-5th}
\end{figure}

Fig.\ref{fig: aorta-outlet-PandQ-5cycles} presents the temporal evolution of volume flow rate and pressure at 
the five outlets over five cardiac cycles. It can be seen that the outlet flow rates closely follow the pattern 
of the inlet waveform. The outlet pressures exhibit physiologically consistent profiles and rapidly reach 
periodic steady states. To further illustrate the hemodynamic behavior at each branch, 
Fig.\ref{fig: aorta-outlet-PandQ-5th} displays the time histories of the pressure and volume flow rate at the 
five outlets during the fifth cardiac cycle. The flow rate curves capture the typical phases of ventricular 
ejection (systole) and relaxation (diastole), while the pressure waveforms demonstrate a slight phase lag 
relative to the peak flow, consistent with the Windkessel model's capacitive response. Based on the waveform 
characteristics, four representative time instants are selected for detailed analysis of flow and structural 
responses: peak flow at 2.73s, peak pressure at 2.83s, onset of diastole at 2.93s and onset of systole at 3.3s.

\begin{figure}[htbp]
    \centering
    \begin{subfigure}[b]{0.49\textwidth}
        \includegraphics[width=\textwidth]{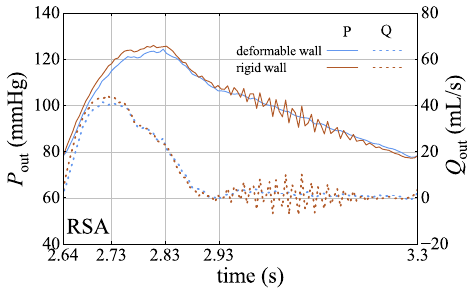}
        \caption{}
    \end{subfigure}
    \begin{subfigure}[b]{0.49\textwidth}
        \includegraphics[width=\textwidth]{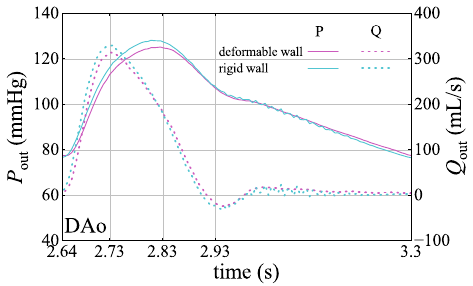}
        \caption{}
    \end{subfigure}
    \caption{Patient-specific aorta case: comparison of volume flow rate and pressure between deformable-wall 
    and rigid-wall assumptions during the fifth cardiac cycle at (a) RSA and (b) DAo.}
    \label{fig: aorta-outlet-PandQ-5th-deformableVSrigid}
\end{figure}

To further evaluate the physiological relevance of the FSI model, a comparative study is conducted between 
deformable-wall and rigid-wall assumptions under identical inflow and outlet boundary conditions. As shown 
in Fig.\ref{fig: aorta-outlet-PandQ-5th-deformableVSrigid}, the volume flow rate and pressure at the outlets 
are compared during the fifth cardiac cycle. For clarity, two representative outlets are selected: RSA, 
which features a smaller cross-sectional area; and DAo, which dominates the downstream flow distribution. 
The rigid-wall model induces noticeably higher-frequency oscillations, particularly in the RSA. 
These oscillations are attributed to the absence of wall compliance, which otherwise buffers pressure wave 
propagation and stabilizes flow fluctuations. This exaggerated pulsatility may lead to non-physiological 
wall shear stresses and unfavorable hemodynamic conditions, potentially contributing to vascular dysfunction 
or remodeling in clinical settings. In contrast, the deformable-wall model yields smoother pressure and 
flow profiles, better aligning with physiological observations.

\begin{figure}[htbp]
    \centering
    \includegraphics[width=15cm]{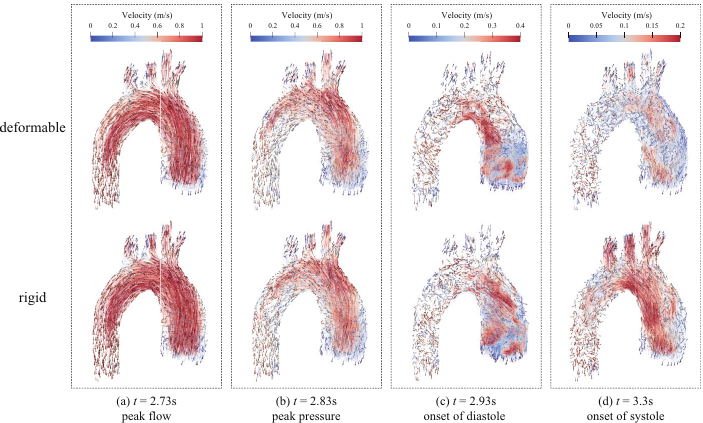}
    \caption{Hemodynamics in the patient-specific aorta: instantaneous velocity streamlines at four representative 
    time instants during the fifth cardiac cycle. Color bars vary between time points to better represent 
    transient flow features.}
    \label{fig: aorta-deformableVSrigid-velocity}
\end{figure}

\begin{figure}[htbp]
    \centering
    \includegraphics[width=15cm]{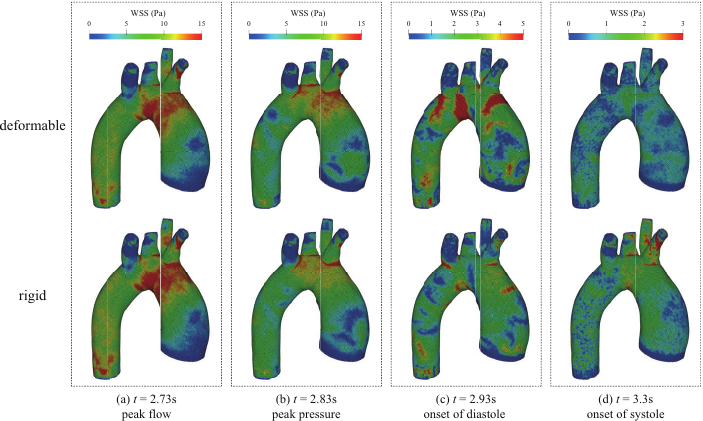}
    \caption{Hemodynamics in the patient-specific aorta: instantaneous WSS at four representative time 
    instants during the fifth cardiac cycle. Color bars vary between time points to better represent 
    transient flow features.}
    \label{fig: aorta-deformableVSrigid-WSS}
\end{figure}

\begin{figure}[htbp]
    \centering
    \includegraphics[width=12cm]{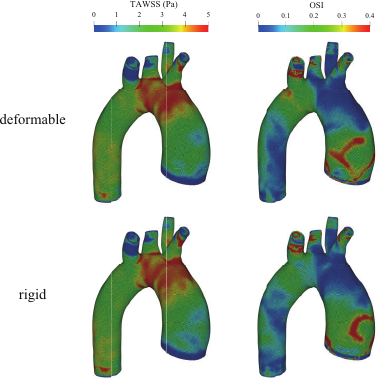}
    \caption{Hemodynamics in the patient-specific aorta: TAWSS and OSI distributions of the fifth cardiac cycle.}
    \label{fig: aorta-deformableVSrigid-TAWSS-OSI}
\end{figure}

Fig.\ref{fig: aorta-deformableVSrigid-velocity} and Fig.\ref{fig: aorta-deformableVSrigid-WSS} present a 
comparative analysis of instantaneous velocity streamlines and WSS distributions between deformable and 
rigid wall assumptions at four representative time instants during the fifth cardiac cycle. To effectively 
capture transient features, the color bar ranges vary across subplots. For the deformable wall case,  
at $t = 2.73\text{s}$, corresponding to peak systole, a strong jet-like flow initiates from the AAo and propagates 
long the outer curvature of the aortic arch. Elevated WSS values are observed near the bifurcations of the 
left and right CCAs and the near aortic arch, indicating strong shear interactions resulting from the 
velocity gradients. At $t = 2.83\text{s}$, as the flow rate declines, both velocity and WSS are reduced, reflecting 
the post-systolic attenuation of flow. At $t = 2.93\text{s}$, the onset of diastole is characterized by flow reversal 
near the AAo inlet. This reversal induces complex secondary flows and prominent vortical structures within 
the aortic arch. Correspondingly, WSS decreases and becomes more localized, especially in regions with strong 
geometric curvature. At $t = 3.3\text{s}$, the diastole phase ends and systole begins anew. The flow remains weak and 
unsteady, with overall low velocity and WSS levels observed throughout the domain. 
Comparatively, the rigid wall case exhibits similar velocity structures at peak systole and peak pressure. 
However, during diastole, the absence of wall compliance results in more intense recirculation and higher 
local velocity magnitudes, indicating reduced damping capacity, as also reflected by the pronounced oscillations 
in flow and pressure waveforms (see Fig.\ref{fig: aorta-outlet-PandQ-5th-deformableVSrigid}). 
The WSS distributions under the rigid assumption reveal distinct discrepancies in both magnitude and spatial 
localization: high WSS regions appear in different anatomical zones, and regional low WSS zones are more 
extensive and disorganized, especially during the diastolic phase. 
Additionally, Fig.\ref{fig: aorta-deformableVSrigid-TAWSS-OSI} illustrates the spatial distributions of TAWSS 
and OSI over the fifth cardiac cycle. In both deformable and rigid wall cases, high TAWSS is observed at 
major bifurcation sites, such as the origins of the carotid and subclavian arteries, where abrupt flow division 
and redirection generate high shear forces. Conversely, regions with low TAWSS and elevated OSI are primarily 
located near the proximal AAo and the origin of the DAo, which are known to correlate with disturbed and 
oscillatory flow patterns. Notably, these hemodynamic features are clinically relevant, as the coexistence of 
low TAWSS and high OSI has been implicated in potential vascular pathologies such as aortic aneurysm and dissection, 
due to their role in promoting endothelial dysfunction and localized wall weakening. Compared to the rigid 
wall model, the deformable wall case differs in TAWSS and OSI distributions. In light of the previously 
observed flow field and WSS variations, these findings underscore the physiological relevance of incorporating 
arterial wall compliance in cardiovascular simulations. The compliant wall modulates flow inertia and attenuates 
shear oscillations during critical phases such as flow deceleration and reversal, whose mechanical environments 
are known to influence endothelial function and mechanotransduction. In contrast, the rigid-wall assumption 
tends to exaggerate hemodynamic extremes, potentially misrepresenting sites at risk for vascular remodeling, 
aneurysm formation, or dissection initiation. Thus, neglecting wall compliance may lead to inaccurate assessment 
of disease-prone regions, limiting the predictive value of such models in clinical and research settings.

\begin{figure}[htbp]
    \centering
    \includegraphics[width=15cm]{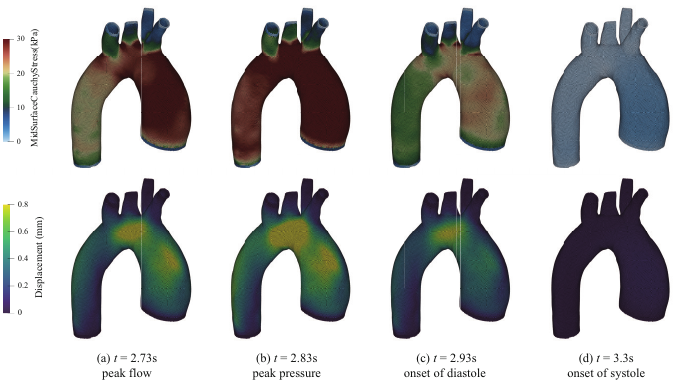}
    \caption{Structural responses of the patient-specific aorta: mid-surface Cauchy stress and 
    displacement distributions in the arterial shell at four representative time instants during the 
    fifth cardiac cycle.}
    \label{fig: aorta-shell-displacement-stress}
\end{figure}

Fig.\ref{fig: aorta-shell-displacement-stress} presents the structural responses of the patient-specific aorta, 
specifically the distributions of mid-surface Cauchy stress and displacement in the shell structure at four key 
time instants during the fifth cardiac cycle. From $t = 2.73\text{s}$ to $2.83\text{s}$, the increase in blood pressure leads 
to a noticeable rise in both stress and displacement magnitudes, with peak values concentrated from the AAo 
inlet to the aortic arch. After the peak pressure phase, these values rapidly decline, reflecting the 
corresponding hemodynamic unloading. 

\section{Conclusions} \label{section: conclusions}

In this study, we extend a reduced-dimensional shell-based SPH model to simulate blood flow in thin-walled 
deformable vessels, targeting applications in cardiovascular hemodynamics.

First, Poiseuille flow in a straight channel is employed as a benchmark to validate the fluid dynamics 
capability of the shell model, with comparisons to the traditional full-dimensional volume model serving as the 
wall boundary. The relative errors in the peak axial velocity at the midsection are 1.14\% and 0.65\% for the 
2D volume and shell models, respectively, and 0.48\% and 0.79\% for their 3D counterparts. These results 
confirm that the shell model can accurately represent the wall boundary in SPH-based fluid simulations. 
In addition, we verify the proper implementation of pressure outlet boundary conditions, including 
resistance-type and Windkessel models, by comparing SPH results with analytical solutions.

To evaluate the performance of the shell model in FSI scenarios, we conduct convergence tests using a T-shaped 
deformable vessel. The results indicate that the shell model achieves faster convergence in solid mechanics 
than the volume model, while maintaining comparable accuracy in both fluid dynamics and structural response. 
Furthermore, we examine the influence of wall compliance by comparing deformable and rigid wall configurations 
with shell model. The resulting differences in flow transition and hemodynamic indices highlight 
the necessity of incorporating FSI effects in cardiovascular modeling, rather than relying on rigid-wall 
assumptions.

Finally, the proposed shell model is applied to two patient-specific vascular geometries. In the carotid artery 
case, rigid-wall simulations using both volume and shell SPH models are compared with FVM results from ANSYS 
Fluent, exhibiting excellent agreement. The wall is then modeled as a deformable shell, and corresponding 
stress and displacement distributions are evaluated. In the second case, the aorta is simulated using the shell 
model with a three-element Windkessel boundary at the outlets. The predicted pressure and flow waveforms at the 
outlets align well with physiological expectations, which further validates the effectiveness of the proposed 
approach for large-scale, patient-specific cardiovascular simulations. Also, a comparative analysis of 
hemodynamic parameters is conducted between the deformable wall and rigid-wall assumptions. The results reveal 
that wall compliance significantly influences the estimation of regions at risk for vascular pathologies. 
These findings highlight the physiological fidelity and clinical relevance of incorporating wall deformability.

\section*{Acknowledgement}
C.X. Zhao is fully supported by the China Scholarship Council (CSC) (No.202206280028). 
D. Wu and X.Y. Hu would like to express their gratitude to Deutsche Forschungsgemeinschaft for their 
sponsorship of this research under grant number DFG HU1527/12-4. 
%
%
\section*{CRediT authorship contribution statement}
{\bfseries  Chenxi Zhao:} Investigation, Methodology, Visualization, Validation, Formal analysis, 
Writing - original draft, Writing - review \& editing;
{\bfseries  Dong Wu:} Investigation, Methodology, Validation;
{\bfseries  Oskar J. Haidn:} Supervision;
{\bfseries  Xiangyu Hu:} Supervision, Methodology, Writing - review \& editing.

%
%
\section*{Declaration of competing interest }
The authors declare that they have no known competing financial interests 
or personal relationships that could have appeared to influence the work reported in this paper.
%
%
\section*{Data availability}
The code is open source on \href{https://github.com/Xiangyu-Hu/SPHinXsys}{https://github.com/Xiangyu-Hu/SPHinXsys}.
%
%
\clearpage
\bibliography{reference}
%
%
\end{document}